\documentclass[12pt]{article}
\usepackage[english]{babel}
\usepackage[utf8]{inputenc}
\usepackage{amsfonts,amssymb,amsmath, epsfig}
\usepackage{color,graphicx,graphics,psfrag}
\usepackage[lofdepth,lotdepth]{subfig}
\usepackage{amsmath,amstext,amssymb,amsfonts, amscd}
\usepackage[title]{appendix}
\usepackage{empheq}
\usepackage{hyperref}   

\def\Box{\leavevmode\vbox{\hrule
     \hbox{\vrule\kern4pt\vbox{\kern4pt}%
           \vrule}\hrule}}
\def\blackbox{\leavevmode\vrule height 5pt width 4pt depth 0pt\relax}
\def\endproof{\null\hfill {$\blackbox$}\bigskip}

\def\paragraph#1{{\bf #1\ }}

\newtheorem{lemma}{Lemma}[section]  
 
\newtheorem{theorem}[lemma]{Theorem}

\newtheorem{corollary}[lemma]{Corollary}

\newtheorem{definition}[lemma]{Definition}

\newtheorem{proposition}[lemma]{Proposition}

\newtheorem{remark}{Remark}[section]

\newtheorem{conject}{Conjecture}[section]

\title{From kinetic to fluid models of liquid crystals by the moment method} 
\author{Pierre Degond$^{1}$, Amic Frouvelle$^{2}$, Jian-Guo Liu$^{3}$} 
\date{} 
\begin{document}

\maketitle

\begin{center}
\emph{In memory of Bob Glassey}
\end{center}

\vspace{0.5cm}

\begin{center}
1- Institut de Mathématiques de Toulouse ; UMR5219 \\
Université de Toulouse ; CNRS\\
UPS, F-31062 Toulouse Cedex 9, France\\
email: pierre.degond@math.univ-toulouse.fr\\
$\mbox{}$ 
\end{center}

\begin{center}
  2- CEREMADE, CNRS, Université Paris-Dauphine\\
  Université PSL, 75016 Paris, France\\
  email: frouvelle@ceremade.dauphine.fr\\
  and\\
  CNRS, Université de Poitiers, UMR 7348\\
  Laboratoire de Mathématiques et Applications (LMA), 86000 Poitiers, France\\
  email: amic.frouvelle@math.univ-poitiers.fr\\
\end{center}

\begin{center}
3- Department of Physics and Department of Mathematics\\
Duke University\\
Durham, NC 27708, USA\\
email: jliu@phy.duke.edu
\end{center}

\vspace{0.5 cm}
\begin{abstract}
This paper deals with the convergence of the Doi-Navier-Stokes model of liquid crystals to the Ericksen-Leslie model in the limit of the Deborah number tending to zero. While the literature has investigated this problem by means of the Hilbert expansion method, we develop the moment method, i.e. a method that exploits conservation relations obeyed by the collision operator. These are non-classical conservation relations which are associated with a new concept, that of Generalized Collision Invariant (GCI). In this paper, we develop the GCI concept and relate it to geometrical and analytical structures of the collision operator. Then, the derivation of the limit model using the GCI is performed in an arbitrary number of spatial dimensions and with non-constant and non-uniform polymer density. This non-uniformity generates new terms in the Ericksen-Leslie model. 
\end{abstract}

\medskip
\noindent
{\bf Acknowledgments:} PD holds a visiting professor association with the Department of Mathematics, Imperial College London, UK. JGL acknowledges support from the Department of Mathematics, Imperial College London, under Nelder Fellowship award and the National Science Foundation under Grants DMS-1812573 and DMS-2106988. AF acknowledges support from the Project EFI ANR-17-CE40-0030 of the French National Research Agency.

\medskip
\noindent
{\bf Key words: } Doi equation, Maier-Saupe interaction, Q-tensor, Deborah number, Ericksen-Leslie equations, Oseen-Franck energy, order parameter, generalized collision invariant, 

\medskip
\noindent
{\bf AMS Subject classification: } 35Q35, 76A10, 76A15, 82C22, 82C70, 82D30, 82D60,
\vskip 0.4cm

%%%%%%%%%%%%%%%%%%%%%%%%%%%%%%%%%%%%%%%%%%%%%%%%%%%%%%%%%%%%%%%%
%%%%%%%%%%%%%%%%%%%%%%%%%%%%%%%%%%%%%%%%%%%%%%%%%%%%%%%%%%%%%%%%
%%%%%%%%%%%%%%%%%%%%%%%%%%%%%%%%%%%%%%%%%%%%%%%%%%%%%%%%%%%%%%%%
%%%%%%%%%%%%%%%%%%%%%%%%%%%%%%%%%%%%%%%%%%%%%%%%%%%%%%%%%%%%%%%%

\setcounter{equation}{0}
\section{Introduction}
%\label{sec:intro}

We consider the Doi kinetic model of liquid crystals coupled with the Navier-Stokes equation for the fluid solvent. We investigate the limit of the Deborah number tending to zero by means of a moment method. The limit model is a system of fluid equations named the Ericksen-Leslie model \cite{E_Zhang_MAA06, Kuzuu_Doi_JPhysSocJapan83, Wang_Zhang_Zhang_CPAM15}. In classical kinetic theory, there are two methods to derive fluid equations, the Hilbert expansion method \cite{Caflisch_fluid_CPAM80, Chapman_kinetic_16/17, Enskog_Uppsala17, Hilbert_Begrundung_1916/17} and the moment method \cite{Bardos_Golse_Levermore_JSP91, Mellet_fractional_2010}. However, for a number of kinetic models including the Doi kinetic model, only the Hilbert method can be used. Indeed, the moment method is subject to a condition on the number of conservation relations satisfied by the collision operator and this condition is not satisfied by the Doi model. This is why the Hilbert expansion method is the only method developed in the literature so far (see e.g. \cite{E_Zhang_MAA06, Kuzuu_Doi_JPhysSocJapan83, Wang_Zhang_Zhang_CPAM15}). In the present work, we address the question whether the moment method can be used for the Doi kinetic model. 

To make this question clearer, let us temporarily consider the Boltzmann equation of rarefied gases for which both methods work. The Boltzmann equation is historically the first kinetic model ever written and the most emblematic one \cite{Boltzmann_1872, Cercignani_Illner_Pulvirenti, Maxwell_1867}. It is schematically written
\begin{equation}
T f^\varepsilon := (\partial_t + v \cdot \nabla_x) f^\varepsilon = \frac{1}{\varepsilon} C(f^\varepsilon), 
\label{eq:Boltzmann}
\end{equation}
where $f^\varepsilon = f^\varepsilon(x,v,t)$ is the distribution function of particles at position $x$, velocity $v$ and time $t$, $\varepsilon \ll 1$ is the dimensionless Knudsen number and $C$ is the collision operator. In the fluid limit $\varepsilon \to 0$, we have (at least formally) $f^\varepsilon \to f^0$ with $C(f^0)=0$. Such equilibria $f^0$ are given by 
\begin{equation}
f^0 = \rho M_{u,T}, 
\label{eq:maxwellian}
\end{equation}
where $(\rho,u,T) \in (0,\infty) \times {\mathbb R}^n \times (0,\infty)$ ($n$ being the dimension) depend on $(x,t)$ and $M_{u,T}$: $v \mapsto M_{u,T}(v) \in (0,\infty)$ is a specific function of $v$ called a Maxwellian. The fluid limit requires finding equations that specify the dependence of $(\rho,u,T)$ with respect to $(x,t)$. 

Finding these equations requires dealing with the singular factor $\frac{1}{\varepsilon}$ in \eqref{eq:Boltzmann}. The most straightforward approach is to expand $f^\varepsilon$ in powers of $\varepsilon$: $f^\varepsilon = f^0 + \varepsilon f^1 + {\mathcal O}(\varepsilon^2)$, insert this expansion in \eqref{eq:Boltzmann} and cancel each power of $\varepsilon$ separately. This is the so-called Hilbert expansion method. The leading order term is $C(f^0)=0$ which recovers that $f^0$ is of the form \eqref{eq:maxwellian}. The next order gives $D_{f^0} C (f^1) = T f^0$, where $D_{f^0} C (f^1)$ is the derivative of $C$ (which is nonlinear) with respect to $f$ at $f^0$ applied to $f^1$. The existence of $f^1$ requires that $T f^0$ be in Im$(D_{f^0} C)$, the image of the operator $D_{f^0} C$. Under spectral properties of $D_{f^0} C$ which are satisfied in a large number of situations and which we will not detail here, we have Im$(D_{f^0} C) = $  ker$(D_{f^0} C^*)^\bot$ where 'ker' denotes the kernel and the exponent '$*$', the adjoint. Thus, the requirement on $Tf^0$ can be written 
\begin{equation}
\int T f^0 \psi dv = 0, \quad \forall \psi \in \mathrm{ker} (D_{f^0} C^*).
\label{eq:orthogonality}
\end{equation}
One can show that ker$(D_{f^0} C^*) = $ Span$\{ 1, \, v, \,|v|^2\}$, so that \eqref{eq:orthogonality} written for $\psi$ successively equal to $1$, $v$, $|v|^2$ gives rise to the system of compressible Euler equations, which thus constitutes the fluid limit of the Boltzmann equation. We note that this system is closed, because there are $n+2$ unknowns $(\rho,u,T)$ and $n+2$ equations (indeed, the dimension of ker$(D_{f^0} C^*)$ is $n+2$). 

However, there is a more direct route, which is to notice that the collision operator satisfies 
\begin{equation} 
\int C(f) \, \psi \, dv = 0, \, \forall f \, \, \Longleftrightarrow \, \,  \psi \in \mathrm{Span }\{ 1, \, v, \,|v|^2\}. 
\label{eq:CI_Boltz}
\end{equation}
A function $\psi$ that satisfies the left-hand side of \eqref{eq:CI_Boltz} is called a collision invariant. Property \eqref{eq:CI_Boltz} states that the only collision invariants are linear combinations of $1$, $v$ and $|v|^2$. Physically, this means that collisions conserve mass, momentum and energy and that these are the only conserved quantities. Thus, multiplying the Boltzmann equation \eqref{eq:Boltzmann} by $\psi \in$ Span$\{ 1, \, v, \,|v|^2\}$ we get $\int T f^\varepsilon \, \psi \, dv = 0$. This removes the $\frac{1}{\varepsilon}$ singularity and allows us to pass to the limit $\varepsilon \to 0$. This leads to $\int T f^0 \, \psi \, dv = 0$ which, again, gives rise the system of compressible Euler equations. Integrals of the type $\int f^0 \, \psi \, dv$ are called ``moments'', hence the name ``moment method'' for this method. This should not be confused with the numerical moment method which consists of approximating the distribution function by a finite number of moments. However, the two are obviously linked. 

We note that the Hilbert expansion method works provided Im$(D_{f^0} C) = $  ker$(D_{f^0} C^*)^\bot$, which is satisfied in a large number of cases. On the other hand, the success of the moment method relies on the requirement that the space of collision invariants has the same dimensions as the number of free parameters in the equilibrium distribution function (here $(\rho,u,T)$). This requirement is not satisfied in general and specifically for the Doi model. So, should we abandon the moment method for such instances? The goal of this paper is to show that the moment method can still be used for the Doi model. However, this necessitates to revisit the concept of collision invariant and to design a weaker concept: the ``generalized collision invariant'' or GCI. 

The GCI concept has first been introduced in \cite{Degond_Motsch_M3AS08} for the Vicsek model \cite{Vicsek_novel_PRL95}, a model of self-propelled particles moving at constant speed and tending to align their direction of motion with their neighbors. Here, the absence of conservation relations beyond the conservation of mass is a consequence of the active character of the particles, i.e. the fact that they sustain a constant speed motion in all circumstances. In \cite{Degond_Motsch_M3AS08}, thanks to the GCI concept, the fluid limit of the Vicsek model is derived and gives rise to a new kind of fluid dynamics model, now referred to as the Self-Organized Hydrodynamic model \cite{Degond_hydrodynamic_MAA13}. Since then, the GCI concept has been applied to a variety of collective dynamics models \cite{Degond_eal_JNLS20, Degond_macroscopic_CMS15, Degond_etal_MMS18, Degond_Merino_M3AS20, Degond_coupled_JMFM19, Frouvelle_M3AS12}. The present work is its first application to visco-elastic fluid models. 

Visco-elastic fluids have been the subject of an abundant literature (see e.g. \cite{Ball_mathematics, Ball_Feireisl_Otto_Springer17, DeGennes_Prost, Doi_Edwards, Giga_Novotny_handbook, Wang_Zhang_Zhang_ActaNumer22} for reviews). The Doi model is one of the most fundamental models of visco-elastic fluids \cite{Doi_Edwards}. It models the dynamics of an assembly of polymer molecules flowing in an incompressible fluid (the solvent). The polymer molecules are assumed to be rigid spheroids mutually interacting through alignment and subject to noise. They are represented by a distribution function of their position and orientation. After Onsager and Maier-Saupe \cite{Maier_Saupe_1959, Onsager_effects_1949}, alignment accounts for the volume exclusion interaction between the molecules. Alignment is supposed to be of nematic type, i.e. invariant if the head and tail of the molecules are flipped. To account for this, following Landau and de Gennes \cite{DeGennes_Prost}, the interaction is written in terms of the so-called Q-tensor which is a quadratic quantity of the orientation and thus, respects this invariance. The fluid solvent is modelled by the incompressible Navier-Stokes equations. Polymer molecules are transported by the fluid and rotated by the fluid gradients. In turn, the polymer molecules influence the fluid through extra-stresses whose expressions involve the polymer distribution function. The mathematical theory of the Doi-Navier-Stokes system has been investigated in \cite{Lin_Liu_Zhang_CPAM07, Otto_Tzavaras_CMS08, Zhang_Zhang_SIMA08} and for active particles, in \cite{Chen_Liu_JDE13}. 

In the Doi model, alignment occurs at a rate characterized by a dimensionless parameter, the Deborah number. When this parameter goes to zero, the distribution of polymer molecule orientations gets a definite profile which has analogies with the Maxwellian velocity distribution of gas dynamics \eqref{eq:maxwellian}. It depends on two parameters, the polymer density $\rho$ and the polymer molecules average orientation $\Omega$ which are functions of space and time. In the case of a constant density $\rho$, it is shown in \cite{E_Zhang_MAA06, Kuzuu_Doi_JPhysSocJapan83, Wang_Zhang_Zhang_CPAM15} that the mean orientation satisfies a transport-diffusion equation. Its coupling with the Navier-Stokes equations leads to the so-called Ericksen-Leslie system \cite{Ericksen_liquid_ARMA91, Leslie_constitutive_1966}. The convergence is formal in \cite{E_Zhang_MAA06, Kuzuu_Doi_JPhysSocJapan83} and rigorous in \cite{Wang_Zhang_Zhang_CPAM15}. In all cases, the method relies on the Hilbert expansion. There is an abundant mathematical literature on the Ericksen-Leslie system per se \cite{Huang_Lin_Wang_CMP14, Lin_Lin_Wang_ARMA10, Lin_Liu_nonparabolic_CPAM95, Lin_Liu_existence_ARMA00, wang2013well}. 

Here, our goal is to provide a formal convergence proof of the Doi model to the Ericksen-Leslie model using the moment method and the new generalized collision invariant concept. Specifically, we will derive the appropriate GCI concept, discuss its rationale and its relation to ker$(D_{f^0} C^*)$ which is the central object in the Hilbert expansion method. There are several motivations to develop a moment method even if a Hilbert expansion theory already exists. The first one is that the GCI concept has an underlying geometrical structure which we will highlight. In view of Noether's theorem relating conservations to invariance under transformation groups, this may lead to new useful structural invariance properties of the Doi collision model. The second reason is that a mathematical theory based on the moment method often requires less regularity than the Hilbert expansion method (compare e.g. \cite{Bardos_Golse_fluid_CPAM93} with \cite{Caflisch_fluid_CPAM80}). This potentially opens the ways to simpler convergence proofs from the Doi to the Ericksen-Leslie models. The third reason is that the moment method naturally leads to the development of efficient numerical methods \cite{Grad_kinetic_CPAM49, levermore1996moment} which might enable us to handle the complexity of the Doi kinetic model in a systematic way.   

Aside to this main goal, we will also pursue two secondary goals. The first one is to provide a treatment of the small Deborah number limit in arbitrary dimension. So far, this has only been done in dimension $3$. This extension is made possible by Wang and Hoffman \cite{Hoffman_Wang_CMS06} who have determined the spatially uniform equilibria in any dimension. Although dimension three is the physically relevant case, there are several reasons for considering an arbitrary dimension. The first one is that the use of dimension $3$ often conceals simple structures under dimension-specific concepts and notations. For instance, in many references, the use of the rotation operator traditionally denoted by  ${\mathcal R}$ whose construction depends on the cross-product and is dimension $3$-specific is unnecessary and cumbersome. As argued in \cite{charbonneau2013dimensional}, the use of an arbitrary dimension often reveals hidden and interesting mathematical properties. Finally, fluid-dynamic equations are based on simple postulates that may be relevant for other objects. For instance, the Doi-Navier-Stokes model could describe flows of different types of information in an abstract space of large dimension. Of course, an information flow model cannot simply be a copy-paste of the Doi-Navier-Stokes model. However, the latter could constitute a good starting point on which further elaboration could be made. 

The second side goal is to investigate the effect of a spatially non-uniform density of polymer molecules. To the best of our knowledge, earlier work on the small Deborah number limit \cite{E_Zhang_MAA06, Kuzuu_Doi_JPhysSocJapan83, Wang_Zhang_Zhang_CPAM15} have assumed the density of polymer molecules to be constant. Investigation of Ericksen-Leslie models with non-uniform order parameter has been made in the literature \cite{Calderer_time_SIMA02, Calderer_Liu_SIAP00, Ericksen_liquid_ARMA91, Lin_nonlinear_CPAM89, Lin_nematic_CPAM91, Lin_global_ARMA15}, but none has explicitly linked this non-uniform order parameter to the non-uniform polymer density (as is should as we will see) and derived these models from kinetic theory. Non-uniform polymer density results in modifications of the equations for the mean director $\Omega$ and for the extra-stresses that will be highlighted in this work. 

The organization of this paper is as follows: Section \ref{sec:kinetic} gives an exposition of the Doi-Navier-Stokes model and the small Deborah number scaling. Section \ref{sec:main_result} is devoted to the statement of the main result, namely the formal convergence of the Doi-Navier-Stokes model to the Ericksen-Leslie model in the zero Deborah number limit. Section \ref{sec:local_equilibria} describes the local equilibria (i.e. the analogs of the Maxwellians \eqref{eq:maxwellian} for the Doi model). Section~\ref{sec:GCI} develops the GCI concept for the Doi model and discusses it. In Section \ref{sec:omega_abstract}, the limiting equations of the Doi model when the Deborah number tends to zero are derived. Conclusions and perspectives are drawn in Section \ref{sec:conclusion}. Auxiliary results stated in Sections~\ref{sec:kinetic}, \ref{sec:main_result}, \ref{sec:GCI} and \ref{sec:omega_abstract} are proved in appendices \ref{sec:app_Doi}, \ref{sec:app_sec_main_res}, \ref{sec:app_GCI} and \ref{sec:app_deriv_Om} respectively.

%%%%%%%%%%%%%%%%%%%%%%%%%%%%%%%%%%%%%%%%%%%%%%%%%%%%%%%%%%%%%%%%
%%%%%%%%%%%%%%%%%%%%%%%%%%%%%%%%%%%%%%%%%%%%%%%%%%%%%%%%%%%%%%%%
%%%%%%%%%%%%%%%%%%%%%%%%%%%%%%%%%%%%%%%%%%%%%%%%%%%%%%%%%%%%%%%%
%%%%%%%%%%%%%%%%%%%%%%%%%%%%%%%%%%%%%%%%%%%%%%%%%%%%%%%%%%%%%%%%

\setcounter{equation}{0}
\section{Kinetic model for rod-like polymer suspensions and scaling}
\label{sec:kinetic}

\subsection{The Doi equation}
%\label{subsec:Doi}

In this paper, we consider the Doi model \cite{DeGennes_Prost, Doi_Edwards, E_Zhang_MAA06, Kuzuu_Doi_JPhysSocJapan83, Otto_Tzavaras_CMS08, Wang_etal_PRE02, Wang_Zhang_Zhang_CPAM15}, where polymer molecules are identified as spheroids. We consider the semi-dilute regime \cite{Doi_Edwards, E_Zhang_MAA06, Wang_etal_PRE02} where a volume-exclusion interaction potential needs to be incorporated. We neglect the inertia of the polymer molecules. Following~\cite{E_Zhang_MAA06, Kuzuu_Doi_JPhysSocJapan83, Wang_etal_PRE02}, we describe the polymer molecules by a kinetic distribution function $f(x,\omega,t)$ where $x \in {\mathbb R}^n$ is the position, $\omega \in {\mathbb S}^{n-1}$ is the molecule orientation and $t\geq 0$ is the time. We let  ${\mathbb S}^{n-1}$ be the unit $(n-1)$-dimensional sphere and since $\omega$ and $-\omega$ refer to the same molecular orientation, we impose 
\begin{equation}
f(x,\omega,t)=f(x,-\omega,t).
\label{eq:symmetry}
\end{equation} 
Let $u(x,t)\in\mathbb{R}^n$ be the fluid velocity. In general, the dimension $n=2$ or $3$ but the theory will be developed for any value of $n$. The equation for $f$ (the so-called Doi equation) reads as follows:
\begin{eqnarray}
&& \hspace{-1cm} \partial_t f + \nabla_x \cdot (u f) + \nabla_\omega \cdot \big(f \, (\Lambda P_{\omega^\perp} E - W ) \omega \big) = D \, \nabla_\omega \cdot ( \nabla_\omega f + \frac{1}{k_B T} \, f \, \nabla_\omega U_f^R ). 
\label{eq:kinetic}
\end{eqnarray}
Here, $D$ denotes the rotational diffusivity, $T$, the fluid temperature and $k_B$, the Boltzmann constant. The tensors $E$ and $W$ are respectively the symmetric and anti-symmetric parts of the velocity gradient, given  by 
\begin{equation}
E=\frac{1}{2}(\nabla_x u + \nabla_x u^T), \qquad W=\frac{1}{2} (\nabla_x u-\nabla_x u^T).
\label{eq:def_WE}
\end{equation}
The symbols $\nabla_x$ and $\nabla_x \cdot$ refer to the spatial gradient and divergence operators while $\nabla_\omega$, $\nabla_\omega \cdot$ to the gradient and divergence operators on the sphere ${\mathbb S}^{n-1}$ respectively. The notation $\nabla_x u$ refers to the gradient tensor of $u$ defined by $(\nabla_x u)_{ij} = \partial_{x_i} u_j$ and the exponent 'T' indicates the transpose. The dimensionless quantity $\Lambda$ is related to the aspect ratio (ratio between the semi-axes) of the spheroidal polymer molecules. Finally, $P_{\omega^\perp} = \mathrm{Id} - \omega \otimes \omega$ for $\omega \in {\mathbb S}^{n-1}$ denotes the projection operator of vectors onto the normal hyperplane to $\omega$. Throughout this paper, $\mathrm{Id}$ denotes the identity matrix and if $u = (u_i)_{i=1,\ldots,n}$ and $v = (v_i)_{i=1,\ldots,n}$ are two vectors, $u \otimes v$ denotes their tensor product, i.e. the $n \times n$ tensor $(u \otimes v)_{ij} = u_i \, v_j$.  For two $n \times n$ tensors $S$ and $S'$, $SS'$ stands for the matrix product of $S$ and $S'$, hence the meaning of $P_{\omega^\perp} E$. The surface measure on the sphere will be normalized, meaning that $\int_{{\mathbb S}^{n-1}} d \omega = 1$.

The quantity $U_f^R$ is the interaction potential stemming from volume exclusion between the polymer molecules. In the Maier-Saupe theory \cite{Maier_Saupe_1959}, this interaction potential reads
\begin{equation}
U_f^R(x,\omega,t) = k_B T \nu \, \int_{{\mathbb R}^n \times {\mathbb S}^{n-1}} \frac{1}{R^n} K \Big( \frac{|x-x'|}{R} \Big) \, \big( 1 - (\omega \cdot \omega')^2 \big) \,  f(x',\omega',t) \, d\omega' \, dx', 
\label{eq:potential_nl}
\end{equation}
where $\nu$ is the potential strength. Following the formalism proposed by \cite{E_Zhang_MAA06, Wang_etal_PRE02}, a spatial non-locality is introduced by means of the kernel $K$: $[0,\infty) \to [0,\infty)$, $\xi \mapsto K(\xi)$ which describes the influence of two neighboring molecules. Specifically, two molecules separated by a distance $\xi$ influence each other with strength $\frac{1}{R^n} K (\frac{\xi}{R})$, where $R$ is the typical interaction range. The kernel $K$ satisfies $ \int_{{\mathbb R}^n} K(|x|) \, dx = 1$. An equivalent expression of $U_f^R$ is 
\begin{equation}
U_f^R(x,\omega,t) = k_B T \nu \rho^R_f \, \big[ - (\omega \cdot Q_f^R \omega) + \frac{n-1}{n} \big] ,
\label{eq:potential_alter}
\end{equation}
where $\rho_f^R$ and $Q_f^R$ are the locally averaged particle density and orientational de Gennes Q-tensor given by
\begin{eqnarray}
\rho_f^R (x,t) &=& \int_{{\mathbb R}^n \times {\mathbb S}^{n-1}} \frac{1}{R^n} K \Big( \frac{|x-x'|}{R}\Big) \, f(x',\omega,t) \, d\omega \, dx', \label{eq:density_nl} \\
(\rho_f^R  \, Q_f^R) (x,t) &=& \int_{{\mathbb R}^n \times {\mathbb S}^{n-1}} \frac{1}{R^n} K \Big( \frac{|x-x'|}{R} \Big) \, \Big(\omega \otimes \omega-\frac{1}{n} \mathrm{Id} \Big) \, f(x',\omega,t) \, d \omega \, dx'. 
\label{eq:orienttensor_nl}
\end{eqnarray}
Note that $Q_f^R$ is a trace-free symmetric matrix obtained by averaging $\omega \otimes \omega-\frac{1}{n} \mathrm{Id}$ over the probability distribution $\rho_f^R (x,t)^{-1} \,  R^{-n} K (|x-x'|/R) \, f(x',\omega,t) \, d \omega \, dx'$. Consequently, thanks to the min-max theorem, its eigenvalues $\lambda$ satisfy the inequality 
\begin{equation}
- \frac{1}{n} \leq \lambda \leq 1 - \frac{1}{n}. 
\label{eq:eigenineq}
\end{equation}
The following fully local versions of the polymer density and orientational tensor:
\begin{eqnarray} 
\rho_f  &=& \int_{{\mathbb S}^{n-1}}  f \, d\omega = \lim_{R \to 0} \rho_f^R, \label{eq:density}\\
 \rho_f Q_f &=& \int_{{\mathbb S}^{n-1}} \Big(\omega \otimes \omega-\frac{1}{n} \mathrm{Id} \Big) \, f \, d \omega = \lim_{R \to 0}  \rho_f^R Q_f^R, \label{eq:orienttensor}
\end{eqnarray}
will also be useful. From \eqref{eq:potential_alter}, it follows that 
$$ \frac{1}{k_B T} \, \nabla_\omega U_f^R (x,\omega,t) = - 2 \nu \rho_f^R \, P_{{\omega}^\bot} Q_f^R \omega, $$
so that an alternate formulation of the Doi equation \eqref{eq:kinetic} is given by 
\begin{eqnarray}
&& \hspace{-1.5cm} \partial_t f + \nabla_x \cdot (u f) + \nabla_\omega \cdot \big(f \, (\Lambda P_{\omega^\perp} E - W ) \omega \big) = D \big( \Delta_\omega f  - 2 \nu \, \rho_f^R \, \nabla_\omega\cdot(f\,P_{\omega^\perp} Q_f^R \, \omega) \big). \label{eq:kinetic_alternate}
\end{eqnarray}

We note that Eq. \eqref{eq:kinetic_alternate} preserves the symmetry constraint \eqref{eq:symmetry}.  The second and third term at the left-hand side of \eqref{eq:kinetic_alternate} model passive transport of the polymer molecules by the fluid: the second term corresponds to translation of the molecules by the fluid velocity and the third term to their rotation by the gradient of the fluid velocity. Here, we assume that the polymer molecules can be described by spheroids, i.e. ellipsoids, in which $n-1$ semi-axes $b$ are equal. The aspect ratio $p$ is the ratio $a/b$ where $a$ is the remaining semi-axis. The quantity $\Lambda$ is related to $p$ by $\Lambda = \frac{p^2-1}{p^2+1}$. In particular, $\Lambda \in [-1,1]$ and $\Lambda = 1$ for infinitely thin rods, $\Lambda = 0$ for spheres, and $\Lambda = -1$ for infinitely flat disks. The rotation operator is derived from Jeffery's equation \cite{Jeffery_motion_1922}. The first term at the right-hand side of \eqref{eq:kinetic_alternate} describes Brownian effects due to rotational diffusion. We neglect translational diffusivity, as it is usually much smaller than rotational diffusivity \cite{DeGennes_Prost}. The second term at the right-hand side of \eqref{eq:kinetic_alternate} takes into account the volume exclusion interaction between the molecules and drives the distribution to that of a system of fully aligned polymer molecules. To measure the degree of alignment of the molecules, one introduces 
\begin{equation}
\chi_f = \frac{n}{n-1} \lambda_f \mbox{ with } \lambda_f= \mbox{ the largest eigenvalue of } Q_f, 
\label{eq:OP}
\end{equation}
where $Q_f$ is given by \eqref{eq:orienttensor}. This quantity can be seen as the order parameter for the distribution $f$. We have $\chi_f \in (0,1)$. If $f$ is close to the uniform distribution on the sphere, which corresponds to a fully disordered distribution of polymer orientations, then $\chi_f$ is close to $0$. By contrast, if $f$ is close to $\frac{1}{2} (\delta_{\Omega} + \delta_{-\Omega})$ where $\Omega$ is any vector on ${\mathbb S}^{n-1}$, which corresponds to a fully aligned distribution of polymer orientations in the direction $\pm \Omega$, then, $\chi_f$ is close to $1$. 

To ensure thermodynamic consistency, one introduces the polymer free energy \cite{E_Zhang_MAA06}:
$$
{\mathcal A}^R(t) = \int_{{\mathbb R}^n \times {\mathbb S}^{n-1}} \big[ k_B T \, (f \, \log f - f)  + \frac{1}{2} U_f^R \, f \big] \, dx \, d \omega. 
$$
From \eqref{eq:potential_nl}, it is easy to check that the quantity $\int_{{\mathbb R}^n \times {\mathbb S}^{n-1}} U_f^R \, g \, dx \, d \omega$ defined for two functions $f$ and $g$ of $(x,\omega)$ is a symmetric bilinear form. Then the functional derivative $\mu_f^R = \frac{\delta {\mathcal A}^R}{\delta f}$, also referred to as the chemical potential, is given by 
\begin{equation}
\mu_f^R = k_B T \log f + U_f^R =  k_B T \, \Big( \, \log f - \nu \rho_f^R \, \big[ (\omega \cdot  Q_f^R \omega) - \frac{n-1}{n} \big] \Big). \label{eq:muf}
\end{equation}
Thus, 
\begin{equation}
\nabla_\omega \mu_f^R = k_B T \, \Big( \frac{\nabla_\omega f}{f} - 2 \nu \rho_f^R \, P_{\omega^\bot} Q_f^R \omega \Big), 
\label{eq:nablaommu}
\end{equation}
so that \eqref{eq:kinetic} can also be written:
\begin{eqnarray}
&& \hspace{-1cm} \partial_t f+ \nabla_x \cdot (u \, f) + \nabla_\omega \cdot \big(f \, (\Lambda P_{\omega^\perp} E - W ) \omega \big) =  \frac{D}{k_B T} \nabla_\omega \cdot \big( f \, \nabla_\omega \mu_f^R \big). 
\label{eq:kinetic_2}
\end{eqnarray}
The right-hand side of \eqref{eq:kinetic_2} can be viewed as describing the steepest descent in the direction of the minimum of the polymer free energy. This is also known as the maximal dissipation principle. Using Green's formula, we have the following identity (provided $f$ vanishes fast enough at infinity), whose proof is sketched in Appendix \ref{sec:virtualwork}: 
\begin{equation}
 \frac{d {\mathcal A}^R}{dt} = \int_{{\mathbb R}^n} \sigma_f^R : \nabla_x u \, dx  
- \int_{{\mathbb R}^n} F_f^R \cdot u \, dx 
- \frac{D}{k_B T} \int_{{\mathbb R}^n \times {\mathbb S}^{n-1}} f \, |\nabla_\omega \mu_f^R|^2 \, dx \, d \omega, 
\label{eq:virtualwork}
\end{equation}
where $\sigma_f^R$ is the extra-stress tensor and $F_f^R$ is a body force, given by :
\begin{equation} 
\sigma_f^R  = \int_{{\mathbb S}^{n-1}} \Big( \Lambda \big( \omega \otimes \nabla_\omega \mu_f^R \big)_s + \big( \omega \otimes \nabla_\omega \mu_f^R \big)_a \Big) \, f \, d \omega, \quad
F_f^R  = -   \int_{{\mathbb S}^{n-1}} \nabla_x \mu_f^R \, f \, d \omega.
\label{eq:sigmae_Fe}
\end{equation}
Here, for two $n \times n$ tensors $S = (S_{ij})_{ij=1,\ldots,n}$ and $S' = (S'_{ij})_{ij=1,\ldots,n}$, we denote by $S:S' = S_{ij} \, S'_{ij}$ their contraction (with the repeated index summation convention) while $S_s$ and $S_a$ are respectively the symmetric and antisymmetric parts of $S$ namely $S_s = \frac{1}{2} (S+S^T)$, $S_a = \frac{1}{2} (S-S^T)$. Contractions and tensor products will be defined and noted similarly for tensors of higher order.

%%%%%%%%%%%%%%%%%%%%%%%%%%%%%%%%%%%%%%%%%%%%%%%%%%%%%%%%%%%%%%%%
%%%%%%%%%%%%%%%%%%%%%%%%%%%%%%%%%%%%%%%%%%%%%%%%%%%%%%%%%%%%%%%%

\subsection{The Navier-Stokes equations}
%\label{subsec:NS}

The Doi equation \eqref{eq:kinetic} (or equivalently, \eqref{eq:kinetic_alternate} or \eqref{eq:kinetic_2}) is coupled to the Navier-Stokes equation for the fluid velocity, which is written \cite{Doi_Edwards, E_Zhang_MAA06, Wang_etal_PRE02}: 
\begin{eqnarray}
&& \hspace{-1cm} \rho_F \big( \partial_t u + u \cdot \nabla_x u \big) + \nabla_x p =  \nabla_x \cdot (\sigma_f^R + \tau_u + {\mathcal T}_{f,u} )  + F_f^R,
\label{eq:NS} \\
&& \hspace{-1cm} \nabla_x \cdot u=0.
\label{eq:divu=0}
\end{eqnarray}
Here $\rho_F$ is the fluid mass density. The extra-stress tensor $\sigma^R_f$ is given by \eqref{eq:sigmae_Fe} while $\tau_u$ and ${\mathcal T}_{f,u}$ are contributions of the fluid and polymer molecules to the viscous stresses respectively given by
$$
\tau_u = 2 \eta \, E, \qquad {\mathcal T}_{f,u} =  \zeta \, \frac{k_B T}{D} \, \rho_f \, {\mathbb T}_f:E,  
$$
with the fourth order orientational tensor ${\mathbb T}_f$ given by
\begin{equation}
\rho_f {\mathbb T}_f = \int_{{\mathbb S}^{n-1}} \omega^{\otimes 4} \, f \, d\omega.  
\label{eq:4tensor}
\end{equation}
For a $n \times n$ tensor $S$, its divergence $\nabla_x \cdot S$ denotes the vector defined by $(\nabla_x \cdot S)_j = \partial_{x_i} S_{ij}$ (using the repeated index summation convention). As above, ${\mathbb T}_f:E$ denotes the contraction of ${\mathbb T}_f$ and $E$ with respect to two indices. Although ${\mathbb T}_f$ is a fourth order tensor, it is symmetric, so which pair of its indices is concerned by the contraction is indifferent. The quantity $\eta$ is the fluid viscosity. Using the divergence-free condition \eqref{eq:divu=0}, we remark that $\nabla_x \cdot \tau_u = \eta \, \Delta_x u$. The quantity $\zeta$ is a dimensionless number. In \cite{Doi_Edwards}, for the dilute polymer regime in dimension $3$, it is shown that $\zeta = \frac{1}{2}$. But this derivation requires the use of the Oseen tensor which has dimensional dependence \cite{charbonneau2013dimensional} and thus, the value of $\zeta$ changes with the dimension. Moreover, even in dimension $3$, in the semi-dilute regime considered here, the value of $\zeta$ may be different from $\frac{1}{2}$ \cite[Section 9.5.1]{Doi_Edwards}. So, we shall consider $\zeta$ as a free parameter of the model.   

We have the following expression for the extra-stress: 
\begin{equation}
\sigma_f^R = n k_B T \Lambda \rho_f Q_f + \int_{{\mathbb S}^{n-1}} \Big[ \frac{\Lambda + 1}{2} \, \omega \otimes \nabla_\omega U_f^R + \frac{\Lambda - 1}{2} \, \nabla_\omega U_f^R \otimes  \omega \Big]\, f \, d \omega . \label{eq:express_sigmaR_1} 
\end{equation}
However, although more complicated, the following expression, which is valid if $f$ is a solution of the Doi equation \eqref{eq:kinetic}, will turn out to be more useful: 
\begin{eqnarray}
\sigma_f^R &=& \frac{k_B T}{D} \frac{\Lambda}{2} \rho_f \big[ \Lambda (E Q_f + Q_f E) + Q_f W - W Q_f + \frac{2 \Lambda}{n} E - 2 \Lambda {\mathbb T}_f:E - D_t Q_f \big] \nonumber \\
&& \hspace{4cm} + \frac{1}{2} \int_{{\mathbb S}^{n-1}} (\omega \otimes \nabla_\omega U_f^R - \nabla_\omega U_f^R \otimes \omega) \, f \, d \omega
, \label{eq:express_sigmaR_2} 
\end{eqnarray}
where 
\begin{equation} 
D_t = \partial_t + u \cdot \nabla_x, 
\label{eq:Dt}
\end{equation}
is the material derivative. Eq. \eqref{eq:express_sigmaR_1} results from the first equation of \eqref{eq:sigmae_Fe} after insertion of \eqref{eq:nablaommu}. Eq.  \eqref{eq:express_sigmaR_2} is obtained by multiplying Doi's equation \eqref{eq:kinetic_2} by $\omega \otimes \omega - \frac{1}{n} \mathrm{Id}$ and integrating with respect to $\omega$, followed by some algebra. These computations have been done in \cite{E_Zhang_MAA06, Kuzuu_Doi_JPhysSocJapan83, Wang_Zhang_Zhang_CPAM15} for $n=3$ and are sketched in Appendix \ref{sec:extrastress} for any $n$. 

The rationale for involving $\sigma_f^R$ and $F_f^R$ in the coupling between the Navier-Stokes equations \eqref{eq:NS} and the Doi equation \eqref{eq:kinetic} is thermodynamical consistency. Indeed, we have the following total free energy dissipation identity (provided spatial boundary terms vanish in the integrations by parts): 
\begin{equation} 
\frac{d}{dt} {\mathcal E}^R + {\mathcal D}^R = 0, 
\label{eq:free_ener_dissip}
\end{equation}
where ${\mathcal E}^R$ is the total free energy (sum of the fluid and polymer free energies):
$$ {\mathcal E}^R (t) = \int_{{\mathbb R}^n} \frac{1}{2} \rho_F |u|^2 \, dx + {\mathcal A}^R, $$
and ${\mathcal D}^R$ is the total free energy dissipation:
$$ {\mathcal D}^R (t) = \frac{D}{k_B T} \int_{{\mathbb R}^n \times {\mathbb S}^{n-1}} f \, \big| \nabla_\omega \mu_f^R \big|^2  \, dx \, d\omega + \int_{{\mathbb R}^n} \big( 2 \eta  \, E : E  + \frac{k_B T \zeta}{D} \, \rho_f {\mathbb T}_f:(E \otimes E) \big) \, dx ,$$
where now, ${\mathbb T}_f:(E \otimes E)$ indicates the contraction of the fourth order tensors ${\mathbb T}_f$ and $E \otimes E$ with respect to all four indices. We have omitted the dependence of $E$ on $u$ for simplicity.

%%%%%%%%%%%%%%%%%%%%%%%%%%%%%%%%%%%%%%%%%%%%%%%%%%%%%%%%%%%%%%%%
%%%%%%%%%%%%%%%%%%%%%%%%%%%%%%%%%%%%%%%%%%%%%%%%%%%%%%%%%%%%%%%%

\subsection{Scaling}
%\label{subsec:scaling}

We now introduce a suitable scaling of this model. Let $x_0$, $t_0$ and $\rho_0$ be space, time and polymer density units and let $u_0 = x_0/t_0$, $f_0 = \rho_0$, $\sigma_0 = k_B T \rho_0$, $p_0 = \rho_F \, u_0^2$, $F_0 = \sigma_0/x_0$, $U_0 = k_B T$ be units for velocity, distribution function, stress tensor, fluid pressure, elastic force and potential respectively. Then, we introduce the following dimensionless quantities: 
$$ \mathrm{De} = \frac{1}{D t_0}, \quad \mathrm{Re} = \frac{u_0 x_0 \rho_F}{\eta}, \quad \mathrm{Er} = \frac{\eta D}{k_B T \rho_0}, \quad \alpha = \nu \rho_0, \quad \bar R = \frac{R}{x_0}.$$
The dimensionless quantities De, Re and Er are the classical Deborah, Reynolds and Ericksen numbers, which respectively encode the relaxation time of the polymer molecular assembly to equilibrium, the ratio of inertial to viscous forces in the fluid and the ratio between the viscous and extra stresses. The parameters $\alpha$ and $\bar R$ are measures of the molecular interaction intensity and range respectively. The other dimensionless parameters of the model are $\zeta$ and $\Lambda$. Introducing scaled variables $x' = x/x_0$, $t' = t/t_0$ and unknowns $f(x,\omega,t) \, dx \, d \omega = \rho_0 \, f'(x',\omega,t') \, dx' \, d\omega$, $u(x,t) = u_0 \, u'(x',t')$, \ldots, we can deduce the following dimensionless form of the Doi model (dropping the primes for clarity): 
\begin{equation}
\partial_t f + \nabla_x \cdot (u f) + \nabla_\omega \cdot \big(f \, (\Lambda P_{\omega^\perp} E - W) \omega \big) = \frac{1}{\mathrm{De}} \, \nabla_\omega \cdot \big( \nabla_\omega f + f \, \nabla_\omega U_f^{\bar R} \big), 
\label{eq:kinetic_sc} 
\end{equation}
with 
$$ 
U_f^{\bar R} = \alpha \rho_f^{\bar R} \big[ - (\omega \cdot Q_f^{\bar R} \omega) + \frac{n-1}{n} \big], 
$$ 
and $\rho_f^{\bar R}$, $Q_f^{\bar R}$ given by \eqref{eq:density_nl}, \eqref{eq:orienttensor_nl} with $R$ replaced by $\bar R$. The polymer free energy is now given by 
$$
{\mathcal A}^{\bar R}(t) = \int_{{\mathbb R}^n \times {\mathbb S}^{n-1}} (f \, \log f - f + \frac{1}{2} \, U_f^{\bar R} \, f ) \, dx \, d \omega 
$$
and the chemical potential $\mu_f^{\bar R} = \frac{\delta {\mathcal A}^{\bar R}}{\delta f}$ by 
$$
\mu_f^{\bar R} = \log f + U_f^{\bar R} = \log f - \alpha \, (\omega \cdot \rho_f^{\bar R} Q_f^{\bar R} \omega) + \alpha \, \frac{n-1}{n} \, \rho_f^{\bar R}. 
$$
Thus, the expression at the right-hand side of \eqref{eq:kinetic_sc} is equivalently written
$$ 
\nabla_\omega \cdot \big( \nabla_\omega f + f \, \nabla_\omega U_f^{\bar R} \big)
= \nabla_\omega \cdot \big( f \, \nabla_\omega \mu_f^{\bar R} \big) =   \Delta_\omega f - 2 \alpha \rho_f^{\bar R} \nabla_\omega (f \, P_{\omega^\bot} Q_f^{\bar R} \omega). 
$$
The scaled Navier-Stokes equation reads as follows
\begin{eqnarray*}
&& \hspace{-1cm} \partial_t u + u \cdot \nabla_x u + \nabla_x p = \frac{1}{\mathrm{Re}} \, \nabla_x \cdot \big( \tau_u + \frac{1}{\mathrm{Er}} \, {\mathcal T}_{f,u} \big)  + \frac{1}{\mathrm{Re} \, \mathrm{Er} \, \mathrm{De}} \, \big( \nabla_x \cdot \sigma_f^{\bar R} + F_f^{\bar R} \big), \\
&& \hspace{-1cm} \nabla_x \cdot u=0,  \\
&& \hspace{-1cm} \tau_u = 2 \, E, \qquad {\mathcal T}_{f,u} =  \zeta \, \rho_f \, {\mathbb T}_f:E,  
\end{eqnarray*}
with $\sigma_f^{\bar R}$, $F_f^{\bar R}$ given by \eqref{eq:sigmae_Fe} with $R$ replaced by $\bar R$ and $\rho_f$, ${\mathbb T}_f$ given by \eqref{eq:density}, \eqref{eq:4tensor}. Expressions \eqref{eq:express_sigmaR_1}, \eqref{eq:express_sigmaR_2} for the stress tensor are scaled into
\begin{eqnarray}
\sigma_f^{\bar R} &=& n \Lambda \rho_f Q_f + \int_{{\mathbb S}^{n-1}} \Big[ \frac{\Lambda + 1}{2} \, \omega \otimes \nabla_\omega U_f^{\bar R} + \frac{\Lambda - 1}{2} \, \nabla_\omega U_f^{\bar R} \otimes \omega \Big] 
\, f \, d \omega. \nonumber \\ 
&=& \mathrm{De}\frac{\Lambda}{2} \rho_f \big[ \Lambda (E Q_f + Q_f E) + Q_f W - W Q_f + \frac{2 \Lambda}{n} E - 2 \Lambda {\mathbb T}_f:E - D_t Q_f \big] \nonumber \\
&& \hspace{2cm} + \frac{1}{2} \int_{{\mathbb S}^{n-1}} (\omega \otimes \nabla_\omega U_f^{\bar R} - \nabla_\omega U_f^{\bar R} \otimes \omega) \, f \, d \omega, \label{eq:express_sigmaR_2_sc} 
\end{eqnarray}
with $Q_f$ still given by \eqref{eq:orienttensor}. The free-energy dissipation identity is still written as \eqref{eq:free_ener_dissip} with ${\mathcal E}^{\bar R}$ and ${\mathcal D}^{\bar R}$ now given by
\begin{eqnarray} 
{\mathcal E}^{\bar R} &=& \frac{1}{2} \int_{{\mathbb R}^n} |u|^2 \, dx + \frac{1}{\mathrm{Re} \, \mathrm{Er} \, \mathrm{De}} \, {\mathcal A}^{\bar R}, \nonumber \\
{\mathcal D}^{\bar R}  &=& \frac{1}{\mathrm{Re}} \int_{{\mathbb R}^n} |\nabla_x u|^2 \, dx  + \frac{1}{\mathrm{Re} \, \mathrm{Er}} \zeta \int_{{\mathbb R}^n} \rho_f {\mathbb T}_f:(E \otimes E)  \, dx \nonumber \\
&&\hspace{3cm} + \frac{1}{\mathrm{Re} \, \mathrm{Er} \, \mathrm{De}^2} \int_{{\mathbb R}^n \times {\mathbb S}^{n-1}} f \, \big| \nabla_\omega \mu_f^{\bar R} \big|^2  \, dx \, d\omega,  
\label{eq:free_ener_dissip_sc}
\end{eqnarray}
where, for a $n \times n$ tensor $S$,  $|S|$ denotes the Frobenius norm of the $S$, i.e. $|S|^2 = \mathrm{Tr} \{S^T S \}$. 

The goal of this article is to investigate the limit of the Deborah number De tending to zero through the use of the new ``generalized collision invariant'' concept. In doing so, we will keep the parameters Re, Er and $\alpha$ of order unity. As for $\bar R$, following \cite{E_Zhang_MAA06, Wang_Zhang_Zhang_CPAM15}, we make the scaling $\bar R = {\mathcal O}(\sqrt{\mathrm{De}})$. This scaling assumption is analogous to the weakly non-local interaction scaling of the Vicsek model \cite{Degond_hydrodynamic_MAA13}. As we may choose the time and space units independently, we assume:
$$\mathrm{De} = \varepsilon, \qquad \bar R = \sqrt{\varepsilon}, \qquad \varepsilon \to 0, $$ 
and assume Re, Er and $\alpha$ independent of $\varepsilon$. A straightforward Taylor expansion shows that 
\begin{eqnarray*}
&&\hspace{-1cm}  \rho_f^{\sqrt{\varepsilon}} = \rho_f + \varepsilon \, \beta \, \Delta_x \rho_f + {\mathcal O}(\varepsilon^2), \qquad \rho_f^{\sqrt{\varepsilon}} Q_f^{\sqrt{\varepsilon}} = \rho_f \, Q_f + \varepsilon \, \beta \, \Delta_x (\rho_f \, Q_f) + {\mathcal O}(\varepsilon^2),
\end{eqnarray*}
where 
\begin{equation}
\beta = \frac{1}{2n} \int_{{\mathbb R}^n} K(|x|) \, |x|^2 \, dx. 
\label{eq:beta}
\end{equation}
Then, we can expand $U_f^{\sqrt{\varepsilon}} = U^0_f + \varepsilon U_f^1 + {\mathcal O}(\varepsilon^2)$, $\mu_f^{\sqrt{\varepsilon}} = \mu^0_f + \varepsilon \mu_f^1 + {\mathcal O}(\varepsilon^2)$
with 
\begin{eqnarray}
U^0_f &=&  \alpha \rho_f \, \big[ - (\omega \cdot Q_f \omega) + \frac{n-1}{n} \big], \qquad \mu^0_f = \log f + U^0_f \label{eq:muf0} \\   
U_f^1 &=& \mu_f^1 = \beta \, \Delta_x U_f^0 . \label{eq:muf1}
\end{eqnarray}
Straightforward computations show that
\begin{equation} 
\int_{{\mathbb S}^{n-1}} \omega \otimes \nabla_\omega U_f^0 \, f \, d \omega = -2 \alpha \rho_f^2 \big[ Q_f^2 + \frac{1}{n} Q_f - {\mathbb T}_f:Q_f \big],
\label{eq:int_omotnaU0}
\end{equation}
so that the left-hand side of \eqref{eq:int_omotnaU0} is a symmetric tensor. We deduce that the integral term in \eqref{eq:express_sigmaR_2_sc} is ${\mathcal O}(\varepsilon)$, so that $\sigma_f^{\sqrt{\varepsilon}}={\mathcal O}(\varepsilon)$. Additionally, similar computations as for \eqref{eq:int_omotnaU0} lead to
$$
\int_{{\mathbb S}^{n-1}} (\omega \otimes \nabla_\omega U_f^1 - \nabla_\omega U_f^1 \otimes \omega) \, f \, d \omega =  2 \alpha \beta \, \rho_f \, 
 \big[ \Delta_x (\rho_f Q_f) Q_f - Q_f \Delta_x (\rho_f Q_f) \big]. 
$$
So, we can write $\sigma_f^{\bar R} = \varepsilon \sigma_f^1 + {\mathcal O}(\varepsilon^2)$ with 
\begin{eqnarray}
\sigma_f^1 &=& \rho_f \frac{\Lambda}{2}  \big[ \Lambda(E Q_f + Q_f E) + Q_f W - W Q_f + \frac{2 \Lambda}{n} E - 2\Lambda {\mathbb T}_f:E - D_t Q_f \big] \nonumber \\
&& \hspace{4cm} 
+ \alpha \beta \, \rho_f \, \big[ \Delta_x (\rho_f Q_f) Q_f - Q_f \Delta_x (\rho_f Q_f) \big].  \label{eq:sigma1}
\end{eqnarray}
We also note that $F_f^{\sqrt{\varepsilon}} = - \nabla_x \varphi_f^0 + \varepsilon F_f^1 + {\mathcal O}(\varepsilon^2)$, with 
\begin{eqnarray}
\varphi_f^0 &=& \rho_f - \frac{\alpha}{2} \rho_f^2 \big[ Q_f:Q_f - \frac{n-1}{n}] , \qquad 
F_f^1 = - \int_{{\mathbb S}^{n-1}} \nabla_x \mu_f^1 \, f \, d \omega. 
\label{eq:F1} 
\end{eqnarray}
We let $\tilde p^\varepsilon = p^\varepsilon + \frac{1}{\varepsilon} \frac{1}{\mathrm{Re Er}} \varphi_f^0$. We will omit the tilde below for simplicity. Since the ${\mathcal O}(\varepsilon^2)$ terms in all these developments have no contribution to the limit model when $\varepsilon \to 0$ (at the leading order), we will just ignore them. 

We finally get the following perturbation problem: 
\begin{eqnarray}
&& \hspace{-1cm} \partial_t f^\varepsilon +\nabla_x\cdot (u^\varepsilon f^\varepsilon)  +\nabla_\omega \cdot (f^\varepsilon \, (\Lambda P_{\omega^\perp} E^\varepsilon - W^\varepsilon) \omega) 
+ 2 \alpha \, \beta \, \nabla_\omega\cdot(f^\varepsilon \,P_{\omega^\perp} \Delta_x (\rho_{f^\varepsilon} \, Q_{f^\varepsilon }) \, \omega ) \nonumber \\
&& \hspace{5cm}
= \frac{1}{\varepsilon} \Big( \Delta_\omega f^\varepsilon   - 2 \alpha \rho_{f^\varepsilon} \nabla_\omega\cdot(f^\varepsilon \,P_{\omega^\perp}Q_{f^\varepsilon } \, \omega) \Big), 
\label{eq:kinetic_eps} \\
&& \hspace{-1cm} \partial_t u^\varepsilon +u^\varepsilon \cdot \nabla_x u^\varepsilon +\nabla p^\varepsilon = \frac{1}{\mathrm{Re}} \Big\{ \Delta_x u^\varepsilon + \frac{1}{\mathrm{Er}} \,  \Big[ \zeta \, \nabla_x \cdot \big( \rho_{f^\varepsilon } {\mathbb T}_{f^\varepsilon } :E^\varepsilon \big)  \nonumber\\
&& \hspace{9cm}
+ \nabla_x \cdot \sigma_{f^\varepsilon}^1 + F_{f^\varepsilon}^1 \Big] \Big\}  ,
\label{eq:NS_eps} \\
&& \hspace{-1cm} \nabla_x \cdot u^\varepsilon =0,
\label{eq:divu=0_eps} 
\end{eqnarray}
where $\sigma_{f^\varepsilon}^1$ is given by \eqref{eq:sigma1} and  $F_{f^\varepsilon}^1$ by \eqref{eq:F1}. 

We define the transport operator $T_u(f)$ (for a given time-dependent vector field $u$: ${\mathbb R}^n \times [0,\infty)  \to {\mathbb R}^n$) and the collision operator $C(f)$ by 
\begin{eqnarray}
T_u (f) &=& \partial_t f +\nabla_x\cdot (u \, f)  + \nabla_\omega \cdot (f \, (\Lambda P_{\omega^\perp} E - W) \omega)  \nonumber \\
&&\hspace{4cm} + 2 \alpha \, \beta \, \nabla_\omega\cdot(f \,P_{\omega^\perp} \Delta_x (\rho_f \, Q_f) \, \omega ), \label{eq:defTu} \\
C (f) &=& \Delta_\omega f   - 2 \alpha \rho_f \nabla_\omega\cdot(f \,P_{\omega^\perp}Q_f \, \omega) = \nabla_\omega \cdot \big( f \, \nabla_\omega \mu_f^0 \big) \label{eq:coll_op} \\
&=& \nabla_\omega \cdot \big( \nabla_\omega f + f \nabla_\omega U^0_f \big), \label{eq:coll_op_alter}
\end{eqnarray}
so that \eqref{eq:kinetic_eps} is written
\begin{equation} 
T_{u^\varepsilon}(f^\varepsilon) = \frac{1}{\varepsilon} C (f^\varepsilon). 
\label{eq:kinetic_abstract}
\end{equation}
We note that $\mu^0_f = \frac{\delta {\mathbb A}^0}{\delta f}$ is the functional derivative of the free energy ${\mathcal A}^0 = \lim_{\varepsilon \to 0} {\mathcal A}^{\sqrt{\varepsilon}}$ given by 
\begin{equation}
{\mathcal A}^0(t) =  \int_{{\mathbb R}^n \times {\mathbb S}^{n-1}} (f \, \log f - f  + \frac{1}{2} U^0_f  \, f ) \, dx \, d\omega, 
\label{eq:def_A^0}
\end{equation}
and recall that $U^0_f$ and $\mu^0_f$ are given by \eqref{eq:muf0}. We refer to \cite{E_Zhang_MAA06} for the formulation of the free energy dissipation identity for the whole model \eqref{eq:kinetic_eps} - \eqref{eq:divu=0_eps}.

%%%%%%%%%%%%%%%%%%%%%%%%%%%%%%%%%%%%%%%%%%%%%%%%%%%%%%%%%%%%%%%%
%%%%%%%%%%%%%%%%%%%%%%%%%%%%%%%%%%%%%%%%%%%%%%%%%%%%%%%%%%%%%%%%
%%%%%%%%%%%%%%%%%%%%%%%%%%%%%%%%%%%%%%%%%%%%%%%%%%%%%%%%%%%%%%%%
%%%%%%%%%%%%%%%%%%%%%%%%%%%%%%%%%%%%%%%%%%%%%%%%%%%%%%%%%%%%%%%%

\setcounter{equation}{0}
\section{Main result}
\label{sec:main_result}

\subsection{Preliminaries}
%\label{subsec_prelim}

The purpose of this paper is to derive the limit of model \eqref{eq:kinetic_eps} - \eqref{eq:divu=0_eps} when $\varepsilon \to 0$. Before stating the result, we need a few preliminaries. We note that $C$ given by \eqref{eq:coll_op} operates on the variable $\omega$ only and leaves $(x,t)$ as parameters. This justifies the definition:

\begin{definition}
A function $f$: ${\mathbb S}^{n-1} \to {\mathbb R}$, $\omega \mapsto f(\omega)$ is called an equilibrium of $C$ if and only if it satisfies
\begin{equation} 
C (f) = 0. 
\label{eq:equi_def}
\end{equation}
%\label{def:equilibrium}
\end{definition}

\begin{remark}
We note that $f$ is an equilibrium if and only if $f$ is a critical point of the free energy functional ${\mathcal A}^0$ given by \eqref{eq:def_A^0} in the spatially homogeneous case  (i.e. when $f$ is a function of $\omega$ only and integration with respect to $x$ in the definition of ${\mathcal A}^0$ is ignored) \cite{Liu_Zhang_Zhang_CMS05, Wang_Zhang_Zhang_CPAM15}. Moreover, such equilibria will be called ``stable'' if they correspond to local minimizers of this free energy (see \cite{Fatkullin_Slastikov_CMS05} for $n=2$, \cite{Fatkullin_Slastikov_nonlinearity05, Liu_Zhang_Zhang_CMS05} for $n=3$ and \cite{Frouvelle_2021} for $n=4$).
\label{rem:crit_point}
\end{remark}

The equilibria will attract the dynamics as $\varepsilon \to 0$ and their determination is of key importance. For this purpose, we introduce the Gibbs distributions: 

\begin{definition} [Gibbs distribution]
Let $S$ be a trace-free symmetric matrix. Then, the Gibbs distribution $G_S$ associated with $S$ is given by: 
\begin{equation}
G_S (\omega) = \frac{1}{Z_S} \, e^{\omega \cdot S \omega}, \quad Z_S = \int_{{\mathbb S}^{n-1}} e^{\omega \cdot S \omega} \, d \omega. 
\label{eq:vonmises2}
\end{equation}
%\label{def:gibbs_distrib}
\end{definition}

Next, we introduce the

\begin{definition} [Normalized prolate uniaxial trace-free tensor]
Let $\Omega \in {\mathbb P}^{n-1}:= {\mathbb S}^{n-1} / \{\pm 1\}$. Then, the normalized prolate uniaxial trace-free tensor in the direction of $\Omega$, $A_\Omega$, is defined by 
\begin{equation}
A_\Omega = \Omega \otimes \Omega - \frac{1}{n} \mathrm{Id}. 
\label{eq:AOm}
\end{equation}
$A_\Omega$ is a traceless symmetric tensor with leading eigenvalue equal to $\frac{n-1}{n}$. 
%\label{def:AOm}
\end{definition}

$A_\Omega$ is called a uniaxial tensor because it has only two eigenvalues with one being simple. The simple eigenvalue has associated normalized eigenvectors $\pm \Omega$. The line spanned by $\Omega$ is called the axis of the uniaxial tensor. It is trace-free and consequently, the two eigenvalues have opposite signs. It is called prolate because the simple eigenvalue is positive (it would be called oblate in the converse case). It is normalized meaning that its leading eigenvalue is exactly $\frac{n-1}{n}$. We note that $A_\Omega$ is invariant by the change $\Omega \to - \Omega$ showing that it actually depends on $\Omega$ seen as an element of the projective space ${\mathbb P}^{n-1} = {\mathbb S}^{n-1} / \{\pm 1\}$.

\begin{proposition}[Gibbs distributions of uniaxial tensors] The Gibbs distributions $G_{\eta \, A_\Omega}$ associated to tensors of the form $\eta \, A_\Omega$ with $\eta >0$ are given by 
\begin{equation}
G_{\eta \, A_\Omega} (\omega) = \frac{1}{Z_\eta} e^{\eta \, (\omega \cdot \Omega)^2}, \qquad Z_\eta = \int_{{\mathbb S}^{n-1}} e^{\, \eta \, (\omega \cdot \Omega)^2} \, d \omega, 
\label{eq:GetaAom}
\end{equation}
where the normalization constant $Z_\eta$ does not depend on $\Omega$ but only on $\eta$. 
\label{prop:gibbs_uniaxial}
\end{proposition}

\noindent
\textbf{Proof.} Eq. \eqref{eq:GetaAom} is obvious from \eqref{eq:AOm}. Defining $\theta \in (0,\pi)$ such that $\cos \theta = (\omega \cdot \Omega)$ and changing $\omega$ to $(\theta, z)$ where $z \in {\mathbb S}^{n-2}$ through $\omega = \cos \theta \, \Omega + \sin \theta \, z$, with $d \omega = C_n \sin^{n-2} \theta \, d \theta \, dz$ ($C_n$ being such that $C_n \int_0^\pi \sin^{n-2} \theta \, d \theta = 1$ and $\int_{{\mathbb S}^{n-2}} dz = 1$), we get:
$$ 
Z_\eta = C_n \int_0^\pi e^{\eta \, \cos^2 \theta} \, \sin^{n-2} \theta \, d \theta,
$$
which does not depend on $\Omega$. 
\endproof

For two functions $g$ and $\varphi$: ${\mathbb S}^{n-1} \to {\mathbb R}$, with $\varphi > 0$ a.e., we define:
$$ \langle g \rangle_{\varphi} 
= \frac{\int_{{\mathbb S}^{n-1}} g(\omega) \, \varphi(\omega) \, d \omega}{\int_{{\mathbb S}^{n-1}}  \varphi(\omega) \, d \omega} 
$$
We introduce the following 

\begin{definition}[Definition of $S_2$ and $S_4$]
The quantities $S_2(\eta)$ and $S_4(\eta)$ are defined by 
\begin{equation}
S_2(\eta) = \langle P_2(\omega \cdot \Omega) \rangle_{G_{\eta A_\Omega}}, \qquad S_4(\eta) = \langle P_4(\omega \cdot \Omega) \rangle_{G_{\eta A_\Omega}}, \label{eq:S2} 
\end{equation}
where $P_2(X)$ and $P_2(X)$ are the polynomials
\begin{eqnarray}
P_2(X) &=& \frac{1}{n-1} (n X^2 - 1), \label{eq:P2}\\
P_4(X) &=& \frac{1}{(n-1)(n+1)} \big[ 3 - 6(n+2) X^2 +(n+2)(n+4) X^4 \big]. 
%\label{eq:P4}
\nonumber
\end{eqnarray}
%\label{def:S2}
\end{definition}

For the same reason as in Proposition \ref{prop:gibbs_uniaxial}, $S_2$ and $S_4$ do not depend on $\Omega$. In dimension $n=3$, the polynomials $P_2$ and $P_4$ are the Legendre polynomials of degree $2$ and $4$ respectively. About $S_2$, we have the following proposition, which will be proved in Appendix \ref{subsec:proof_prop_3.6}. 

\begin{proposition}[Properties of $S_2$]
(i) We have 
\begin{equation}
Q_{G_{\eta \, A_\Omega}} = S_2(\eta) \, A_\Omega.
\label{eq:Q_equib}
\end{equation}

\noindent
(ii) The order parameter \eqref{eq:OP} of the distribution $\rho G_{\eta A_\Omega}$ is 
$\chi_{\rho G_{\eta A_\Omega}} = S_2(\eta). $

\noindent
(iii) $S_2$ is a non-decreasing function from $(0,\infty)$ onto $(0,1)$, i.e. $S_2(0) = 0$ and $S_2 \to 1$ as $\eta \to \infty$. 
\label{prop:OP}
\end{proposition}

We note that, when $\eta \to 0$, $G_{\eta A_\Omega}$ converges to the uniform probability distribution  on ${\mathbb S}^{n-1}$. Likewise, when $\eta \to \infty$, $G_{\eta A_\Omega}$ concentrates on two Dirac deltas $\frac{1}{2} (\delta_{\Omega} + \delta_{-\Omega})$ which characterizes fully aligned distributions of molecules in the direction $\Omega$. Therefore,~$S_2$ takes the value $0$ on fully disordered distributions and the value $1$ on fully ordered ones. As $\eta$ increases, $G_{\eta A_\Omega}$ shows increasing order evidenced by the increase of the order parameter $S_2$. Now, we have the following

\begin{proposition}[Implicit definition of $\eta(\rho)$]
The implicit equation 
\begin{equation}
\frac{\eta}{\alpha \, \rho} = S_2(\eta), 
\label{eq:eta}
\end{equation}
has at least a root $\eta$ if and only if $\rho \in (\rho^*, +\infty)$ where $\rho^* >0$. It has at most two roots. By choosing the largest root (which is necessarily nonnegative), it defines a smooth non-decreasing function $(\rho^*, +\infty) \to (\eta^*, +\infty)$, $\rho \mapsto \eta(\rho)$, where $\eta^*= \lim_{\rho \to \rho^*} \eta(\rho) \geq 0$. 
\label{prop:eta(rho)_defin}
\end{proposition}

This proposition is a consequence of the result of Wang and Hoffman \cite{Hoffman_Wang_CMS06} which will be recalled in Section \ref{sec:local_equilibria}. With this, we formulate the following conjecture, which has been verified in dimension $n=2$ \cite{Fatkullin_Slastikov_CMS05}, $n=3$ \cite{Fatkullin_Slastikov_nonlinearity05, Liu_Zhang_Zhang_CMS05} and $n=4$ \cite{Frouvelle_2021}.

\begin{conject} [Stable anisotropic equilibria]
The set ${\mathcal E}$ of stable anisotropic equilibria (in the sense of Remark \ref{rem:crit_point}) is given by 
$$
{\mathcal E} = \{ \rho \, G_{\eta(\rho) A_\Omega} \, \, | \, \, \rho \in (\rho^*,+\infty), \, \, \Omega \in {\mathbb P}^{n-1} \}. 
$$
\label{conject:stable_0}
\end{conject}

We will only consider anisotropic equilibria, i.e. belonging to the set ${\mathcal E}$ above. Stable isotropic equilibria (i.e. such that $f=\rho$ is independent of $\omega$) do exist but will not be used here. 

\begin{remark}
In the case $n=3$, using the change of variables $z=\cos \theta$ and an integration by parts, Eq. \eqref{eq:eta} can be recast as 
$$ \frac{3e^{\eta}}{\int_0^1 e^{\eta \, z^2} \, dz} = 3 + 2\eta + \frac{4 \eta^2}{\alpha \, \rho}. $$
Upon changing $\eta$ into $-\eta$ and making $\rho = 1$, we recover Eq. (1.9) of \cite{Liu_Zhang_Zhang_CMS05} and Eq. (3.2) of \cite{Wang_Zhang_Zhang_CPAM15} (up to a typo in the latter: a factor $4$ is missing in front of the $\eta^2$ term). 
%\label{rem:link_LZZ}
\end{remark}

Now, we introduce the molecular interaction potential at equilibrium $U^0_{\rho \, G_{\eta A_\Omega}}$ where $U^0_f$ is given by \eqref{eq:muf0}. Thanks to \eqref{eq:Q_equib}, \eqref{eq:eta}, we have 
\begin{equation}
\alpha \rho Q_{G_{\eta A_\Omega}} = \eta(\rho) A_\Omega.
\label{eq:alpharhoQ=etaA}
\end{equation}
Thus, introducing $\theta \in [0,\pi]$ such that $\omega \cdot \Omega = \cos \theta$, straightforward computations give
\begin{eqnarray} 
U^0_{\rho \, G_{\eta A_\Omega}} &=& - \eta (\omega \cdot A_{\Omega} \omega) + \frac{n-1}{n} \alpha \rho = -  \eta \, \big( (\omega \cdot \Omega)^2 - \frac{1}{n} \big)  + \frac{n-1}{n} \alpha \rho \label{eq:def_U0_0} \\
&=& -  \eta \, \big(\cos^2 \theta - \frac{1}{n} \big)  + \frac{n-1}{n} \alpha \rho  =: \tilde U_0(\theta) , \nonumber
\end{eqnarray}
so defining the function $\tilde U_0(\theta)$. We note that 
\begin{equation}
\frac{d \tilde U_0}{d \theta} = 2 \eta \, \cos \theta \, \sin \theta.
\label{eq:dU0dthet}
\end{equation} 
For two functions $\varphi$ and $\psi$ defined on $[0,\pi]$ with $\psi >0$, a.e., we define
$$ \langle \hspace{-0.8mm} \langle \varphi \rangle \hspace{-0.8mm} \rangle_\psi = \frac{\int_0^\pi \varphi(\theta) \, \psi (\theta) \, \sin^{n-2} \theta \, d \theta}{\int_0^\pi \psi (\theta) \, \sin^{n-2} \theta \, d \theta}. $$
Thanks to these notations, we can state the

\begin{definition}[Auxiliary function $g$]
The function $g$: $[0,\pi] \to {\mathbb R}$, $\theta \mapsto g(\theta)$, is the unique solution (in a sense made precise in Section \ref{sec:GCI}) of the elliptic equation 
\begin{equation}
 \frac{1}{\sin^{n-2} \theta} \frac{d}{d \theta} \Big( \sin^{n-2} \theta \frac{d g}{d \theta} \Big) - \frac{d \tilde U_0}{d \theta}  \frac{d g}{d \theta} - (n-2) \frac{g}{\sin^2 \theta} = - \frac{d \tilde U_0}{d \theta}.  
\label{eq:g}
\end{equation}
%\label{def:auxi_fct_g}
\end{definition}

Note that, in the special case $n =3$, \eqref{eq:g} coincides with Eq. (5.31) of \cite{Kuzuu_Doi_JPhysSocJapan83}. Thanks to $g$ we have the following proposition, proved in Section \ref{subsec:GCI_first_derivation}:

\begin{proposition}[Constant $c$]
Assume $\Lambda \not = 0$. Then, the constant $c$  given by 
\begin{equation}
c = \frac{\displaystyle (n-1) \Lambda S_2(\eta)}{\displaystyle \Big\langle \hspace{-1.6mm} \Big\langle g  \, \frac{d \tilde U_0}{d \theta} \Big\rangle \hspace{-1.6mm} \Big\rangle_{\exp(\eta \cos^2 \theta)}}, 
\label{eq:def_c}
\end{equation}
is such that $c/\Lambda >0$. 
\label{prop:const_c}
\end{proposition}

In dimension $n=3$, this formula coincides with formula (5.33) of \cite{Kuzuu_Doi_JPhysSocJapan83}. We now introduce the following definitions

\begin{definition}[Definition of the Leslie constants $\alpha_k$, $k=1, \ldots, 6$] The Leslie constants $\alpha_k$, $k=1, \ldots, 6$ are defined by 
\begin{eqnarray}
\alpha_1 &=& (\zeta - \Lambda^2) S_4, \quad \alpha_2 = - \frac{\Lambda S_2}{2} \big( \frac{1}{c} + 1 \big), \quad  \alpha_3 = \frac{\Lambda S_2}{2} \big( \frac{1}{c} - 1 \big) , \label{eq:al123} \\
\alpha_4 &=& \frac{2 (\zeta - \Lambda^2)}{(n+2)(n+4)} S_4 - \frac{2}{n} \Big( \frac{\Lambda^2}{2} + \frac{2(\zeta-\Lambda^2)}{n+4} \Big) S_2 + \frac{1}{n} \Big( \Lambda^2 + \frac{2(\zeta-\Lambda^2)}{n+2} \Big)
, \label{eq:al4} \\
\alpha_5 &=& - \frac{2 (\zeta - \Lambda^2)}{n+4} S_4 + \Big( \frac{\Lambda}{2} + \frac{\Lambda^2}{2} + \frac{2 (\zeta - \Lambda^2)}{n+4} \Big) S_2, \label{eq:al5} \\
\alpha_6 &=& - \frac{2 (\zeta - \Lambda^2)}{n+4} S_4 + \Big( - \frac{\Lambda}{2} + \frac{\Lambda^2}{2} + \frac{2 (\zeta - \Lambda^2)}{n+4} \Big) S_2, \label{eq:al6} 
\end{eqnarray}
where $S_2$ and $S_4$ are given by \eqref{eq:S2} and their dependence on $\eta$ has been omitted for simplicity, and where $c$ is given by \eqref{eq:def_c}. We note Parodi's relation: $\alpha_6 - \alpha_5 = \alpha_2 + \alpha_3$. 
\end{definition}

%%%%%%%%%%%%%%%%%%%%%%%%%%%%%%%%%%%%%%%%%%%%%%%%%%%%%%%%%%%%%%%%
%%%%%%%%%%%%%%%%%%%%%%%%%%%%%%%%%%%%%%%%%%%%%%%%%%%%%%%%%%%%%%%%

\subsection{Main result: statement and comments}
%\label{subsec:main_result}

Now, our aim is to prove the following formal result: 

\begin{theorem}[Formal limit of model \eqref{eq:kinetic_eps} - \eqref{eq:divu=0_eps}]
We assume $n\geq 2$, $\Lambda \not = 0$. For $n \geq 5$, we assume that Conjecture \ref{conject:stable_0} is true (for $2 \leq n \leq 4$, this conjecture is a theorem \cite{Fatkullin_Slastikov_CMS05, Fatkullin_Slastikov_nonlinearity05, Frouvelle_2021, Liu_Zhang_Zhang_CMS05}). When $\varepsilon \to 0$, we assume that $(f^\varepsilon, u^\varepsilon) \to (f, u)$ as smoothly as needed, where $f (x, \cdot, t)$ is a stable anisotropic local equilibrium for all $(x,t)$. Then, on the open set 
\begin{equation}
{\mathcal B} = \{ (x,t) \in {\mathbb R}^n \times [0,\infty) \, \, | \, \, \rho_f(x,t) > \rho^* \},
\label{def:set_B}
\end{equation}
(where $\rho^*$ is defined at Proposition \ref{prop:eta(rho)_defin}), we have 
\begin{equation}
f(x,\omega,t) = \rho(x,t) G_{\eta(\rho(x,t)) A_{\Omega(x,t)}}(\omega),  
\label{eq:localequ}
\end{equation}
where the function $(\rho^*, \infty) \ni~\rho \mapsto \eta(\rho) \in [0,\infty)$ is defined by \eqref{eq:eta}. The functions $(x,t) \mapsto (\rho, \Omega, u)(x,t)$ satisfy the following system of partial differential equations (called the Ericksen-Leslie system): 
\begin{eqnarray}
&& \hspace{-1cm} 
\partial_t \rho  + \nabla_x \cdot (\rho u)  = 0, \label{conservation-mass} \\
&& \hspace{-1cm} 
\partial_t \Omega + u \cdot \nabla_x \Omega + W \Omega - c \, P_{\Omega^\perp} \big( E \Omega + \frac{2 \beta}{\Lambda} \, \Delta_x (\eta \Omega) \big)  = 0,
\label{eq:Omega} \\
&& \hspace{-1cm} \partial_t u + u \cdot \nabla_x u +\nabla p = \frac{1}{\mathrm{Re}} ( \Delta_x u + \frac{1}{\mathrm{Er}} \,  \nabla_x \cdot \sigma ),
\label{eq:NS_final} \\
&& \hspace{-1cm} \nabla_x \cdot u =0,
\label{eq:divu=0_final} \\
&& \hspace{-1cm} 
\sigma = \sigma_L + \sigma_E, 
\label{eq:sigma_equi} \\
&& \hspace{-1cm} 
\sigma_L = \rho \big\{ \alpha_1 \big(E:(\Omega \otimes \Omega) \big) \, \Omega \otimes \Omega + \alpha_2 \Omega \otimes N + \alpha_3 N \otimes \Omega \nonumber \\
&& \hspace{3cm} + \alpha_4 E + \alpha_5 (\Omega \otimes \Omega) E + \alpha_6 E (\Omega \otimes \Omega) \big\}, \label{eq:sigmae_equi} \\
&& \hspace{-1cm} 
\sigma_E = -\frac{2 \beta}{\alpha} \, \nabla_x (\eta \Omega) \nabla_x (\eta \Omega)^T \nonumber \\
&& \hspace{1cm} + \frac{(n+1) \beta}{n \alpha} \, \nabla_x \eta  \otimes \nabla_x \eta + \frac{(n-1) \alpha \beta}{n} \, \nabla_x \rho \otimes \nabla_x \rho , \label{eq:sigma_elas_equi}
\end{eqnarray}
where $W$ and $E$ are given by \eqref{eq:def_WE}, $\beta$ by \eqref{eq:beta}, $c$ by \eqref{eq:def_c}, $\alpha_k$, $k=1, \ldots, 6$ by \eqref{eq:al123}-\eqref{eq:al6}, and $N$ by 
\begin{equation}
N = D_t \Omega + W \Omega, 
\label{eq:defN}
\end{equation}
with $D_t$ given by \eqref{eq:Dt}.
\end{theorem}

\begin{remark}
Using \eqref{eq:eta}, we have the following equivalent expression of $\sigma_E$: 
$$
\sigma_E = -\frac{2 \beta}{\alpha} \, \nabla_x \Omega \nabla_x \Omega^T - \frac{(n-1) \beta}{n \alpha} \big[ 1 - \frac{1}{S_2^2} \big( 1 - \eta \frac{S'_2}{S_2} \big)^2 \big] \nabla_x \eta \otimes \nabla_x \eta, 
$$
where $S'_2$ denotes the derivative of $S_2$ with respect to $\eta$. In particular, this formula shows that the contribution of the density gradient to $\sigma_E$ is a rank-1 tensor (which is not obvious from \eqref{eq:sigma_elas_equi}; on the other hand, \eqref{eq:sigma_elas_equi} has more symmetry between $\nabla_x \rho$ and $\nabla_x \eta$). 
%\label{rem:Ericksen_stresses}
\end{remark}

\begin{remark}
In the case $\Lambda = 0$, the result is still valid, except that \eqref{eq:Omega}  must be replaced by 
$$ \partial_t \Omega + u \cdot \nabla_x \Omega + W \Omega - 2 \beta \tilde c \, P_{\Omega^\perp}  \Delta_x (\eta \Omega)  = 0, $$
where $\tilde c = (n-1) S_2(\eta) / \langle \hspace{-1mm} \langle g  \, \frac{d \tilde U_0}{d \theta} \rangle \hspace{-1mm} \rangle_{\exp(\eta \cos^2 \theta)}$.
\end{remark}

In the literature \cite{Kuzuu_Doi_JPhysSocJapan83, Wang_Zhang_Zhang_CPAM15}, Eq. \eqref{eq:Omega} is written differently. For this we need the 

\begin{definition}[Molecular field and $\gamma$-constants]
We define 
\begin{eqnarray}
\gamma_1 &=& \frac{\Lambda S_2}{c} = \alpha_3 - \alpha_2, \qquad \gamma_2 = - \Lambda S_2 = \alpha_6 - \alpha_5 = \alpha_2 + \alpha_3,\label{eq:gamma_EL} \\
H &=& 2 \beta S_2 \Delta_x (\eta \Omega).  \nonumber
%\label{eq:H_EL}
\end{eqnarray}
The quantity $H$ is called the molecular field.
%\label{def:molecfield}
\end{definition}

Then, we have the following proposition, whose proof is immediate:

\begin{proposition}[Equivalent form of Eq. \eqref{eq:Omega}]
Eq. \eqref{eq:Omega} is equivalent to 
\begin{equation}
P_{\Omega^\bot} \big( H - \gamma_1 N - \gamma_2 E \Omega \big) = 0.
\label{eq:Omega_alter}
\end{equation}
\label{prop:Omega_eq_alter}
\end{proposition}

We compare System \eqref{conservation-mass}-\eqref{eq:sigma_elas_equi} with the literature. Ref. \cite{Kuzuu_Doi_JPhysSocJapan83} considers a spatially homogeneous model in dimension $n=3$ with $\zeta = 0$. Spatial homogeneity means that $\rho$ and $\Omega$ do not depend on $x$, and so $H=0$, $\sigma_E = 0$ and $N = \partial_t \Omega + W \Omega$ while $E$ and $W$ are constant. In this case, our model reduces to \eqref{eq:Omega_alter} (with $H=0$) and $\sigma = \sigma_L$ with $\sigma_L$ given by \eqref{eq:sigmae_equi}, which are the two equations obtained in \cite{Kuzuu_Doi_JPhysSocJapan83}, provided the external magnetic field considered in \cite{Kuzuu_Doi_JPhysSocJapan83} is set to $0$. Finally, formulas \eqref{eq:al123}-\eqref{eq:al6} for $n=3$ and $\zeta = 0$ are identical with Formula (6.2) of \cite{Kuzuu_Doi_JPhysSocJapan83}. So, our model is consistent with \cite{Kuzuu_Doi_JPhysSocJapan83}. 

Then, Refs. \cite{E_Zhang_MAA06, Wang_Zhang_Zhang_CPAM15} consider a spatially non-homogeneous setting, but still with a constant and uniform $\rho$ (we easily see that $\rho =$ Constant is consistent with both the kinetic model \eqref{eq:kinetic_eps} and the fluid one \eqref{conservation-mass} due to the incompressibility conditions \eqref{eq:divu=0_eps} and \eqref{eq:divu=0_final}). Their setting is $n=3$, $\zeta = \frac{1}{2}$ and $\Lambda = 1$. In this case, we see that formulas \eqref{eq:al123}-\eqref{eq:al6} are identical with Formulas (2.6), (2.7) of \cite{Wang_Zhang_Zhang_CPAM15}. If $\rho = $ Constant, then, $\eta = $ Constant as well. So, the Ericksen stresses and molecular field reduce to 
\begin{equation}
\sigma_E = - k \nabla_x \Omega \nabla_x \Omega^T, \quad \rho H = k \Delta_x \Omega, \, \, \mbox{ with } \, \, k= \frac{2 \beta}{\alpha} \eta^2, 
\label{eq:sigE_H_rhocst}
\end{equation}
which are the corresponding expressions (see top of p. 7) of \cite{Wang_Zhang_Zhang_CPAM15}. With these expressions, our model reduces to \eqref{prop:Omega_eq_alter} coupled with \eqref{eq:NS_final}-\eqref{eq:sigmae_equi} and \eqref{eq:sigE_H_rhocst}. It is identical with the model obtained in \cite{Wang_Zhang_Zhang_CPAM15}.

So, our model is consistent with the literature but has two additional features: the consideration of an arbitrary dimension $n \geq 2$ and the spatial non-homogeneity of $\rho$ (and consequently, of $\eta$) which brings additional components to the elastic stresses and, as we will see below, to the elastic energy. Non-uniform $\eta$ has been previously considered in \cite{Calderer_Liu_SIAP00, Calderer_time_SIMA02, Ericksen_liquid_ARMA91, Lin_nonlinear_CPAM89, Lin_nematic_CPAM91, Lin_global_ARMA15}, but to the best of our knowledge, none has explicitly linked it to the polymer density and to kinetic theory.  

A well-posedness theory of System \eqref{conservation-mass}-\eqref{eq:sigma_elas_equi} is outside the scope of this paper (see e.g. \cite{Huang_Lin_Wang_CMP14, Lin_Lin_Wang_ARMA10, Lin_Liu_nonparabolic_CPAM95, Lin_Liu_existence_ARMA00, wang2013well} for existence results of the Ericksen-Leslie system in a variety of forms). Note however that a condition for the well-posedness of the parabolic equation \eqref{eq:Omega} is that $\frac{\beta c}{\Lambda}>0$. This is indeed ensured by Prop. \ref{prop:const_c}. 

The main objective of this paper is to provide a (formal) derivation of Eqs. \eqref{conservation-mass}, \eqref{eq:Omega} using the moment method and the generalized collision invariant (GCI) concept. Prior to this, in Section \ref{sec:local_equilibria}, we will return to the determination of the stable equilibria of the Doi model and provide support to Conjecture \ref{conject:stable_0} and to Formula \ref{eq:eta} linking $\rho$ and $\eta$. Then, in Section \ref{sec:GCI}, we develop the GCI concept and discuss its rationale and how it can be linked to the Hilbert expansion procedure. The derivation of \eqref{eq:Omega} itself will be performed in Section \ref{sec:omega_abstract}. The second main objective of the paper is to provide expressions for the Leslie and Ericksen stresses in arbitrary dimension and for spatially inhomogeneous densities, which, to the best of our knowledge, has not been considered before. As these computations are lengthy, they are deferred to Appendix \ref{sec:app_sec_main_res}. Other auxiliary results can be found in this appendix and in the subsequent ones, Appendices \ref{sec:app_GCI} and \ref{sec:app_deriv_Om}.

%%%%%%%%%%%%%%%%%%%%%%%%%%%%%%%%%%%%%%%%%%%%%%%%%%%%%%%%%%%%%%%%
%%%%%%%%%%%%%%%%%%%%%%%%%%%%%%%%%%%%%%%%%%%%%%%%%%%%%%%%%%%%%%%%

\subsection{Energetics of the Ericksen-Leslie system}
%\label{subsec:energetics}

Next, we define the following energies:

\begin{definition}[Oseen-Franck and Ericksen-Leslie energies]
(i) The Oseen-Franck energy is defined by:
\begin{eqnarray}
{\mathcal E}_F &=& \frac{2 \beta}{\alpha} \int_{{\mathbb R}^n} \frac{|\nabla_x (\eta \Omega)|^2}{2} \, dx - \alpha \beta \frac{n-1}{n} \int_{{\mathbb R}^n} \frac{|\nabla_x \rho|^2}{2} \, dx - \beta \frac{n+1}{n \alpha} \int_{{\mathbb R}^n} \frac{|\nabla_x \eta|^2}{2} \, dx \nonumber \\
&=:& {\mathcal E}_F^\Omega + {\mathcal E}_F^\rho + {\mathcal E}_F^\eta. 
\label{eq:Franck_ener}
\end{eqnarray}
(ii) The Ericksen-Leslie energy is defined by 
$$
{\mathcal E}_{EL} = \int_{{\mathbb R}^n} |u|^2 \, dt + \frac{1}{\mathrm{Re} \mathrm{Er}} {\mathcal E}_F. 
$$
%\label{def:Franck_EL_ener} 
\end{definition}

\begin{remark}
(i) If $\rho$ is uniformly constant (and hence, $\eta$ too), ${\mathcal E}_F$ reduces to 
$$
{\mathcal E}_F = \frac{2 \beta \eta^2}{\alpha} \int_{{\mathbb R}^n} \frac{|\nabla_x \Omega|^2}{2} \, dx,
$$ 
which is the classical Oseen-Franck elastic energy \cite {E_Zhang_MAA06, Wang_Zhang_Zhang_CPAM15}. The additional terms ${\mathcal E}_F^\rho$ and ${\mathcal E}_F^\eta$ make up for the non-uniformity of $\rho$ and $\eta$. 

\noindent 
(ii) Using \eqref{eq:eta}, we find an alternate expression of ${\mathcal E}_F$: 
$$ {\mathcal E}_F = \frac{2 \beta}{\alpha} \int_{{\mathbb R}^n} \eta^2 \, \frac{|\nabla_x \Omega|^2}{2} \, dx +  \frac{(n-1) \beta}{n \alpha} \int_{{\mathbb R}^n}  \Big( 1 - \frac{1}{S_2^2} \big( 1 - \eta \frac{S'_2}{S_2} \big)^2 \Big) \frac{|\nabla_x \eta|^2}{2} . 
$$
In particular, we see that this energy is positive if the following relation holds
$$  1 - \frac{1}{S_2^2} \big( 1 - \eta \frac{S'_2}{S_2} \big)^2 \geq 0. $$
The investigation of this property is left to future work. 
\label{rem:Franck_ener}
\end{remark}

Now, we have the following proposition, which relates the molecular field to the derivative of the Franck energy with respect to the orientation field $\Omega$. 

\begin{proposition}[Relation between the Franck energy and the molecular field]
We have the following relation:
\begin{equation} 
\rho H = - \frac{\delta {\mathcal E}_F}{\delta \Omega} (\eta, \Omega) = \frac{2 \beta}{\alpha} \, \eta \, \Delta_x (\eta \Omega), 
\label{eq:rel_H_EF}
\end{equation}
where $\frac{\delta {\mathcal E}_F}{\delta \Omega}(\eta, \Omega)$ is the functional derivative of ${\mathcal E}_F$ with respect to the field $\Omega$ evaluated at the pair $(\eta, \Omega)$. 
%\label{prop:rel_H_EF}
\end{proposition} 

\medskip
\noindent
\textbf{Proof.} For a $n \times n$ tensor $S$, we introduce the following energy density
$$ e_F^\Omega (S) = \frac{2 \beta}{\alpha} \, \frac{|S|^2}{2}, $$
so that we can write 
$$ {\mathcal E}_F^\Omega = \int_{{\mathbb R}^n} e_F^\Omega  \big( \nabla_x (\eta \Omega) \big) \, dx. $$
Now, straightforward computations show that the functional derivative $\frac{\delta {\mathcal E}_F}{\delta \Omega}$ is given by 
$$ \frac{\delta {\mathcal E}_F}{\delta \Omega} (\eta, \Omega) = \frac{\delta {\mathcal E}_F^\Omega}{\delta \Omega} (\eta, \Omega) = - \eta \nabla_x \cdot \big( \frac{\partial e_F^\Omega}{\partial S} \big( \nabla_x (\eta \Omega) \big) \Big) = - \frac{2 \beta}{\alpha} \, \eta \, \Delta_x (\eta \Omega) = - \rho H, $$
where the first equality is due to the fact that the energies ${\mathcal E}_F^\rho$ and ${\mathcal E}_F^\eta$ do not depend on $\Omega$, and the last one, to \eqref{eq:eta}. Then, Eq. \eqref{eq:rel_H_EF} follows. \endproof

The following proposition gives the energy identity for the Ericksen-Leslie system. Its proof is developed in Appendix \ref{sec:ener_EL}

\begin{proposition}[Energy identity for the Ericksen-Leslie system] 
We have the following identity:
\begin{eqnarray}
&& \hspace{-1cm} 
\frac{d {\mathcal E}_{EL}}{dt} + {\mathcal D}_{EL} = 0, \label{eq:ener_EL} \\
&& \hspace{-1cm} 
{\mathcal D}_{EL} =  \frac{1}{\mathrm{Re}} \int_{{\mathbb R}^n} |\nabla_x u|^2 \, dx + \frac{1}{\mathrm{Re} \mathrm{Er}} \int_{{\mathbb R}^n} \rho \Big\{ \Big( \alpha_1 + \frac{\gamma_2^2}{\gamma_1} \Big) \big( E:(\Omega \otimes \Omega) \big)^2 + \alpha_4 |E|^2 \nonumber \\
&& \hspace{5cm} 
+ \Big( \alpha_5 + \alpha_6 - \frac{\gamma_2^2}{\gamma_1} \Big) |E \Omega|^2 + \frac{1}{\gamma_1} | P_{\Omega^\bot} H |^2 \Big\} \, dx . \nonumber
%\label{eq:ener_dissip_EL}
\end{eqnarray}
%\label{prop:ener_EL}
\end{proposition}

\begin{remark}
(i) The use of this energy identity to derive a priori bounds for the solution of the Ericksen-Leslie equations is subject to two conditions: first, that the Oseen-Franck energy is positive as already mentioned in Remark \ref{rem:Franck_ener}; second, that the dissipation functional ${\mathcal D}_{EL}$ is positive as well, which is not obvious given that the coefficients are not all positive. In \cite{Wang_Zhang_Zhang_CPAM15}, it is shown that, in the case $n=3$, $\zeta = \frac{1}{2}$ and $\Lambda = 1$, ${\mathcal D}_{EL}$ is positive. Besides, conditions for the positive-definiteness of ${\mathcal D}_{EL}$ with coefficients which are not necessarily linked with a microscopic model can be found in \cite{wang2013well}. The inspection of the positivity of ${\mathcal E}_F$ and ${\mathcal D}_{EL}$ for the present model is left to future work. 

(ii) It is expected that this energy identity is the limit as $\varepsilon \to 0$ of the free-energy dissipation identity \eqref{eq:free_ener_dissip} of the Doi-Navier-Stokes system. This is indeed formally shown in \cite{E_Zhang_MAA06}. However, due to the presence of the square of the Deborah number at the denominator of \eqref{eq:free_ener_dissip_sc}, we expect that the limiting free-energy dissipation identity will involve the first order correction $f^1 = \lim_{\varepsilon \to 0} \frac{f^\varepsilon - f^0}{\varepsilon}$. Showing that the terms involving $f^1$ eventually vanish is not obvious and left to future work. 
%\label{rem:EL_energy}
\end{remark}

%%%%%%%%%%%%%%%%%%%%%%%%%%%%%%%%%%%%%%%%%%%%%%%%%%%%%%%%%%%%%%%%
%%%%%%%%%%%%%%%%%%%%%%%%%%%%%%%%%%%%%%%%%%%%%%%%%%%%%%%%%%%%%%%%
%%%%%%%%%%%%%%%%%%%%%%%%%%%%%%%%%%%%%%%%%%%%%%%%%%%%%%%%%%%%%%%%
%%%%%%%%%%%%%%%%%%%%%%%%%%%%%%%%%%%%%%%%%%%%%%%%%%%%%%%%%%%%%%%%

\setcounter{equation}{0}
\section{Local equilibria}
\label{sec:local_equilibria}

In this section, we develop the rationale for Conjecture \ref{conject:stable_0}. Since we aim at formal convergence results only, we suppose that the solution $f^\varepsilon$ to \eqref{eq:kinetic_abstract} satisfies  
$$f^\varepsilon \to f \quad \mbox{ as } \quad \varepsilon \to 0 \quad \mbox{ as smoothly as needed.}$$ Then, from \eqref{eq:kinetic_abstract}, it follows that $f$ should satisfy \eqref{eq:equi_def}, i.e. should be an equilibrium for any $(x,t)$. Eq. \eqref{eq:equi_def} leaves the dependence of $f$ on $(x,t)$ undetermined. Such an equilibrium is called 'local' (by contrast to a global equilibrium where $f$ should not depend on $(x,t)$). 

In this section, our goal is to determine the stable equilibria. Indeed, we anticipate that only stable equilibria can lead to a long time dynamics described by hydrodynamic equations. First, we should note that local equilibria are known in any dimension $n$ \cite{Hoffman_Wang_CMS06} (see also \cite{constantin2004note, Fatkullin_Slastikov_CMS05, Luo_Zhang_Zhang_Nonlinearity04} for the case $n=2$ and \cite{constantin2004asymptotic, Fatkullin_Slastikov_nonlinearity05, Liu_Zhang_Zhang_CMS05, Zhang_Zhang_PhysD07, Zhou_Wang_Forest_Wang_Nonlinearity05} for the case $n=3$). However, the stability of these equilibria is not known for general dimension $n$ but only for $n=2$ \cite{Fatkullin_Slastikov_CMS05}, $n=3$ \cite{Fatkullin_Slastikov_nonlinearity05, Liu_Zhang_Zhang_CMS05} and $n=4$ \cite{Frouvelle_2021}. These results strongly support a conjecture about the stable equilibria in general dimension $n$ that we will make below and whose rigorous investigation is deferred to future work. We first need to introduce a set of notations and intermediate results.  

\begin{definition} [Auxiliary operator]
Let $S$ be a trace-free symmetric matrix. Then, the auxiliary operator $L_S$ is given by 
\begin{equation}
L_S f = \nabla_\omega \cdot \Big[ G_S \nabla_\omega \big( \frac{f}{G_S} \big) \Big]. 
\label{eq:def_LSig}
\end{equation}
with $G_S$ given by \eqref{eq:vonmises2}.
\end{definition}

The relation between the collision operator $C(f)$ and the auxiliary operator $L_S$ is given by the following lemma. Note that $L_S$ is NOT the linearization of $C$ about $G_S$. 

\begin{lemma} [Relation between $C$ and $L$]
We have 
\begin{equation}
C(f)  =  L_{\alpha \, \rho_f \, Q_f} f. 
\label{eq:VM1}
\end{equation}
\label{lem:relCLSig}
\end{lemma}

\noindent
{\bf Proof of Lemma \ref{lem:relCLSig}.} 
We can write
\begin{eqnarray*}  
L_S f &=& \nabla_\omega \cdot \Big[ \nabla_\omega f - f \nabla_\omega \big( \log G_S \big) \Big] . 
\end{eqnarray*}  
But $- \log G_S = - \omega \cdot S \omega + \log Z_S$. So, $- \log G_{\alpha \rho_f Q_f} = U_f^0 + \tilde Z(f)$ where $\tilde Z(f)$ does not depend on $\omega$. Thus, $- \nabla_\omega ( \log G_{\alpha \rho_f Q_f} ) = \nabla_\omega U_f^0$ and so,  $L_{\alpha \rho_f Q_f} = C(f)$,
%, and $ \nabla_\omega ( \log ( \frac{1}{G_S} ) ) = - 2 P_{\omega^\bot} S \omega$. Therefore, 
%$$
%L_{\alpha \, \rho_f \, Q_f} f = \nabla_\omega \cdot \Big[ \nabla_\omega f - 2 \alpha \, \rho_f \, f P_{\omega^\bot} Q_f \omega \Big] = C(f) , 
%$$
thanks to \eqref{eq:coll_op_alter}.  \endproof

Now, we have a first result:

\begin{lemma} [First step towards a characterization of the equilibria] 
(i) Let $f \geq 0$, $f \not = 0$ be an equilibrium. Then, there exists $\rho >0$ and a trace-free symmetric matrix $Q$ such that 
\begin{equation}
f = \rho \, G_{\alpha \, \rho \, Q}. 
\label{eq:equi_1}
\end{equation}
(ii) Reciprocally, let $f$ be given by \eqref{eq:equi_1}. Then, $f$ is an equilibrium if and only if $Q$ satisfies the fixed-point equation also known as the compatibility equation:
\begin{equation}
Q = Q_{\rho \, G_{\alpha \, \rho \, Q}}, 
\label{eq:fixed_point}
\end{equation}
where we recall that for a distribution $f$, $Q_f$ is given by \eqref{eq:orienttensor}. 
%\label{lem:char_equi}
\end{lemma}

\noindent
\textbf{ Proof.} 
(i) Suppose $C(f)=0$. Letting $S = \alpha \, \rho_f \, Q_f$, \eqref{eq:VM1} implies $L_S f= 0$. Multiplying \eqref{eq:def_LSig} by $f/G_S$, integrating over ${\mathbb S}^{n-1}$ and using Green's formula leads to 
$$ \int_{{\mathbb S}^{n-1}} G_S \, \Big| \nabla_\omega \Big( \frac{f}{G_S} \Big) \Big|^2 \, d \omega = 0. $$
Since the quantity inside the integral is nonnegative, and $G_S >0$, this implies $\nabla_\omega ( \frac{f}{G_S} ) = 0$. So, there exists $\rho > 0$ such that $f = \rho \, G_S$ which leads to \eqref{eq:equi_1}. 

\medskip
\noindent
(ii) Let $f$ be given by \eqref{eq:equi_1}. Then, since $G_{\alpha \, \rho \, Q}$ is a probability density, we have $\rho_f = \rho$. Now, from the proof of Part (i), if $f$ is an equilibrium, then $f= \rho_f G_{\alpha \, \rho_f \, Q_f}$. We deduce that $G_{\alpha \, \rho \, Q_f} = G_{\alpha \, \rho \, Q}$, and, by taking the logarithm, that
$$  \omega \cdot (Q_f-Q) \omega = \frac{1}{\alpha \, \rho} \, \big( \log Z_{\alpha \, \rho \, Q_f} - \log Z_{\alpha \, \rho \, Q} \big)=:\mu, $$
where $\mu$ is a constant, independent of $\omega$. So, $Q_f-Q - \mu \, \mathrm{Id}$ is the matrix of a quadratic form which is zero on ${\mathbb S}^{n-1}$ and so, by homogeneity, on ${\mathbb R}^n$. Thus, $Q_f-Q - \mu \, \mathrm{Id} = 0$ and, owing to the fact that $Q_f$ and $Q$ are trace-free, we have $\mu = 0$. It follows that $Q_f = Q$. Replacing $f$ by its expression \eqref{eq:equi_1}, we get \eqref{eq:fixed_point}. \endproof

To complete the characterization of the equilibria, we  need to solve the compatibility equation \eqref{eq:fixed_point}. As pointed out above, this has been done in any dimension $n$ in \cite{Hoffman_Wang_CMS06} (see also \cite{Luo_Zhang_Zhang_Nonlinearity04} for $n=2$ and \cite{Fatkullin_Slastikov_nonlinearity05, Liu_Zhang_Zhang_CMS05, Zhou_Wang_Forest_Wang_Nonlinearity05} for $n=3$). This result is summarized without proof in the following lemma

\begin{lemma} [Final characterization of the equilibria \cite{Hoffman_Wang_CMS06}]
Let $f$ be an an equilibrium. Then $Q_f$ has at most two distinct eigenvalues. 
\begin{itemize}
\item If all eigenvalues of $Q_f$ are identical, then $Q_f = 0$ and $f = \rho$ is a uniform equilibrium. 
\item If $Q_f$ has exactly two distinct eigenvalues, denote by $\lambda_f$ its largest eigenvalue and by ${\mathcal Y}_f$ the associated eigenspace, supposed of dimension $d$ such that $1 \leq d \leq n-1$. Then, $0<\lambda_f <\frac{1}{d} - \frac{1}{n}$ and $Q_f$ is written
\begin{equation}
Q_f = B_{\lambda_f, {\mathcal Y}_f} := \lambda_f \big( P_{{\mathcal Y}_f} - \frac{d}{n-d} P_{{\mathcal Y}_f^\bot} \big), 
\label{eq:equi_Qtens}
\end{equation}
where $ P_{{\mathcal Y}_f}$ and $P_{{\mathcal Y}_f^\bot}$ are the orthogonal projections of ${\mathbb R}^n$ onto ${\mathcal Y}_f$ and ${\mathcal Y}_f^\bot$ respectively. Then, $f$ is of the form
$$
f = \rho_d^n(\lambda_f) \, G_{\alpha \, \rho_d^n(\lambda_f) \, B_{\lambda_f, {\mathcal Y}_f}}, 
$$
where $\rho_d^n$: $[0,\frac{1}{d} - \frac{1}{n}) \to [0,\infty)$, $\lambda \mapsto \rho_d^n(\lambda)$ is a specific function (not detailed here except for the case $d=1$, see below). Furthermore, $\lambda_f$ is a root of the equation
\begin{equation}
\rho_d^n (\lambda) = \rho. 
\label{eq:rhodn_lmbda}
\end{equation}
The existence and number of classes of equilibria such that $\rho_f = \rho$ are determined by the existence and number of roots $\lambda$ of Eq. \eqref{eq:rhodn_lmbda}. A given root $\lambda$ gives rise to a family of equilibria parametrized by the Grassmann manifold Gr$(k,n)$ of $d$-dimensional vector subspaces ${\mathcal Y}$ of ${\mathbb R}^n$.   
\end{itemize}
\label{lem:equilibria}
\end{lemma}

\begin{figure}[ht!]
\centering
\subfloat[$n=2$]{\includegraphics[trim={3.5cm 9.cm 2.cm 4.5cm},clip,width=7cm]{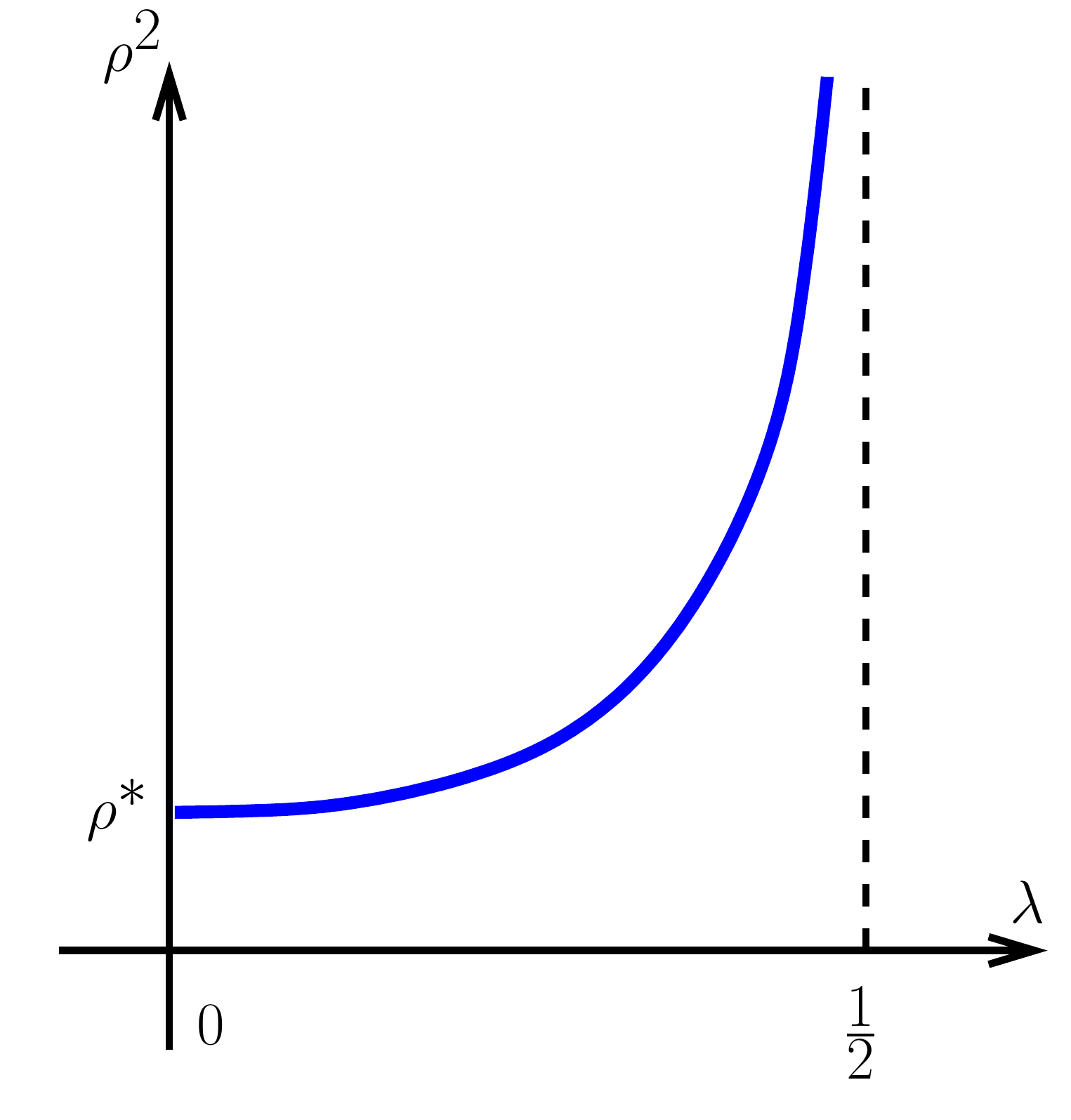}\label{fig:phase_diag_n=2}}
\hspace{0.2cm}
\subfloat[$n \geq 3$]{\includegraphics[trim={3.5cm 9.cm 2.cm 4.5cm},clip,width=7cm]{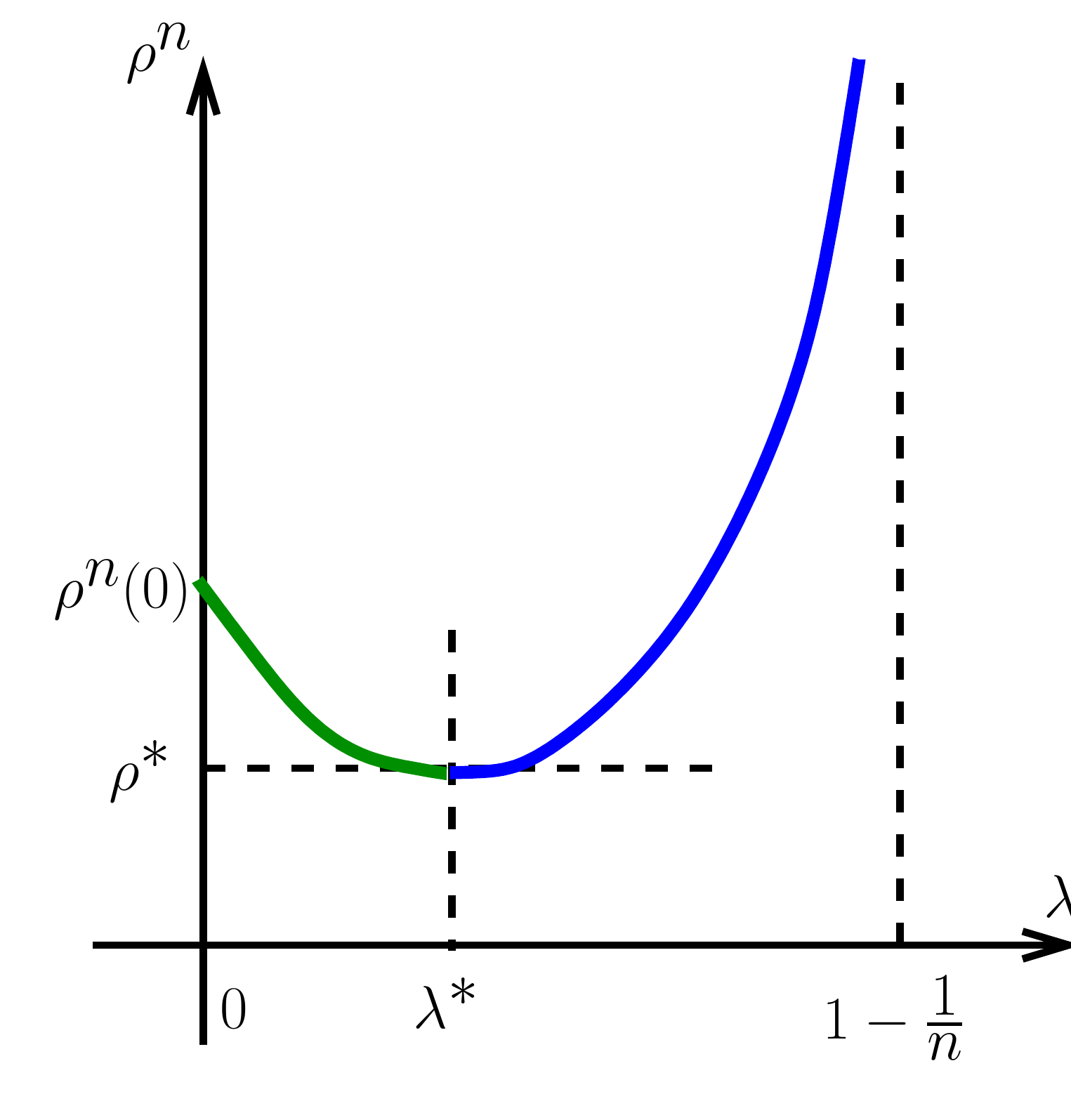}\label{fig:phase_diag_ngeq3}}
\caption{Graphical representation of the function $\lambda \mapsto \rho^n(\lambda)$ (after \cite{Hoffman_Wang_CMS06}). (a) case $n=2$. (b) case $n \geq 3$. The portions of the curves that correspond to stable equilibria are in blue, the unstable ones, in green. }
%\label{fig:phase_diag}
\end{figure}

Here, we are only interested in the case $d=1$ as we will conjecture that this is the only case which includes stable equilibria (see conjecture \ref{conject:stable} below). For simplicity, $\rho^n$ stands for the function $\rho_1^n$. In the case $n=2$, $\rho^2$ is monotonously increasing and maps $[0, \frac{1}{2})$ onto the interval $[\rho^*, +\infty)$ with $\rho^* = \rho^2(0)$ (see Fig. \ref{fig:phase_diag_n=2}). In the case $n \geq 3$, $\rho^n$ is decreasing in the interval $[0,\lambda^*]$ and increasing in $[\lambda^*, 1 - \frac{1}{n})$. Thus $\rho^*=\rho(\lambda^*)$ is a global minimum of $\rho^n$ (see Fig. \ref{fig:phase_diag_ngeq3}). In all cases, the equation $\rho_n(\lambda) = \rho$ has a solution if and only if $\rho \geq \rho^*$ and this solution is unique in the case $n=2$ while, in the case $n \geq 3$, there are two solutions if $\rho \in (\rho^*, \rho^n(0)]$, and one solution  if $\rho \in \{ \rho^* \} \cup (\rho^n(0),\infty)$ (see \cite{Hoffman_Wang_CMS06} for details). 

As already stated, for general~$n$, the stability of the equilibria described in Lemma \ref{lem:equilibria} is not known yet. However, their stability is known for $n=2$ \cite{Fatkullin_Slastikov_CMS05}, $n=3$ \cite{Fatkullin_Slastikov_nonlinearity05, Liu_Zhang_Zhang_CMS05} and $n=4$ \cite{Frouvelle_2021}. Based on these results, we formulate the following conjecture for any dimension $n \geq 2$ and refer to the above-mentioned references for details on the notion of stability involved.  

\begin{conject} [Stable anisotropic equilibria]
For any dimension $n\geq 2$, the branch of solutions to the equation $\rho^n(\lambda) = \rho$ (which corresponds to $d=1$) with largest $\lambda$, which is defined for $\rho \in (\rho^*, \infty)$, corresponds to the unique class of stable anisotropic equilibria.   
\label{conject:stable}
\end{conject}

We denote  by the function $\lambda$: $(\rho^*,+\infty) \to (\lambda^*, 1 - \frac{1}{n})$, $\rho \mapsto \lambda(\rho)$, the largest solution to $\rho^n(\lambda) = \rho$. With Conjecture \ref{conject:stable}, the stable equilibria $f$ correspond to the class of equilibria described in Lemma \ref{lem:equilibria}, Case 2, with $d=1$ and $\lambda_f = \lambda(\rho)$. In this case, ${\mathcal Y}_f$ is one-dimensional and thus, spanned by a unique normalized vector (up to a sign) $\Omega \in {\mathbb P}^{n-1}$. Hence, we have $P_{{\mathcal Y}_f} = \Omega \otimes \Omega$ and $P_{{\mathcal Y}_f^\bot} = P_{\Omega^\bot}$. Then by \eqref{eq:equi_Qtens}, 
\begin{equation}
Q_f = \frac{n}{n-1} \lambda(\rho) \, A_\Omega, 
\label{eq:Qtens_equi}
\end{equation}
where $A_\Omega$ is the normalized uniaxial tensor given by \eqref{eq:AOm}. 
Defining 
\begin{equation}
\eta(\rho) = \frac{n}{n-1} \, \alpha \, \rho \, \lambda(\rho),
\label{eq:eta=alrholam}
\end{equation}
from \eqref{eq:fixed_point} we get that the equilibria are of the form $f=\rho \, G_{\eta(\rho) A_\Omega}$where $\rho$ is arbitrary as long as $\lambda(\rho)$ is defined, i.e. $\rho \in (\rho^*, \infty)$, and where $\Omega$ is arbitrary in ${\mathbb P}^{n-1}$. Hence, Conjecture \ref{conject:stable_0} is a direct consequence of Conjecture \ref{conject:stable}, provided we show that the function $\rho \mapsto \eta(\rho)$ is the one given by Proposition \ref{prop:eta(rho)_defin}, which we do now:

\medskip
\noindent
\textbf{Proof of Proposition \ref{prop:eta(rho)_defin}.}
Equating \eqref{eq:Qtens_equi} with \eqref{eq:Q_equib} and using \eqref{eq:eta=alrholam}, we get \eqref{eq:eta}. The root with the largest $\eta$ must be chosen because this corresponds to the choice of largest $\lambda$ in Conjecture \ref{conject:stable} (as $\lambda$ is proportional to $\eta$ by \eqref{eq:eta=alrholam}). \endproof 

From \eqref{eq:kinetic_abstract} and Conjecture \ref{conject:stable}, we deduce the:

\begin{corollary} [Local equilibria] Let $f$ be the formal limit of $f^\varepsilon$ as $\varepsilon \to 0$ and suppose that $u^\varepsilon \to u$ smoothly. On the open set ${\mathcal B}$ defined by \eqref{def:set_B}, $f$ is given by \eqref{eq:localequ}
where $\rho=\rho_f$: $(x,t) \in {\mathbb R}^n \times [0,\infty) \mapsto [\rho^*,\infty)$ and $\Omega$: $(x,t) \in {\mathbb R}^n \times [0,\infty) \mapsto {\mathbb P}^{n-1}$ are functions such that $f$ satisfies
\begin{equation}
T_u(f) = \lim_{\varepsilon \to 0} \frac{C(f^\varepsilon)}{\varepsilon} . \label{eq:kinetic_lim}
\end{equation}
\label{cor:localequ}
\end{corollary}

Note that $\rho = \rho_f$ is the local density associated to $f$, while $\Omega(x,t)$ if the axis of the uniaxial Q-tensor $Q_f$ thanks to \eqref{eq:Q_equib}. The restriction to the set ${\mathcal B}$ is needed to ensure that $\eta(\rho(x,t))$ is well-defined. The determination of the functions $(\rho,\Omega)$ such that \eqref{eq:kinetic_lim} holds is quite challenging, due to the presence of $\varepsilon$ in the denominator at the right-hand side. It will require the Generalized Collision Invariant concept as detailed below.

%%%%%%%%%%%%%%%%%%%%%%%%%%%%%%%%%%%%%%%%%%%%%%%%%%%%%%%%%%%%%%%%
%%%%%%%%%%%%%%%%%%%%%%%%%%%%%%%%%%%%%%%%%%%%%%%%%%%%%%%%%%%%%%%%
%%%%%%%%%%%%%%%%%%%%%%%%%%%%%%%%%%%%%%%%%%%%%%%%%%%%%%%%%%%%%%%%
%%%%%%%%%%%%%%%%%%%%%%%%%%%%%%%%%%%%%%%%%%%%%%%%%%%%%%%%%%%%%%%%

\setcounter{equation}{0}
\section{Generalized collision invariants}
\label{sec:GCI}

\subsection{Collision invariant}
%\label{subsec:CI}

We first recall the notion of Collision Invariant (CI). The goal is to eliminate the singular right hand side of \eqref{eq:kinetic_lim} by using integration against appropriate test functions. More precisely we have:

\begin{definition}
A Collision Invariant (CI) $\psi(\omega)$ is a function such that 
$$ \int_{{\mathbb S}^{n-1}} C(f) \, \psi \, d \omega = 0, \qquad \forall f. $$
%\label{def:CI}
\end{definition}

Here, we do not specify any regularity requirement on $\psi$ since our goal is to develop a formal theory only. If $\psi$ is a CI, using it as a test function for \eqref{eq:kinetic_abstract}, we have, after integration with respect to $\omega$ and omitting $\varepsilon$ as the identity is valid for any $\varepsilon$:
\begin{eqnarray}
&&\hspace{-1cm}
\partial_t \Big( \int_{{\mathbb S}^{n-1}} f \psi \, d \omega \Big)  + \nabla_x \cdot \Big(  u \int_{{\mathbb S}^{n-1}} f \psi \, d \omega \Big) - \int_{{\mathbb S}^{n-1}} \nabla_\omega \psi \cdot (\Lambda P_{\omega^\bot} E - W) \omega  \, f \, d\omega \nonumber \\
&&\hspace{4cm}
- 2 \alpha \, \beta \int_{{\mathbb S}^{n-1}} \nabla_\omega \psi \cdot P_{\omega^\perp} \Delta_x (\rho_f \, Q_f)\, \omega \, f \, d \omega = 0, 
\label{eq:moments}
\end{eqnarray}
which is an evolution equation for the moment $\int_{{\mathbb S}^{n-1}} f \psi \, d \omega$. Since this equation does not depend on $\varepsilon$, it is still verified by the solution of \eqref{eq:kinetic_lim}. 
We have an obvious CI, namely, $\psi = 1$, which leads to the mass conservation (or continuity) equation 
\begin{equation}
\partial_t \rho_f  + \nabla_x \cdot ( \rho_f u )  = 0. 
\label{eq:conserv_mass_general}
\end{equation}  
In particular, taking the limit $\varepsilon \to 0$, it shows \eqref{conservation-mass}. 
As $u$ is divergence free thanks to \eqref{eq:divu=0_eps}, \eqref{eq:conserv_mass_general} can be equivalently written 
\begin{equation}
D_t \rho_f = 0,  
\label{transport-mass}
\end{equation}
with $D_t$ given by \eqref{eq:Dt}.

Any odd function $\psi$ of $\omega$ is also a CI. However, it is not invariant when $\omega$ is changed into $-\omega$, a condition that has been enforced throughout this work (see e.g. \eqref{eq:symmetry}). Indeed, Eq. \eqref{eq:moments} with odd functions $\psi$ have all their terms identically zero and do not provide any useful information. We do not have any other obvious CI. Therefore, we are lacking an equation for $\Omega$. In order to overcome this problem, we use the concept of ``Generalized Collision Invariant (GCI)'' introduced in \cite{Degond_Motsch_M3AS08} and adapted to the present context.

%%%%%%%%%%%%%%%%%%%%%%%%%%%%%%%%%%%%%%%%%%%%%%%%%%%%%%%%%%%%%%%%
%%%%%%%%%%%%%%%%%%%%%%%%%%%%%%%%%%%%%%%%%%%%%%%%%%%%%%%%%%%%%%%%

\subsection{Generalized collision invariant: definition and characterization}
\label{subsec:GCI}

\medskip
To introduce the GCI concept, we first need some additional notations and definitions. 

\begin{definition} [and notations]
(i) $\mathcal{S}^0_n$ is the vector space of symmetric trace free $n\times n$ matrices. \\
(ii) $\mathcal{U}^0_n$ is the subset of $\mathcal{S}^0_n$ consisting of tensors whose leading eigenvalue is equal to $\frac{n-1}{n}$ and is simple. \\
(iii) We denote by $\lambda_f$ the leading eigenvalue of $Q_f$ and by $\eta_f$ the following quantity:
\begin{equation} 
\eta_f = \alpha \, \rho_f \, \frac{n}{n-1} \lambda_f. 
\label{eq:etaf}
\end{equation}
From \eqref{eq:eigenineq}, we have
$$ 0 \leq \frac{n}{n-1} \lambda_f \leq 1. $$
Note that in general, $\lambda_f$ may not be simple. \\
(iv) If $Q_f \not = 0$, then $\lambda_f \not = 0$ and  we define the ``Normalized Q-Tensor (NQT) of $f$'',  $\Sigma_f$ by
\begin{equation} 
\Sigma_f = \frac{n-1}{n} \, \frac{Q_f}{\lambda_f}. 
\label{eq:defSigmaf}
\end{equation}
$\Sigma_f \in \mathcal{S}^0_n$. Its leading eigenvalue is $\frac{n-1}{n}$ which, again, may not be simple. \\
(vi) Let $\Sigma \in \mathcal{U}^0_n$. We denote by $\Omega_\Sigma \in {\mathbb P}^{n-1}$ the normalized eigenvector (up to a sign) associated with the simple eigenvalue $\frac{n-1}{n}$ of $\Sigma$. Note that  the tensor $A_{\Omega_\Sigma}$ is uniquely defined, irrespective of the choice of the sign of $\Omega_\Sigma$. \\
(v) Suppose $\Sigma_f \in \mathcal{U}^0_n$. Then, $\Omega_{\Sigma_f}$ is simply denoted by $\Omega_f$. 
%\label{def:NQT}
\end{definition}

\begin{remark}
From \eqref{eq:Q_equib}, we get that $\Sigma_{\rho \, G_{\eta A_\Omega}} = A_{\Omega}$ meaning that the NQT's of the stable anisotropic equilibria are all equal to $A_\Omega$. 
\label{rem:order_parameter}
\end{remark}

\medskip
We recall that the auxiliary operator $L_S$ for $S \in \mathcal{S}^0_n$ is defined by \eqref{eq:def_LSig}. The GCI are now defined in the following

\begin{definition}
Let $(\eta,\Sigma) \in (0,\infty) \times {\mathcal U}^0_n$. A Generalized Collisional Invariant (GCI) associated to the pair $(\eta,\Sigma)$ is a function $\psi$ such that
\begin{equation}
\label{def-GCI-S}
\int_{{\mathbb S}^{n-1}} (L_{\eta \Sigma} f) \, \psi \, d\omega =0 \quad \mbox{ for all } \, f \, \text{ such that  } \,  P_{\Omega_\Sigma^\bot} ( Q_f \Omega_\Sigma) =0.
\end{equation}
The set of GCI associated to a given pair $(\eta,\Sigma) \in (0,\infty) \times {\mathcal U}^0_n$ is a linear vector space and is denoted by~$\mathcal{C}_{\eta \Sigma}$. 
\label{def:GCI}
\end{definition}

\noindent
There is a rationale for this definition, which is developed in Section \ref{subsec:ratGCI} below.

\medskip
The following lemma gives the equation satisfied by the GCI: 

\begin{lemma}
Let $(\eta,\Sigma) \in (0,\infty) \times {\mathcal U}^0_n$. Then $\psi \in \mathcal{C}_{\eta \Sigma}$ if and only if there exists $V \in \{ \Omega_\Sigma \}^\bot$ such that 
\begin{equation}
\nabla_\omega \cdot \big( G_{\eta \Sigma}(\omega) \nabla_\omega \psi \big) = (\omega \cdot \Omega_\Sigma) \, (\omega \cdot V) \, G_{\eta \Sigma}(\omega), \quad \forall \omega \in {\mathbb S}^{n-1}. 
\label{eq:GCI1}
\end{equation}
%\label{lem:GCI1}
\end{lemma}

\noindent
{\bf Proof.} For $\Omega \in {\mathbb S}^{n-1} / \{ \pm 1 \}$, we define the following space of functions: 
\begin{equation}
{\mathcal X}_\Omega = \{ {\mathbb S}^{n-1} \ni \omega \mapsto (\Omega \cdot \omega) \, (V \cdot \omega) \in {\mathbb R} \, \, | \, \, V \in \{\Omega\}^\bot \}, 
\label{eq:defmathcalXOm}
\end{equation}
The space ${\mathcal X}_\Omega$ is a finite-dimensional subspace of $L^2({\mathbb S}^{n-1})$. We first note that for any $f \in L^2({\mathbb S}^{n-1})$, we have 
\begin{eqnarray}
P_{\Omega^\bot} ( Q_f \Omega) = 0 & \Longleftrightarrow & \int_{{\mathbb S}^{n-1}} f(\omega) \, (\omega \cdot V) \, (\omega \cdot \Omega) \, d \omega = 0, \quad \forall V \in \{\Omega\}^\bot \nonumber \\
& \Longleftrightarrow & f \in {\mathcal X}_\Omega^\bot, \label{eq:PTNQ=0}
\end{eqnarray}
where the orthogonality is meant with respect to the standard $L^2$-product on $L^2({\mathbb S}^{n-1})$.

On the other hand, we note that $\int_{{\mathbb S}^{n-1}} (L_{\eta \Sigma} f) \, \psi \, d\omega =0$ is equivalent to saying that $f \in \{ L_{\eta \Sigma}^* \psi \}^\bot$ where again, the orthogonality is meant with respect to the standard $L^2$-product on $L^2({\mathbb S}^{n-1})$ and where $L_{\eta \Sigma}^*$ is the formal $L^2$-adjoint of $L_{\eta \Sigma}$, i.e. 
$$
L_{\eta \Sigma}^* \psi = \frac{1}{G_{\eta \Sigma}} \, \nabla_\omega \cdot ( G_{\eta \Sigma} \, \nabla_\omega \psi). 
$$
Therefore, thanks to \eqref{eq:PTNQ=0}, Condition \eqref{def-GCI-S} is equivalent to saying that 
$$  f \in {\mathcal X}_{\Omega_\Sigma}^\bot \, \Longrightarrow \, f \in \{ L_{\eta \Sigma}^* \psi \}^\bot, $$
or in other words, that ${\mathcal X}_{\Omega_\Sigma}^\bot \subset \{ L_{\eta \Sigma}^* \psi \}^\bot$. Taking the orthogonal to this relation and noting that both ${\mathcal X}_{\Omega_\Sigma}$ and $\mathrm{Span} \{L_{\eta \Sigma}^* \psi \}$ (where for a subset $B$ of a vector space, Span $B$ denotes the subspace generated by $B$) are finite-dimensional, hence, closed subspaces of $L^2({\mathbb S}^{n-1})$, we get $\mathrm{Span} \{L_{\eta \Sigma}^* \psi \} \subset {\mathcal X}_{\Omega_\Sigma}$. In particular, this implies that there exists $V \in \{\Omega_\Sigma\}^\bot$ such that $L_{\eta \Sigma}^* \psi(\omega)= (\omega \cdot \Omega_\Sigma) \, (\omega \cdot V)$, which, upon multiplying by $G_{\eta \Sigma}$, gives \eqref{eq:GCI1}. The converse is straightforward. \endproof

Now, we give an existence theory for the solutions of \eqref{eq:GCI1}. We denote by $H^1({\mathbb S}^{n-1})$ the space of square integrable functions of ${\mathbb S}^{n-1}$ into ${\mathbb R}$ whose derivatives are square integrable and introduce
$$ \dot H^1({\mathbb S}^{n-1}) = \Big\{ u \in H^1({\mathbb S}^{n-1}) \, \Big| \, \int_{{\mathbb S}^{n-1}} u(\omega) \, d \omega = 0 \Big\}. $$
Then we have the

\begin{proposition}
Let $(\eta,\Sigma) \in (0,\infty) \times {\mathcal U}^0_n$ and $V \in \{\Omega_\Sigma\}^\bot$. Then, there exists a unique solution of \eqref{eq:GCI1} in $\dot H^1({\mathbb S}^{n-1})$ denoted by $\psi_{\eta \Sigma,V}$. 
The  linear vector space ${\mathcal C}_{\eta \Sigma}$ of GCI associated with $(\eta,\Sigma)$ is given by 
\begin{equation}
{\mathcal C}_{\eta \Sigma}= \big\{C_0+\psi_{\eta \Sigma,V} \, \, | \, \,  C_0\in {\mathbb R}, \, \, V \in \{\Omega_\Sigma\}^\bot \big\}.
\label{struct-GCI}
\end{equation}
%\label{prop_GCI1}
\end{proposition}

\noindent
{\bf Proof.} We look for solutions of \eqref{eq:GCI1} in variational form. Those solutions read as follows: find $\psi \in H^1({\mathbb S}^{n-1})$ such that 
\begin{equation}
\int_{{\mathbb S}^{n-1}} G_{\eta \Sigma} \, \nabla_\omega \psi \cdot \nabla_\omega \theta \, d \omega = - \int_{{\mathbb S}^{n-1}} G_{\eta \Sigma} \, (\omega \cdot \Omega_\Sigma) \, (\omega \cdot V) \, \theta \, d \omega, \quad \forall \theta \in H^1({\mathbb S}^{n-1}). 
\label{eq:varform}
\end{equation}
By Poincaré inequality and the fact that $G_{\eta \Sigma}$ is smooth and bounded from above and below, the bilinear form $\int G_{\eta \Sigma} \, \nabla_\omega \psi \cdot \nabla_\omega \theta \, d \omega $ is continuous and coercive on $\dot H^1({\mathbb S}^{n-1})$. Therefore, by Lax-Milgram theorem, the variational formulation \eqref{eq:varform} has a unique solution in $\dot H^1({\mathbb S}^{n-1})$ denoted by $\psi_{\eta \Sigma,V}$ when $\theta$ is restricted to belong to $\dot H^1({\mathbb S}^{n-1})$. To show that this is a solution for all $\theta \in H^1({\mathbb S}^{n-1})$, it is enough  to show that it satisfies  \eqref{eq:varform} for $\theta = 1$, i.e. that the following holds: 
\begin{equation}
\int_{{\mathbb S}^{n-1}} G_{\eta \Sigma} \, (\omega \cdot \Omega_\Sigma) \, (\omega \cdot V) \, \, d \omega=0, \quad \forall V \in \{\Omega_\Sigma\}^\bot. 
\label{eq:compat}
\end{equation}

Let $(e_1, \ldots, e_n)$ with $e_n = \Omega_\Sigma$ be an ortho-normal basis of ${\mathbb R}^n$ consisting of eigenvectors of $\Sigma$. Let $\lambda_1$, \ldots, $\lambda_n$ be the associated eigenvalues. Let $\omega = \sum_{k=1}^n \omega_k \, e_k$ be the decomposition of $\omega$ in this basis. It is enough to show \eqref{eq:compat} for $V=e_j$ with $j \in \{1, \ldots, n-1\}$. Then, we have 
$$ \int_{{\mathbb S}^{n-1}} G_{\eta \Sigma} \, (\omega \cdot \Omega_\Sigma) \, (\omega \cdot e_j) \, \, d \omega = \frac{1}{Z_{\eta \Sigma}} \int_{{\mathbb S}^{n-1}} e^{\eta (\lambda_1 \omega_1^2 + \ldots \lambda_n \omega_n^2)} \, \omega_j \, \omega_n \, \, d \omega = 0, $$
thanks to the change of $\omega_n$ into $-\omega_n$. This shows \eqref{eq:compat} and so, the existence and uniqueness of a solution of \eqref{eq:GCI1} in $\dot H^1({\mathbb S}^{n-1})$ is proved.

Now, all solutions in $H^1({\mathbb S}^{n-1})$ of \eqref{eq:varform} are of the form $\psi_{\eta \Sigma,V} + C_0$ where $C_0$ is any constant. Collecting all the solutions for all the possible $V \in \{\Omega_\Sigma\}^\bot$  leads to \eqref{struct-GCI} and ends the proof. \endproof

\begin{remark}
We note that if $\Omega_\Sigma$ is changed into $-\Omega_\Sigma$, $\psi_{\eta \Sigma,V}$ must be changed into $\psi_{\eta \Sigma,-V}$. It follows that \eqref{struct-GCI} remains unchanged.  
%\label{rem:gauge_invar}
\end{remark}

We now define a vector-valued GCI $\vec \psi_{\eta \Sigma}$ in the following way 

\begin{definition}
Given $(\eta,\Sigma) \in (0,\infty) \times {\mathcal U}^0_n$, we introduce the function $\vec \psi_{\eta \Sigma}$: ${\mathbb S}^{n-1} \to {\mathbb R}^n$, defined as the unique solution (in $\dot H^1({\mathbb S}^{n-1})$) of the following vector-valued equation: 
$$
\nabla_\omega \cdot \big( G_{\eta \Sigma}(\omega) \nabla_\omega \vec \psi_{\eta \Sigma} \big) = (\omega \cdot \Omega_\Sigma) \, P_{\Omega_\Sigma^\bot} \omega \, \, G_{\eta \Sigma}(\omega), \quad \forall \omega \in {\mathbb S}^{n-1}. 
$$
%\label{def:vecpsi}
\end{definition}

We note that 
$$\psi_{\eta \Sigma,V} = \vec \psi_{\eta \Sigma} \cdot V, \quad \forall V \in \{\Omega_\Sigma\}^\bot \quad \mbox{ and } \quad \vec \psi_{\eta \Sigma} \cdot \Omega_\Sigma = 0,$$ 
and that $\vec \psi_{\eta \Sigma}$ is changed into $-\vec \psi_{\eta \Sigma}$ if $\Omega_\Sigma$ is changed into $- \Omega_\Sigma$. 

\medskip
We can provide an explicit expression of $\vec \psi_{\eta A_\Omega}$, for all $(\eta, \Omega) \in (0,\infty) \times {\mathbb S}^{n-1}$ as the next proposition shows. Let us first define the following space: 
\begin{eqnarray*}
 {\mathcal H} &=& \Big\{ h: \, (-1,1) \to {\mathbb R} \, \, \Big| \, \, \int_{-1}^1 (1-r^2)^{\frac{n-1}{2}} \, |h(r)|^2 \, dr < \infty, \\
&&\hspace{6cm}  \int_{-1}^1 (1-r^2)^{\frac{n+1}{2}} \, |h'(r)|^2 \, dr < \infty \Big\}, 
\end{eqnarray*}
where $h'$ denotes the derivative of $h$. 

\begin{proposition}
Let $(\eta, \Omega) \in (0,\infty) \times {\mathbb S}^{n-1}$ be given. We have 
\begin{equation}
\vec \psi_{\eta A_\Omega} (\omega) = h_\eta (\omega \cdot \Omega) \, \omega_\bot, 
\label{eq:vecpsiAom}
\end{equation}
where $\omega_\bot = P_{\Omega^\bot} \omega$ and $h_\eta$ is the unique solution in ${\mathcal H}$ of the following equation:
\begin{eqnarray}
&&\hspace{-1cm} - (1-r^2)^{\frac{n-1}{2}} \, e^{\eta r^2} \, \big( 2 \eta \, r^2 + n-1 \big) \, h_\eta \nonumber \\
&&\hspace{4cm} + \frac{d}{dr} \Big[ (1-r^2)^{\frac{n+1}{2}} \, e^{\eta r^2} \, \frac{dh}{dr} \Big] = r \, (1-r^2)^{\frac{n-1}{2}} \, e^{\eta r^2}. 
\label{eq:eq_eta}
\end{eqnarray}
Furthermore, $h_\eta$ is odd and $h_\eta(r) \leq 0$ for $r \geq 0$. 
\label{prop:expressGCIAOm}
\end{proposition}

\noindent
{\bf Proof.} We apply \cite{Degond_Merino_M3AS20}, Proposition 4.2 (ii) (with the following changes: $u \to \Omega$, $\frac{\kappa}{2} \to \eta$, $d \to n$, $\bar \Gamma^*(\psi,u) \to L_{\eta A_\Omega}^* \psi$). Note that these techniques were first developed in \cite{Degond_etal_MMS18, Frouvelle_M3AS12}. \endproof

\begin{remark}
Formula \eqref{eq:vecpsiAom} shows that the vector GCI $\vec \psi_{\eta A_\Omega}$ is invariant under rotations leaving $\Omega$ fixed. This is a consequence of the fact that $A_\Omega$ is uniaxial with axis $\Omega$. No simple formula like \eqref{eq:vecpsiAom} is available for more general vector GCI $\vec \psi_{\eta \Sigma}$, when $\Sigma \in {\mathcal U}^0_n$ is not uniaxial. However, while we will need vector GCI for general $\Sigma \in {\mathcal U}^0_n$, we will only need an explicit expression of them in the case of a uniaxial tensor $\Sigma = A_\Omega$. So, Prop. \ref{prop:expressGCIAOm} is enough for our purpose. 
%\label{rem:vecGCI_rotinvar}
\end{remark}

The following proposition provides an alternate equation satisfied by $h_\eta$ in terms of the function $g$ defined in \eqref{eq:g}. Its proof is easy and is sketched in Appendix \ref{sec:proofpropalternh} for the reader's convenience. 

\begin{proposition}[Alternate equation for $h_\eta$] 
For $\theta \in [0,\pi]$, we define the function
\begin{equation} 
g(\theta) = - 2 \eta \, h_\eta(\cos \theta) \, \sin \theta. 
\label{eq:defg}
\end{equation}
Then $g$ satisfies the equation \eqref{eq:g}. 
\label{prop:altern_express_h}
\end{proposition}

Finally, the following proposition will have important consequences for the derivation of the macroscopic model: 

\begin{proposition}
Let $f$: ${\mathbb S}^{n-1} \to {\mathbb R}$ be twice continuously differentiable such that $Q_f \not = 0$ and $\Sigma_f \in {\mathcal U}_n^0$. Then, the vector GCI $\vec \psi_{\eta_f \Sigma_f}$ is well-defined and we have 
\begin{equation}
\int_{{\mathbb S}^{n-1}} C(f) \, \vec \psi_{\eta_f \Sigma_f} \, d \omega = 0. 
\label{eq:GCIcoll}
\end{equation}
\label{prop:GCIcoll}
\end{proposition}

\begin{remark}
Proposition \ref{prop:GCIcoll} expresses an important structural property of $C$. Let $(\eta,\Sigma) \in (0,\infty) \times {\mathcal U}^0_n$. The GCI $\vec \psi_{\eta \Sigma}$ cancels the collision operator acting on all functions $f$ which satisfy $(\eta_f, \Sigma_f)=(\eta,\Sigma)$. 
%\label{rem:collgeom}
\end{remark}

\noindent
{\bf Proof.} We show that $P_{\Omega_f^\bot} (Q_f \Omega_f) = 0$. Indeed, if this is the case, from \eqref{def-GCI-S}, we get
$$\int_{{\mathbb S}^{n-1}} L_{\eta_f \Sigma_f} f \, \vec \psi_{\eta_f \Sigma_f} \, d \omega = 0,$$
and using \eqref{eq:VM1}, \eqref{eq:etaf} and \eqref{eq:defSigmaf}, this shows \eqref{eq:GCIcoll}. But, by definition, $\Omega_f$ is the leading eigenvector of $Q_f$ with eigenvalue $\lambda_f$. So, $Q_f \Omega_f = \lambda_f \, \Omega_f$ and thus $P_{\Omega_f^\bot} (Q_f \Omega_f) =  0$, which ends the proof. \endproof

\medskip
Thanks to the GCI, we can now find how \eqref{eq:kinetic_lim} translates into an equation for the Q-tensor principal direction $\Omega$. This will be done below but first we provide some discussion of the GCI concept.

%%%%%%%%%%%%%%%%%%%%%%%%%%%%%%%%%%%%%%%%%%%%%%%%%%%%%%%%%%%%%%%%
%%%%%%%%%%%%%%%%%%%%%%%%%%%%%%%%%%%%%%%%%%%%%%%%%%%%%%%%%%%%%%%%

\subsection{Discussion of the GCI concept}
\label{subsec:ratGCI}

\subsubsection{Rationale for Definition \ref{def-GCI-S}} 
%\label{subsubsec:ratGCI}

First, let us note that the condition $P_{\Omega_\Sigma^\bot} ( Q_f \Omega_\Sigma) =0$ involved in Definition \ref{def:GCI} simply means that $\Omega_\Sigma$ is an eigenvector of $Q_f$. We now try to provide a geometric interpretation of Condition \eqref{def-GCI-S}. First let us introduce a few additional notations. We endow $\mathcal{S}^0_n$ with the inner-product $S:P = \mathrm{Tr} \{ SP \}$ and for a subset $B$ of $\mathcal{S}^0_n$, its orthogonal with respect to this inner-product is denoted by $B^\bot$. We recall that $B^\bot$ is a linear subspace of $\mathcal{S}^0_n$ and that $(B^\bot)^\bot = \mathrm{Span} (B)$. 

We now define the submanifold ${\mathcal N}$ of $\mathcal{U}^0_n$ which consists of normalized prolate uniaxial Q-tensors i.e. 
$$
{\mathcal N} = \{ A_{\Omega} \, \, | \, \, \Omega \in {\mathbb P}^{n-1} \} = \{ \Omega \otimes \Omega - \frac{1}{n} \mathrm{Id} \, \, | \, \, \Omega \in {\mathbb P}^{n-1} \}.
$$
Note that ${\mathcal N}$ is the manifold spanned by the NQT's of the equilibria (see Remark \ref{rem:order_parameter}). The mapping ${\mathbb P}^{n-1}  \ni \Omega \mapsto A_{\Omega} \in {\mathcal N}$ is a diffeomorphism. The tangent space of ${\mathcal N}$ at $A_\Omega$ is given by:
\begin{equation} 
T_{A_\Omega}{\mathcal N} = \{ \Omega \otimes V + V \otimes \Omega  \, \, | \, \, V \in \{ \Omega \}^\bot \}. 
\label{eq:tangentN}
\end{equation}
Indeed, for $V  \in T_{\Omega} {\mathbb P}^{n-1} = \{ \Omega \}^\bot$, consider a curve $I \ni t \mapsto \xi(t) \in {\mathbb P}^{n-1}$ where $I$ is an open interval of ${\mathbb R}$ containing $0$, such that $\xi(0) = \Omega$ and $\xi'(0) = V$. Then, $ \frac{d}{dt} (A_{\xi(t)})  = \Omega \otimes V + V \otimes \Omega$, showing the claim. We denote by $P_{T_{A_\Omega}{\mathcal N}}$ the orthogonal projection of  ${\mathcal S}^0_n$ on  $T_{A_\Omega}{\mathcal N}$ for the inner product defined just above. 

We have a mapping $p$: ${\mathcal U}^0_n \to {\mathcal N}$, $\Sigma \mapsto A_{\Omega_\Sigma}$. For any $\Omega \in {\mathbb P}^{n-1}$, the pre-image $p^{-1} (\{ A_\Omega \})$ is denoted by ${\mathcal F}_\Omega$. All these pre-images are homeomorphic to one-another. Let us choose one of them and denote it by ${\mathcal F}$. This endows ${\mathcal U}^0_n$ of a fiber bundle structure of base ${\mathcal N}$ and fiber ${\mathcal F}$. Now, we have the following lemma:

\begin{lemma}
Let $\Omega \in {\mathbb P}^{n-1}$ be given. \\
(i) Let $Q \in {\mathcal S}_0^n$. Then, $ 
P_{\Omega^\bot} ( Q \Omega) =0 \, \Longleftrightarrow \, Q \in \big(T_{A_\Omega}{\mathcal N} \big)^\bot$. \\
(ii) ${\mathcal F}_\Omega$ is a subset of $\big( T_{A_\Omega} {\mathcal N} \big)^\bot$. 
\label{lem:geom_interpret}
\end{lemma}

\noindent
\textbf{Proof.} 
(i) Using the symmetry of $Q$ and \eqref{eq:tangentN}, we have: 
\begin{eqnarray*} 
P_{\Omega^\bot} ( Q \Omega) =0 & \Longleftrightarrow & (Q \Omega) \cdot V = 0, \quad \forall V \in \{\Omega\}^\bot \\
& \Longleftrightarrow & Q: (\Omega \otimes V) = 0, \quad \forall V \in \{\Omega\}^\bot \\
& \Longleftrightarrow & Q: (\Omega \otimes V + V \otimes \Omega) = 0, \quad \forall V \in \{\Omega\}^\bot  \\
& \Longleftrightarrow & Q: B = 0, \quad \forall B \in T_{A_\Omega}{\mathcal N}  \, \, \Longleftrightarrow \, \, Q \in \big( T_{A_\Omega} {\mathcal N} \big)^\bot, 
\end{eqnarray*}
which shows (i). 

(ii) Suppose $\Sigma \in {\mathcal F}_\Omega$. Then $A_{\Omega_\Sigma} = A_\Omega$ which implies $\Omega_\Sigma = \Omega$ (in ${\mathbb P}^{n-1}$). Thus, $\Omega$ is an eigenvector of $\Sigma$ i.e. $P_{\Omega^\bot} ( \Sigma \Omega) =0$. Hence, by (i), $\Sigma \in \big( T_{A_\Omega} {\mathcal N} \big)^\bot$. \endproof

\begin{figure}[ht!]
\centering
\includegraphics[trim={3.5cm 15.cm 1.cm 4.5cm},clip,width=12cm]{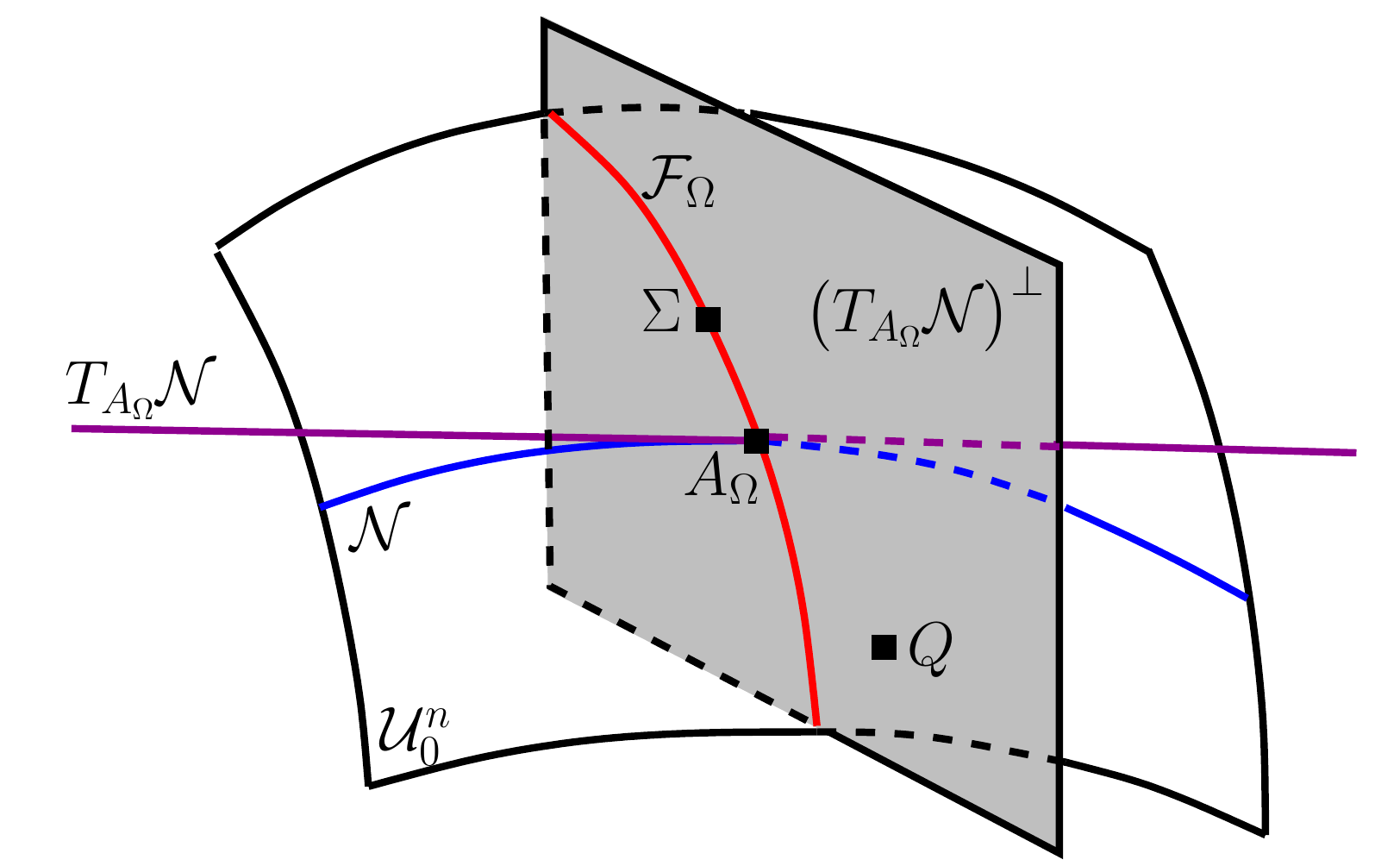}
\caption{Graphical representation of Condition \eqref{def-GCI-S_2}. The ambient three-dimensional space in the figure represents the flat space ${\mathcal S}_0^n$ in which ${\mathcal U}_0^n$ is an imbedded manifold represented by a surface. ${\mathcal N}$ is a submanifold of ${\mathcal U}_0^n$ depicted as the curvy blue line. It endows ${\mathcal U}_0^n$ of a fiber bundle structure of base ${\mathcal N}$. Let $\Sigma \in {\mathcal U}_0^n$. It projects (in the bundle sense) onto $A_\Omega \in {\mathcal N}$ and so, belongs to the fiber ${\mathcal F}_\Omega$ represented by the curvy red line. The tangent space to ${\mathcal N}$ at $A_\Omega$, $T_{A_\Omega} {\mathcal N}$ is represented by the magenta straight line. Its orthogonal $(T_{A_\Omega} {\mathcal N})^\bot$ is the gray-shaded plane on the figure. It contains ${\mathcal F}_\Omega$ by virtue of Lemma \ref{lem:geom_interpret} (ii). Then, condition  \eqref{def-GCI-S_2} means that the GCI associated with $(\eta,\Sigma)$ are the functions $\psi$ that cancel $L_{\eta \Sigma} f$ for all $f$ whose Q-tensor $Q_f$ (represented by the point~Q on the figure) belongs to $(T_{A_\Omega} {\mathcal N})^\bot$.}
\label{fig:geometry}
\end{figure}

So, Eq. \eqref{def-GCI-S} can be equivalently written:
\begin{equation}
\label{def-GCI-S_2}
\int_{{\mathbb S}^{n-1}} (L_{\eta \Sigma} f) \, \psi \, d\omega =0 \quad \mbox{ for all } \, f \, \text{ such that  } \,  Q_f \in \big( T_{A_{\Omega_\Sigma}} {\mathcal N} \big)^\bot .
\end{equation}
This can be geometrically interpreted as follows: to any $\Sigma \in {\mathcal U}_n^0$ we consider its projection (in the fiber bundle sense) $p(\Sigma) = A_{\Omega_\Sigma}$ onto ${\mathcal N}$.
Then, \eqref{def-GCI-S_2} means that the GCI associated to $(\eta,\Sigma)$ are all the functions $\psi$ whose integrals against $L_{\eta \Sigma} f$ cancel when $Q_f$ belongs to the orthogonal of the tangent space to ${\mathcal N}$ at $A_{\Omega_S}$. This is illustrated in Fig. \ref{fig:geometry}. It is likely that this geometrical structure persists with other collision operators as it seems to express some intrinsic geometrical constraint. This point will be further developed in future work.

%%%%%%%%%%%%%%%%%%%%%%%%%%%%%%%%%%%%%%%%%%%%%%%%%%%%%%%%%%%%%%%%
\subsubsection{Relation between the GCI and the linearized collision operator}
%\label{subsubsec:GCI_linearized}

Let $D_f C$ the linearization of the collision operator $C$ about the distribution function $f$ and let $D_f C^*$ be its formal $L^2$-adjoint. For a distribution function $f$, we call $(\eta_f, \Sigma_f)$ the 'moments' of $f$. In this section, we show the following: suppose $(\eta,\Sigma) \in (0,\infty) \times {\mathcal U}_0^n$ is the moment of an equilibrium distribution function, i.e. $(\eta, \Sigma) = (\eta(\rho), A_\Omega)$ where $(\rho,\Omega) \in (\rho^*,\infty) \times {\mathbb S}^{n-1} / \{ \pm 1 \}$ and denote by $f^0 = \rho G_{\eta(\rho) A_\Omega}$ the corresponding equilibrium. Then, we have 
\begin{equation}
{\mathcal C}_{\eta(\rho) A_\Omega} = \mathrm{ ker } (D_{f^0} C^*).
\label{eq:Cequi=kerDC*}
\end{equation}
On the other hand,  if $(\eta,\Sigma)$ is not the moment of an equilibrium, then, although there exist  Gibbs distributions $f = \rho  G_{\eta \Sigma}$ associated with $(\eta,\Sigma)$, in general, we have 
\begin{equation}
{\mathcal C}_{\eta \Sigma} \not = \mathrm{ ker } (D_{f} C^*).
\label{eq:CnonequiisnotkerDC*}
\end{equation}
Thus, a GCI associated to an arbitrary moment $(\eta,\Sigma)$ is in general not in the kernel of the adjoint linearized collision operator about the corresponding Gibbs distribution. It is only so if $(\eta,\Sigma)$ is the moment of an equilibrium in the above sense. Consequently, GCI are different and truly more general concepts than elements of such kernels. Likewise, Eq.~\eqref{eq:defg} linking the GCI to the auxiliary function $g$ given by \eqref{eq:g} is only valid for moments $(\eta(\rho), A_\Omega)$ related to equilibria. Observe however that we will not need to explicit the form of the GCI for general moments, but only for those corresponding to an equilibrium (see Section \ref{sec:omega_abstract} below).

Formula \eqref{eq:Cequi=kerDC*} is unsurprising. Indeed, Eq. \eqref{eq:Omega} has been shown in \cite{Kuzuu_Doi_JPhysSocJapan83, Wang_Zhang_Zhang_CPAM15} using the Hilbert expansion method. This method corresponds to inserting the Hilbert expansion $f^\varepsilon = f^0 + \varepsilon f^1 + {\mathcal O}(\varepsilon^2)$ into the kinetic equation \eqref{eq:kinetic_abstract} and matching identical powers of~$\varepsilon$. We get 
$$
C(f^0) = 0, \qquad DC_{f^0} f^1 = T_{u^0} f^0,
$$ 
for the terms of order $\varepsilon^{-1}$ and $\varepsilon^0$ respectively (note that we also need to Hilbert-expand the velocity $u^\varepsilon$). Now, the first equation implies that $f^0$ is an equilibrium $f^0 = \rho G_{\eta(\rho) A_\Omega}$. Then, one looks for a necessary and sufficient condition for the existence of a solution $f^1$ to the second equation. Assuming that Im~$DC_{f^0}$ = (ker~$DC_{f^0}^*)^\bot$ (which can be proved via a careful study of the spectral properties of $DC_{f^0}$, see \cite{Wang_Zhang_Zhang_CPAM15}), such a condition is 
$$ \int_{{\mathbb S}^{n-1}} T_{u^0} f^0 \, \psi \, d \omega = 0, \quad \forall \psi \in \mathrm{ ker } DC_{f^0}^*. 
$$
Since this is also what we get when $\psi$ ranges in ${\mathcal C}_{\eta(\rho) A_\Omega}$ (see Eq. \eqref{GCI-AOmega} below), Eq. \eqref{eq:Cequi=kerDC*} must be true. However, it would be desirable to have a direct proof of \eqref{eq:Cequi=kerDC*}. This is our goal here. As a by-product, we will also see why we have \eqref{eq:CnonequiisnotkerDC*}. We first compute the adjoint linearized collision operator. 

\begin{lemma}[Adjoint linearized collision operator]
Let $\rho \in (0,\infty)$, $S \in {\mathcal S}_0^n$. We have
\begin{equation}
D_{\rho G_S}C^* g (\omega) = L_{\alpha \rho Q_{\rho G_S}}^* g (\omega) - \alpha \rho \,  (\rho Q)_{G_S L_S^* g} : \omega \otimes \omega  ,
\label{eq:adjlinear_gene}
\end{equation}
where $G_S$ is defined by \eqref{eq:vonmises2}, the auxiliary operator $L$ by \eqref{eq:def_LSig} and $L^*$ is its formal $L^2$-adjoint. Here $(\rho Q)_{G_S L_S^* g}$ stands for the right-hand side of \eqref{eq:orienttensor} with $f$ replaced by $G_S L_S^* g$ (note that $\rho_{G_S L_S^* g} = 0$ so that $Q_{G_S L_S^* g}$ is not defined but $(\rho Q)_{G_S L_S^* g}$ itself is well-defined). 
%\label{lem:linearized}
\end{lemma}

\medskip
\noindent
{\bf Proof.} From \eqref{eq:coll_op_alter} and the fact that $U^0_f$ depends linearly on $f$, we get 
\begin{equation} 
D_{\rho G_S} C f = \nabla_\omega \cdot \big( \nabla \omega f + f \nabla_\omega U^0_{\rho G_S} + \rho G_S \nabla_\omega U^0_f \big). 
\label{eq:linear_gene_comput1}
\end{equation}
We note that $  \nabla_\omega U^0_{\rho G_S} = - \nabla_\omega \big( \log G_{\alpha \rho Q_{\rho G_S}} \big)$. Inserting this into \eqref{eq:linear_gene_comput1}, we get 
\begin{equation}
D_{\rho G_S}C f = L_{\alpha \rho Q_{\rho G_S}} f + \rho \, G_S \, L_S^* U^0_f.  
\label{eq:linear_gene} 
\end{equation}
Thanks to \eqref{eq:muf0}, we also note that $L_S^* U^0_f = L_S^* \breve U^0_f$ with $ \breve U^0_f = - \alpha (\omega \cdot \rho_f Q_f \omega)$. Thus, using \eqref{eq:linear_gene}, Stokes formula and that $L_S(G_S g) = G_S L_S^* g$, we get 
$$
\int_{{\mathbb S}^{n-1}} D_{\rho G_S} C f \, g \, d \omega = \int_{{\mathbb S}^{n-1}} f \, L_{\alpha \rho Q_{\rho G_S}}^* g \, d \omega  + \rho \int_{{\mathbb S}^{n-1}} \breve U^0_f \, G_S \, L_S^* g \, d \omega. 
$$
Inserting the expression of $\breve U^0_f$ into this formula, using the expression \eqref{eq:orienttensor} of $\rho_f Q_f$ and exchanging $\omega$ and $\omega'$ in the resulting integral, we are led to \eqref{eq:adjlinear_gene}. \endproof

Now, in the case of an equilibrium, we compute the kernel of the adjoint linearized collision operator: 

\begin{lemma}[kernel of $D_{f^0}C^*$ when $f^0$ is an equilibrium]
Let $\rho \in (\rho^*,\infty)$ and $\Omega \in {\mathbb S}^{n-1} / \{ \pm 1 \}$. Let $f^0 = \rho G_{\eta(\rho) A_\Omega}$ be an equilibrium of $C$, where the function $\rho \mapsto \eta(\rho)$ is defined in Prop. \ref{prop:eta(rho)_defin}. Define $\tilde {\mathcal X}_{\rho, \Omega}$ to be the space of functions $\varphi: \omega \mapsto \varphi(\omega)$ which satisfy
\begin{equation}
\varphi(\omega) = \alpha \rho \, (\rho Q)_{G_{\eta(\rho) A_\Omega} \varphi} : \omega \otimes \omega, \quad \forall \omega \in {\mathbb S}^{n-1} . 
\label{eq:defXetaAom}
\end{equation}
Then we have 
\begin{equation}
g \in \mathrm{ker} \, \big( D_{\rho G_{\eta(\rho) A_\Omega}}C^* \big) \, \Longleftrightarrow \, \int_{{\mathbb S}^{n-1}}
L_{\eta(\rho) A_\Omega} f \, g \, d\omega = 0, \quad \forall f \in \tilde {\mathcal X}_{\rho,\Omega}^\bot, 
\label{eq:kerDf0C*_equi}
\end{equation}
where the orthogonality is with respect to the standard $L^2({\mathbb S}^{n-1})$-inner product. 
\label{lem:ker_adjoint_equi}
\end{lemma}

\medskip
\noindent
{\bf Proof.} Defining $S = \eta(\rho) A_\Omega$, we have 
\begin{equation}
\alpha \rho Q_{\rho G_S} = \alpha \rho Q_{\rho G_{\eta A_\Omega}} = \eta A_\Omega = S, 
\label{eq:keyprop}
\end{equation}
thanks to \eqref{eq:Q_equib} and \eqref{eq:eta}. Thus, thanks to \eqref{eq:adjlinear_gene}, we are led to 
\begin{equation}
D_{\rho G}C^* g (\omega) = L^* g (\omega) - \alpha \rho \,  (\rho Q)_{GL^* g} : \omega \otimes \omega, 
\label{eq:adjlinear_equi}
\end{equation}
where here and in the remainder of the proof, we omit the dependence of $\eta$ on $\rho$, as well as the index $\eta A_\Omega$ on $L^*$ and $G$ and the indices $\rho,\, \Omega$ on $\tilde {\mathcal X}$ for clarity.  

For any smooth enough function $f$, we have by the Stokes formula:
$$ 
\int_{{\mathbb S}^{n-1}} L f \, g \, d \omega =  \int_{{\mathbb S}^{n-1}} f \, L^* g \, d\omega = \int_{{\mathbb S}^{n-1}} f \, \varphi \, d \omega, 
$$ 
with $\varphi = L^* g$. Thanks to \eqref{eq:adjlinear_equi} and the fact that $g \in \mathrm{ ker } D_{\rho G}C^*$, $\varphi$ satisfies \eqref{eq:defXetaAom}, so $\varphi \in \tilde {\mathcal X}$. If $f \in \tilde {\mathcal X}^\bot$, we deduce that $\int L f \, g \, d \omega = 0 $, which shows the left-to-right implication of \eqref{eq:kerDf0C*_equi}. 

Conversely suppose that $g$ is such that $\int L f \, g \, d\omega = 0$, $\forall f \in \tilde {\mathcal X}^\bot$, i.e.
$$
f \in \tilde {\mathcal X}^\bot \, \Longrightarrow \,  f \in \{ L^* g \}^\bot.
$$ 
Taking the orthogonals, we get 
$$ 
\mathrm{Span } \{ L^* g \} \subset \tilde {\mathcal X}.
$$ 
Indeed, both Span $\{ L^* g \}$ and $\tilde {\mathcal X}$ are finite-dimensional, hence closed. This is obvious for the former which is one-dimensional. For the latter, by \eqref{eq:defXetaAom}, $\tilde {\mathcal X}$ is included in the space of quadratic polynomials in $\omega$, which is a finite-dimensional space. So, defining $\varphi = L^* g$, we have $\varphi \in \tilde {\mathcal X}$. Replacing  $\varphi$ by its expression in terms of $g$ in \eqref{eq:defXetaAom}, we get $D_{\rho G}C^* (g) = 0$, which shows the right-to-left implication of \eqref{eq:kerDf0C*_equi} and ends the proof. \endproof

Next, we prove an alternate characterization of the space ${\mathcal X}_{\rho, \Omega}$.

\begin{lemma}
Let $\rho$, $\Omega$, $f^0$ and $\eta$ as in Lemma \ref{lem:ker_adjoint_equi}. Then, 
\begin{equation}
\tilde {\mathcal X}_{\rho, \Omega} = {\mathcal X}_\Omega,
\label{eq:equivX}
\end{equation}
where ${\mathcal X}_\Omega$ is defined by \eqref{eq:defmathcalXOm}.
\label{lem:equivX}
\end{lemma}

\medskip
\noindent
{\bf Proof.} Let $\varphi \in \tilde {\mathcal X}$ (using the simplified notations of the previous proof). From \eqref{eq:defXetaAom}, we have $\varphi(\omega) = K:\omega \otimes \omega$ where $K= \alpha \rho \, (\rho Q)_{G \varphi}$. Hence, $K$ satisfies the fixed point equation
\begin{equation}
K = \alpha \rho \,  \big( \rho Q \big)_{G \, K:\omega \otimes \omega} ,  
\label{eq:definK}
\end{equation}
which implies that 
\begin{equation}
\mathrm{Tr} K = 0. 
\label{eq:traceK=0}
\end{equation}
Using \eqref{eq:orienttensor}, \eqref{eq:4tensor} and \eqref{eq:traceK=0}, we can develop \eqref{eq:definK} into:
\begin{equation}
K = \alpha \rho \,  {\mathbb T}_{G_{\eta A_\Omega}} : K . 
\label{eq:Kfixedpoint}
\end{equation}
According to \eqref{eq:Q4_comput_2}, there are three real numbers $a_k$, $k=1, \ldots, 3$, such that 
\begin{equation}
{\mathbb T}_{G_{\eta A_\Omega}} = a_1 \, \Omega^{\otimes 4} + 6 a_2 \, \big( \Omega \otimes \Omega \otimes \mathrm{Id} \big)_s + 3 a_3 \big( \mathrm{Id} \otimes \mathrm{Id} \big)_s. 
\label{eq:TGetaOm_simple}
\end{equation}
We uniquely define $V \in \{\Omega\}^\bot$ and $r \in {\mathbb R}$ by $K \Omega = r \Omega + V$. inserting \eqref{eq:TGetaOm_simple} into \eqref{eq:Kfixedpoint} and using \eqref{eq:traceK=0}, we get 
\begin{equation}
\Big( \frac{1}{2 \alpha \rho} - a_3 \Big) K = a_2 \, (\Omega \otimes V + V \otimes \Omega) + \frac{1}{2} \big( (a_1 + 4 a_2) r \, \Omega \otimes \Omega + a_2 r \, \mathrm{Id} \big)
\label{eq:deterK}
\end{equation}
We now state the following lemma, whose proof can be found in Appendix \ref{sec:prooflem:a1a2a3}

\medskip
\begin{lemma}
We have
\begin{eqnarray}
& & \frac{1}{2 \alpha \rho} - a_3  = a_2 \not = 0, \label{eq:alrhoa23} \\
& & a_1 + (n+4) a_2 = S_2(\eta). \label{eq:a1A2}
\end{eqnarray}
\label{lem:a1a2a3}
\end{lemma}

Using \eqref{eq:alrhoa23}, Eq. \eqref{eq:deterK} leads to
$$
K = \Omega \otimes V + V \otimes \Omega + \frac{1}{2 a_2} \big[ (a_1 + 4 a_2) r \, \Omega \otimes \Omega  + a_2 r \, \mathrm{Id} \big]. 
$$
With \eqref{eq:traceK=0}, we get 
$$ 
0 = \mathrm{Tr} K = \frac{1}{2 a_2} \big[ (a_1 + (n+4) a_2) \big] \, r, 
$$ 
which, with \eqref{eq:a1A2} and the fact that $S_2(\eta) \not = 0$ (see Prop. \ref{prop:OP} (iii)),  leads to $r=0$ and 
$$
K = \Omega \otimes V + V \otimes \Omega. 
$$
Thus, 
\begin{equation}
\varphi = 2 \, (\Omega \cdot \omega) \, (V \cdot \omega). 
\label{eq:deterphi}
\end{equation}
Reciprocally, by similar but simpler computations, we easily get that $\varphi$ given by \eqref{eq:deterphi} with arbitrary $V \in \{ \Omega \}^\bot$ satisfies \eqref{eq:definK}. In the end, we find
$$ 
\tilde {\mathcal X} = \{ (\Omega \cdot \omega) \, (V \cdot \omega) \, \, \big| \, \, V \in \{\Omega\}^\bot \} = {\mathcal X}_\Omega, 
$$
which ends the proof. \endproof

We can now state the following

\begin{theorem}
Let $f^0 = \rho G_{\eta(\rho) A_\Omega}$ be an equilibrium of $C$. Then, we have 
$$
{\mathcal C}_{\eta(\rho) A_\Omega} = \mathrm{ker} \, \big( D_{\rho G_{\eta(\rho) A_\Omega}}C^*  \big), 
$$
where ${\mathcal C}_{\eta(\rho) A_\Omega}$ is the space of GCI associated with the equilibrium moments $(\eta(\rho), A_\Omega)$ (see Definition \ref{def:GCI}).  
\label{thm:GCI=kerDf0C*}
\end{theorem}

\textbf{Proof.} Indeed, we have the sequence of equivalences: 
\begin{eqnarray*}
&&\hspace{-1cm}
\psi \mbox{ is a GCI associated with } (\eta(\rho), A_\Omega) \, \Longleftrightarrow \\
&& \hspace{3cm} \Longleftrightarrow \,  \Big( f \in {\mathcal X}_\Omega^\bot \, \Longrightarrow \, \int_{{\mathbb S}^{n-1}} L_{\eta(\rho) A_\Omega} f \, \psi \, d\omega = 0 \Big) \\
&& \hspace{3cm} \Longleftrightarrow \,  \Big( f \in \tilde {\mathcal X}_{\rho,\Omega}^\bot \, \Longrightarrow \, \int_{{\mathbb S}^{n-1}} L_{\eta(\rho) A_\Omega} f \, \psi \, d\omega = 0 \Big) \\
&& \hspace{3cm} \Longleftrightarrow \,   \psi \in \mathrm{ker} \, \big( D_{\rho G_{\eta(\rho) A_\Omega}}C^*  \big), 
\end{eqnarray*}
where the first equivalence comes from \eqref{def-GCI-S} and \eqref{eq:PTNQ=0}, the second one from \eqref{eq:equivX} and the third one, from \eqref{eq:kerDf0C*_equi}. This ends the proof. \endproof

The key property which led to Theorem \ref{thm:GCI=kerDf0C*} in the case where $f^0$ is an equilibrium is \eqref{eq:keyprop}. It gave rise to the structure 
\begin{equation}
D_{\rho G}C^* g = \varphi (\omega) - \alpha \rho \,  (\rho Q)_{G \varphi} : \omega \otimes \omega, 
\label{eq:keystruct}
\end{equation} 
with $\varphi = L^* g$ which led to the definition of the space $\tilde {\mathcal X}_{\rho, \Omega}$. Now, if $(\eta, \Sigma)$ is not a moment of an equilibrium, we have $\alpha \rho Q_{\rho G_S} \not = S$ as the equality is a characterization of the moments of equilibria. Then, by inspection of \eqref{eq:adjlinear_gene}, we see that the structure \eqref{eq:keystruct} is lost and the proof cannot be continued. These considerations strongly support \eqref{eq:CnonequiisnotkerDC*}. Indeed, we have the following counter-example in dimension $n=3$ whose proof can be found in Appendix \ref{sec:CnonequiisnotkerDC*}. 

\begin{proposition}
Let $n=3$. Let $f = \rho G_{\eta A_{\Omega}}$ where $\eta \not = \eta(\rho)$ (in other words, in spite of being a Gibbs distribution, $f$ is not an equilibrium). Then we have \eqref{eq:CnonequiisnotkerDC*} (with $\Sigma = A_{\Omega}$). 
\label{prop:CnonequiisnotkerDC*}
\end{proposition}

So, the space of GCI ${\mathcal C}_{\eta \Sigma}$ is related to important structural properties of $C$ such as Prop. \ref{eq:GCIcoll}. By contrast, the space ker$\,(D_{f} C^*)$ does not play any particular role. The exception is when the Gibbs distribution $\rho G_{\eta \Sigma}$ is an equilibrium, in which case the two spaces are equal. This shows that GCI are a more relevant and general concept than the space ker$\,(D_{f} C^*)$ which appears in the Hilbert method.

%%%%%%%%%%%%%%%%%%%%%%%%%%%%%%%%%%%%%%%%%%%%%%%%%%%%%%%%%%%%%%%%
%%%%%%%%%%%%%%%%%%%%%%%%%%%%%%%%%%%%%%%%%%%%%%%%%%%%%%%%%%%%%%%%
%%%%%%%%%%%%%%%%%%%%%%%%%%%%%%%%%%%%%%%%%%%%%%%%%%%%%%%%%%%%%%%%
%%%%%%%%%%%%%%%%%%%%%%%%%%%%%%%%%%%%%%%%%%%%%%%%%%%%%%%%%%%%%%%%

\setcounter{equation}{0}
\section{Equation for the Q-tensor axis direction $\Omega$}
\label{sec:omega_abstract}

\subsection{Abstract derivation}
%\label{subsec:abstract}

In this section, we provide an abstract set of equations allowing us to determine the evolution equation for the Q-tensor axis direction $\Omega$. We recall the expression \eqref{eq:defTu} of $T_u(f)$. We have the:

\begin{proposition}
Let $f = \lim_{\varepsilon \to 0} f^\varepsilon$ with $f(x,\omega,t) = \rho(x,t) \ G_{\eta(\rho(x,t)) A_{\Omega(x,t)}}(\omega)$ for all $(x,t) \in {\mathcal B}$ where ${\mathcal B}$ is given by \eqref{def:set_B} and the function $\rho \mapsto \eta(\rho)$ is defined in Prop. \ref{prop:eta(rho)_defin}. Then, we have 
\begin{equation}
\int_{{\mathbb S}^{n-1}} T_{u} \big( \rho(x,t) \, G_{\eta(\rho(x,t)) A_{\Omega(x,t)}} \big) \, \vec \psi_{\eta(\rho(x,t))  A_{\Omega(x,t)}}(\omega) \, d \omega =0,
\label{GCI-AOmega}
\end{equation}
where $\vec \psi_{\eta(\rho(x,t))  A_{\Omega(x,t)}}$ is the vector GCI associated with $(\eta(\rho(x,t)), A_{\Omega(x,t)})$ (see Section \ref{subsec:GCI}). 
%\label{prop:intGCITu}
\end{proposition}

\begin{remark}
We note that \eqref{GCI-AOmega} is unchanged if $\Omega(x,t)$, and consequently $\vec \psi_{A_{\Omega(x,t)}}$, are changed in their opposites. 
%\label{rem:change_sign}
\end{remark}

\medskip
\noindent
{\bf Proof.} Let $(x,t) \in {\mathcal B}$ be given. For simplicity, in the proof, we omit the variables $(x,t)$. We also denote $\rho^\varepsilon := \rho_{f^\varepsilon}$, $Q^\varepsilon := Q_{f^\varepsilon}$, $\lambda^\varepsilon := \lambda_{f^\varepsilon}$, etc. and $\rho := \rho_f$, $Q := Q_f$, $\lambda := \lambda_f$, etc. By the fact that $f^\varepsilon \to f$, we get $\rho^\varepsilon Q^\varepsilon \to \rho Q=  \rho \, \frac{n}{n-1} \lambda \,  A_\Omega$, with $\frac{n}{n-1} \lambda = \frac{\eta(\rho)}{\alpha \rho}$. Since $\rho \not = 0$ (because $(x,t) \in {\mathcal B}$) and $\lambda$ is a simple eigenvalue of $Q$, then, for $\varepsilon$ small enough, $\rho^\varepsilon \not = 0$, $Q^\varepsilon \to Q$ and $\lambda^\varepsilon$ is a simple eigenvalue of $Q^\varepsilon$ such that $\lambda^\varepsilon \to \lambda$ (because the subset of ${\mathcal S}_n^0$ of matrices which have simple leading eigenvalue is an open set). Thus, $\Sigma^\varepsilon = \frac{n-1}{n \, \lambda^\varepsilon} Q^\varepsilon$ is defined, belongs to ${\mathcal U}_n^0$ and is such that $\Sigma^\varepsilon \to \Sigma = A_\Omega$ as $\varepsilon \to 0$. 

By the smoothness of $\Sigma^\varepsilon$ with respect to $\varepsilon$, we can find a smooth lifting of $\Omega_{\Sigma^\varepsilon} \in {\mathbb P}^{n-1}$ into $\Omega^\varepsilon \in {\mathbb S}^{n-1}$. Thus, we can form the GCI $\vec \psi_{\eta^\varepsilon \Sigma^\varepsilon}$ using this smooth determination of $\Omega_{\Sigma^\varepsilon}$ (remember that we need to fix the sign of $\Omega_{\Sigma^\varepsilon}$ because the sign of $\vec \psi_{\eta^\varepsilon \Sigma^\varepsilon}$ depends on it). This makes $\vec \psi_{\eta^\varepsilon \Sigma^\varepsilon}$ a smooth function of $\varepsilon$ (because $\vec \psi_{\eta S}$ is a smooth function of $(\eta,S) \in [0,\infty) \times {\mathcal U}_0^n$) such that $\vec \psi_{\eta^\varepsilon \Sigma^\varepsilon} \to \vec \psi_{\eta A_\Omega}$ when $\varepsilon \to 0$. 

Thanks to \eqref{eq:GCIcoll}, we have
$$ \int_{{\mathbb S}^{n-1}} C(f^\varepsilon) \, \vec \psi_{\eta^\varepsilon \Sigma^\varepsilon} \, d \omega = 0. $$
So, multiplying \eqref{eq:kinetic_abstract} by $\vec \psi_{\eta^\varepsilon \Sigma^\varepsilon}$, integrating the resulting expression with respect to~$\omega$ leads to 
$$\int_{{\mathbb S}^{n-1}} T_{u^\varepsilon}(f^\varepsilon) \, \vec \psi_{\eta^\varepsilon \Sigma^\varepsilon} \, d \omega=0.$$
Now letting $\varepsilon \to 0$, with $u^\varepsilon \to u$, $f^\varepsilon \to \rho \, G_{\eta(\rho)  A_\Omega}$, $\eta^\varepsilon \to \eta(\rho)$, $\Sigma^\varepsilon \to A_\Omega$, $\vec \psi_{\eta^\varepsilon \Sigma^\varepsilon} \to \vec \psi_{\eta(\rho)  A_{\Omega}}$, we get \eqref{GCI-AOmega}. 
This ends the proof. \endproof

%%%%%%%%%%%%%%%%%%%%%%%%%%%%%%%%%%%%%%%%%%%%%%%%%%%%%%%%%%%%%%%%
%%%%%%%%%%%%%%%%%%%%%%%%%%%%%%%%%%%%%%%%%%%%%%%%%%%%%%%%%%%%%%%%

\subsection{Derivation of the equation for $\Omega$}
\label{subsec:GCI_first_derivation}

In this section, we derive the explicit equation for $\Omega$ by inserting expression \eqref{eq:vecpsiAom} into the abstract formulation \eqref{GCI-AOmega} and compute the integral explicitly. This is summarized in the following

\begin{proposition}
Let $f = \lim_{\varepsilon \to 0} f^\varepsilon = \rho(x,t) G_{\eta(\rho(x,t)) A_{\Omega(x,t)}}$ as given in Corollary \ref{cor:localequ}. Then, $\Omega$ satisfies \eqref{eq:Omega} 
\label{prop:eq_Omega_ffam}
\end{proposition}

\noindent
{\bf Proof of Proposition \ref{prop:eq_Omega_ffam}.} 
For simplicity, we omit the dependencies of $\eta$ and $\lambda$ on $\rho$, of $h_\eta$ on $(\omega \cdot \Omega)$, of $G_{\eta A_\Omega}$ on $\omega$ and of $\rho$ and $\Omega$ on $(x,t)$. Inserting \eqref{eq:vecpsiAom} into \eqref{GCI-AOmega}, we get:
\begin{equation}
V_\Omega := \int_{{\mathbb S}^{n-1}} T_{u}(\rho \, G_{\eta A_\Omega}) \, h_\eta \, \omega_\bot \,d\omega=0.
\label{eq:Vomega}
\end{equation}
We define
\begin{eqnarray*}
D_t&=&\partial_t+u\cdot\nabla_x, \\
A f &=& \nabla_\omega \cdot \big( f \, (\Lambda P_{\omega^\bot} E - W) \omega \big), \\
B f &=& 2 \alpha \, \beta \, \nabla_\omega \cdot \big( f \,P_{\omega^\perp} \Delta_x (\rho_f \, Q_f) \, \omega \big), 
\end{eqnarray*}
so that $T_u(f) = D_t f + A f + B f$ and 
\begin{equation}
V_\Omega = \int_{{\mathbb S}^{n-1}} (D_t  + A  + B )(\rho \, G_{\eta A_\Omega}) \, h_\eta \, \omega_\bot \, d\omega = V_\Omega^{(1)} + V_\Omega^{(2)} + V_\Omega^{(3)}. 
\label{eq:V}
\end{equation}

Using ~\eqref{transport-mass} which gives~$D_t\rho=0$ and $D_t \eta = \eta' D_t\rho = 0$, where $\eta'$ is the derivative of $\eta$ with respect to $\rho$, we get 
\begin{eqnarray*} 
D_t(\rho \, G_{\eta A_\Omega}) &=& 
\rho \, G_{\eta A_\Omega} \, 2 \eta \, (\omega \cdot \Omega) \, (P_{\Omega^\bot} \omega) \cdot D_t \Omega, 
\end{eqnarray*}
where we have used that the denominator of \eqref{eq:GetaAom} does not depend on $\Omega$. Then, we apply \eqref{eq:moments2} and the fact that $D_t\Omega$ is orthogonal to $\Omega$ and get 
\begin{equation}
V_\Omega^{(1)} =\tilde \gamma_1 \, D_t\Omega, 
\label{eq:V1}
\end{equation}
with 
\begin{equation}
\tilde \gamma_1 = \frac{2 \, \eta \, \rho }{n-1} \int_{{\mathbb S}^{n-1}} G_{\eta A_\Omega} \, h_\eta
\, \, (\omega \cdot \Omega) \, (1- (\omega \cdot \Omega)^2) \, d\omega. 
\label{eq:gamma1}
\end{equation}

Next, we have 
\begin{eqnarray*}
A (\rho \, G_{\eta A_\Omega}) &=& \nabla_\omega \cdot \big(\rho \, G_{\eta A_\Omega} \, (\Lambda P_{\omega^\bot} E - W) \omega \big) \\
&=& \rho \, G_{\eta A_\Omega} \, \big[ \nabla_\omega (\log G_{\eta A_\Omega}) \cdot (\Lambda P_{\omega^\bot} E - W) \omega + \nabla_\omega \cdot \big( (\Lambda P_{\omega^\bot} E - W) \omega \big) \big] .
\end{eqnarray*}
First, we compute $\nabla_\omega \cdot \big( (\Lambda P_{\omega^\bot} E - W) \omega \big)$. Let $X=\Lambda E - W$ for simplicity and let $(e_i)_{i=1, \ldots, n}$ be the canonical basis of ${\mathbb R}^n$. Define ${\mathcal X}_i = \sum_{j=1}^n X_{ij} \, e_j$. Then, we can write $X = \sum_{i=1}^n e_i \otimes {\mathcal X}_i$. Then, $P_{\omega^\bot} X \, \omega =  \sum_{i=1}^n ({\mathcal X}_i \cdot \omega) \, P_{\omega^\bot} e_i$. We note that $\nabla_\omega \cdot P_{\omega^\bot} e_i = \Delta_\omega (\omega \cdot e_i) = - (n-1) \, (\omega \cdot e_i)$ because $(\omega \cdot e_i)$ is a spherical harmonic of degree $1$ hence an eigenfunction of the spherical laplacian associated to the eigenvalue $-(n-1)$. Thus, 
\begin{eqnarray*} 
&&\hspace{-1cm}
\nabla_\omega \cdot (P_{\omega^\bot} X \, \omega) = \sum_{i=1}^n \Big[ P_{\omega^\bot} {\mathcal X}_i  \cdot P_{\omega^\bot} e_i - (n-1) \, ({\mathcal X}_i \cdot \omega) \, (\omega \cdot e_i) \Big] \\
&&\hspace{1cm}
 = \sum_{i, \, j=1}^n X_{ij} \, ( P_{\omega^\bot} e_i \cdot P_{\omega^\bot} e_j - (n-1) \,  \omega_i \, \omega_j) ) =  \sum_{i, \, j=1}^n X_{ij} (\delta_{ij} - n \, \omega_i \, \omega_j) \\
&&\hspace{1cm}= \mathrm{Tr} X - n X:(\omega \otimes \omega), 
\end{eqnarray*}
where $\delta_{ij}$ is the Kronecker symbol and $\mathrm{Tr} X$ is the trace of $X$. Now, with $X=\Lambda E - W$, owing to the facts that $\mathrm{Tr} X = \Lambda \nabla_x \cdot u = 0$ and remembering that $E$ is symmetric and $W$, antisymmetric, we get 
\begin{eqnarray*}
&&\hspace{-1cm}
A (\rho \, G_{\eta A_\Omega}) = \rho \, G_{\eta A_\Omega} \, \big[ 2 \eta \, (\omega \cdot \Omega) \, P_{\omega^\bot} \Omega \cdot (\Lambda P_{\omega^\bot} E - W)  \omega - n \Lambda \, E: (\omega \otimes \omega) \big] \\
&& \hspace{0cm}
= \rho \, G_{\eta A_\Omega} \Big\{ \Lambda \big[ 2 \, \eta  \, (\omega \cdot \Omega) \, P_{\omega^\bot} \Omega \otimes \omega - n \, \omega \otimes \omega \big] : E - 2 \, \eta  \, (\omega \cdot \Omega) \, (P_{\omega^\bot} \Omega \otimes \omega):W \Big\}. 
\end{eqnarray*}
Using the decomposition \eqref{eq:omdecomp}, we get $P_{\omega^\bot} \Omega = (1 - (\omega \cdot \Omega)^2) \, \Omega - (\omega \cdot \Omega) \, \omega_\bot$, and so, 
\begin{eqnarray*}
&&\hspace{-1cm}
A (\rho \, G_{\eta A_\Omega}) = \rho \, G_{\eta A_\Omega} \, 2 \eta (\omega \cdot \Omega)  \, \Big[ \Lambda \Big( 1 - \frac{n}{\eta} - 2 (\omega \cdot \Omega)^2 \Big) \, (\omega_\bot \otimes \Omega):E + (\omega_\bot \otimes \Omega):W \Big] \\
&&\hspace{9cm}
 + \mbox{ even tensor powers of } \omega_\bot. 
\end{eqnarray*}
Now, multiplying by $h  \, \omega_\bot$ and integrating over $\omega$, the resulting odd tensor powers of $\omega_\bot$ vanish in the integration thanks to \eqref{eq:moments1}. Thanks to \eqref{eq:moments2}, we find that 
\begin{equation}
V_\Omega^{(2)} = \tilde \gamma_1 \,  W \Omega + \tilde \gamma_3 \Lambda \, P_{\Omega^\bot} E \Omega, \label{eq:V2}
\end{equation}
with 
\begin{eqnarray}
\tilde \gamma_3 &=& \big( 1 - \frac{n}{\eta} \big) \, \tilde \gamma_1 - 2 \, \tilde \gamma_2 , \label{eq:gamma3} \\
\tilde \gamma_2 &=& \frac{2 \, \eta \, \rho}{n-1} \int_{{\mathbb S}^{n-1}} G_{\eta A_\Omega} \, h_\eta
\, \, (\omega \cdot \Omega)^3 \, (1- (\omega \cdot \Omega)^2) \, d\omega. \label{eq:gamma2} 
\end{eqnarray}

The computation of $V_\Omega^{(3)}$ is the same as that of $V_\Omega^{(2)}$ with $\Lambda E - W$ replaced by $2 \alpha \beta \, \Delta_x (\rho Q_{G_{\eta A_\Omega}})$. Since $\Delta_x (\rho Q_{G_{\eta A_\Omega}})$ is a symmetric trace-free tensor, we get from \eqref{eq:V2}: 
$$
V_\Omega^{(3)} = 2 \alpha \beta \tilde \gamma_3 \, P_{\Omega^\bot} \big( \Delta_x (\rho Q_{G_{\eta A_\Omega}}) \Omega \big). 
$$
With \eqref{eq:alpharhoQ=etaA}, we get
\begin{eqnarray}
\alpha \Delta_x (\rho Q_{G_{\eta A_\Omega}}) = \Delta_x (\eta A_\Omega) &=&  \Delta_x \eta \, A_\Omega + 4  \Big[ \Big( \big( \nabla_x \eta \cdot \nabla_x \big) \Omega \Big) \otimes \Omega \Big]_s \nonumber \\
&& \hspace{1cm} + 2 \eta \, \big( (\nabla_x \Omega)^T (\nabla_x \Omega) + \big[(\Delta_x \Omega ) \otimes \Omega \big]_s \big), 
\label{eq:DeltaetaA}.
\end{eqnarray}
where the index $s$ means the symmetric part of a tensor (i.e. $S_s = \frac{1}{2} (S+S^T)$ for an $n \times n$ matrix $S$). Then, owing to the fact that any derivative of $\Omega$ is orthogonal to $\Omega$, we have
$$
\alpha \Delta_x (\rho Q_{G_{\eta A_\Omega}}) \Omega =  \Delta_x \eta \, \Omega + 2 \big( \nabla_x \eta \cdot \nabla_x \big) \Omega  +  \eta \, \big( \Delta_x \Omega +  (\Omega \cdot \Delta_x \Omega) \, \Omega \big),  
$$
and with \eqref{eq:eta=alrholam}, 
$$
\alpha P_{\Omega^\bot} \big( \Delta_x (\rho Q_{G_{\eta A_\Omega}}) \Omega \big) =  2 \big( \nabla_x \eta \cdot \nabla_x \big) \Omega  +  \eta \, P_{\Omega^\bot} \Delta_x \Omega =  P_{\Omega^\bot} \Delta_x (\eta \Omega).   
$$
It follows that 
\begin{equation}
V_\Omega^{(3)} = 2 \beta \tilde \gamma_3 \, P_{\Omega^\bot} \Delta_x (\eta \Omega). 
\label{eq:V3}
\end{equation}

Inserting \eqref{eq:V1}, \eqref{eq:V2}, \eqref{eq:V3}, into \eqref{eq:V},  we get 
$$
V_\Omega = \tilde \gamma_1 \, \big( D_t \Omega +  W \Omega \big) + \tilde \gamma_3 \, P_{\Omega^\bot} \big( \Lambda E \Omega + 2\beta \, \Delta_x (\eta \Omega) \big).  
$$
So, with \eqref{eq:Vomega} and \eqref{eq:gamma3}, we get \eqref{eq:Omega} with 
\begin{equation} 
c= - \Lambda \frac{\tilde \gamma_3}{\tilde \gamma_1} =  \Lambda \Big( \frac{n}{\eta} -1 + 2 \frac{\tilde \gamma_2}{\tilde \gamma_1} \Big).
\label{eq:express_c}
\end{equation}
Now, the following formulas are shown in the Appendix \ref{sec:proofgamalter}: 
\begin{equation}
\tilde \gamma_3 = \frac{\rho S_2(\eta)}{2 \eta}, \qquad \tilde \gamma_1 = - \frac{\rho}{2 \eta (n-1)} \, \langle \hspace{-0.8mm} \langle g  \, \frac{d \tilde U_0}{d \theta} \rangle \hspace{-0.8mm} \rangle_{e^{\eta \cos^2 \theta}}. 
\label{eq:gamalter}
\end{equation}
Thus, \eqref{eq:express_c} leads to \eqref{eq:def_c} and ends the proof. \endproof 

We now investigate under which conditions $c$ is non-negative:

\medskip
\noindent
\textbf{Proof of Proposition \ref{prop:const_c}.} From Prop. \ref{prop:OP} (iii), we know that the $(n-1) S_2(\eta) >0$. Now, Prop. \ref{prop:expressGCIAOm} and Eqs. \eqref{eq:defg} and \eqref{eq:dU0dthet} show that both $g(\theta)$ and $\frac{d \tilde U_0}{d \theta}(\theta)$ have the sign of $\cos \theta$. This implies that $g(\theta) \frac{d \tilde U_0}{d \theta}(\theta)$ is positive on $[0,\pi]$ and consequently, that the denominator of \eqref{eq:def_c} is positive. Altogether, this shows that $\frac{c}{\Lambda}>0$ and ends the proof.  \endproof

%%%%%%%%%%%%%%%%%%%%%%%%%%%%%%%%%%%%%%%%%%%%%%%%%%%%%%%%%%%%%%%%
%%%%%%%%%%%%%%%%%%%%%%%%%%%%%%%%%%%%%%%%%%%%%%%%%%%%%%%%%%%%%%%%
%%%%%%%%%%%%%%%%%%%%%%%%%%%%%%%%%%%%%%%%%%%%%%%%%%%%%%%%%%%%%%%%
%%%%%%%%%%%%%%%%%%%%%%%%%%%%%%%%%%%%%%%%%%%%%%%%%%%%%%%%%%%%%%%%

\setcounter{equation}{0}
\section{Conclusion}
\label{sec:conclusion}

We have investigated the passage from the Doi-Navier-Stokes model of liquid crystals to the Ericksen-Leslie system when the Deborah number goes to zero. By contrast to previous literature, we have developed a moment method, exploiting the conservations satisfied by the collision operator. These conservations are of a non-classical type and have required the development of a new concept, the generalized collision invariants. Their link to geometrical and analytical structures of the collision operator has been discussed and their use for the derivation of the limit model has been detailed. This derivation has been achieved in arbitrary dimensions and assuming a full spatio-temporal dependence of the polymer molecule density. The latter generates additional terms in the Ericksen stresses that have not been previously described in the literature. 

This works open many research directions. The first one is the development of a rigorous convergence result using this moment method. This is a quite challenging task but one may hope that, if successful, it would lead to a result in a weaker setting than the currently available results. The energetic properties of the limit model must be investigated. A proof that the extra terms appearing in the Oseen-Franck energy due to the spatio-temporal dependence of the polymer molecule density lead to a positive energy is missing at the present time. This would be a necessary step for a well-posedness theory for the resulting Ericksen-Leslie system. In spite of using Q-tensors as auxiliary quantities, the Doi model and its limit, the Ericksen-Leslie system are, in essence, vector models, i.e. models for polymer orientations only. Currently, attempts are being made to build truly tensorial models in association with Landau-de Gennes energies i.e. energies depending on the local average Q-tensor and its gradients. This is clearly an interesting playground to test the applicability of the GCI concept to more general situations.

%%%%%%%%%%%%%%%%%%%%%%%%%%%%%%%%%%%%%%%%%%%%%%%%%%%%%%%%%%%%%%%%
%%%%%%%%%%%%%%%%%%%%%%%%%%%%%%%%%%%%%%%%%%%%%%%%%%%%%%%%%%%%%%%%
%%%%%%%%%%%%%%%%%%%%%%%%%%%%%%%%%%%%%%%%%%%%%%%%%%%%%%%%%%%%%%%%
%%%%%%%%%%%%%%%%%%%%%%%%%%%%%%%%%%%%%%%%%%%%%%%%%%%%%%%%%%%%%%%%

%%%%%%%%%%%%%%%%%%%%%%%%%%%%%%%%%%%%%%%%%%%%%%%%%%%%%%%%%%%%%%%%
%%%%%%%%%%%%%%%%%%%%%%%%%%%%%%%%%%%%%%%%%%%%%%%%%%%%%%%%%%%%%%%%
%%%%%%%%%%%%%%%%%%%%%%%%%%%%%%%%%%%%%%%%%%%%%%%%%%%%%%%%%%%%%%%%
%%%%%%%%%%%%%%%%%%%%%%%%%%%%%%%%%%%%%%%%%%%%%%%%%%%%%%%%%%%%%%%%

\newpage 

\begin{center}
\bf\LARGE Appendices
\end{center}

\appendix

\setcounter{equation}{0}
\section{Appendix to Section \ref{sec:kinetic} on Doi's model}
\label{sec:app_Doi}

\subsection{Proof of the virtual work principle \eqref{eq:virtualwork}}
\label{sec:virtualwork}

We have, with \eqref{eq:kinetic_2}:
\begin{eqnarray*}
&&\hspace{-1cm}
\frac{d {\mathcal A}^R}{dt} = \Big \langle \frac{\delta {\mathcal A}^R}{\delta f} (f) , \frac{\partial f}{\partial t} \Big \rangle = \int_{{\mathbb R}^n \times {\mathbb S}^{n-1}} \mu_f^R \, \frac{\partial f}{\partial t} \, dx \, d \omega \\
&&\hspace{-0.8cm}
= \int_{{\mathbb R}^n \times {\mathbb S}^{n-1}} \mu_f^R \Big\{ - \nabla_x \cdot (u \, f) - \nabla_\omega \cdot \big(f \, (\Lambda P_{\omega^\perp} E - W ) \omega \big) +  \frac{D}{k_B T} \nabla_\omega \cdot \big( f \, \nabla_\omega \mu_f^R \big) \Big\} dx \, d \omega \\
&&\hspace{-0.8cm}
=: \mathrm{I} + \mathrm{II} + \mathrm{III}
\end{eqnarray*}
Using Stokes's formula, assuming that all terms vanish at infinity and with \eqref{eq:sigmae_Fe}, we find
$$
\mathrm{I} = - \int_{{\mathbb R}^n}  F_f^R \cdot u \, dx, \quad 
\mathrm{III} =  - \frac{D}{k_B T} \int_{{\mathbb R}^n \times {\mathbb S}^{n-1}} f \, |\nabla_\omega \mu_f^R |^2 \,  dx \, d \omega. 
$$
Then, using Stokes's formula, the fact that $\nabla_\omega \mu_f^R \cdot \omega = 0$ and straightforward tensor algebra, we have 
\begin{eqnarray*}
\mathrm{II} &=& \int_{{\mathbb R}^n \times {\mathbb S}^{n-1}} f \,  (\Lambda E - W) \omega \cdot \nabla_\omega \mu_f^R \, \, dx \, d \omega \\
&=&  \int_{{\mathbb R}^n} \Big( \int_{{\mathbb S}^{n-1}} f \, \big( \omega \otimes \nabla_\omega \mu_f^R \big) \, d \omega \Big) : (\Lambda E + W) \, dx \\
&=&  \int_{{\mathbb R}^n} \Big( \int_{{\mathbb S}^{n-1}} f \, \Big[ \frac{\Lambda + 1}{2} \big( \omega \otimes \nabla_\omega \mu_f^R \big) + \frac{\Lambda - 1}{2} \big( \nabla_\omega \mu_f^R \otimes \omega \big) \Big] \, d \omega \Big) : \nabla_x u \, dx \\
&=&  \int_{{\mathbb R}^n} \Big( \int_{{\mathbb S}^{n-1}} f \, \Big[ \Lambda \big( \omega \otimes \nabla_\omega \mu_f^R \big)_s  -  \big( \omega \otimes \nabla_\omega \mu_f^R \big)_a  \Big] \, d \omega \Big) : \nabla_x u \, dx \\
&=& \int_{{\mathbb R}^n} \sigma_f^R : \nabla_x u \, dx. 
\end{eqnarray*}
This leads to \eqref{eq:virtualwork}. \endproof

%%%%%%%%%%%%%%%%%%%%%%%%%%%%%%%%%%%%%%%%%%%%%%%%%%%%%%%%%%%%%%%%
%%%%%%%%%%%%%%%%%%%%%%%%%%%%%%%%%%%%%%%%%%%%%%%%%%%%%%%%%%%%%%%%
\subsection{Proofs of Formulas \eqref{eq:express_sigmaR_1} and \eqref{eq:express_sigmaR_2} for the extra-stresses}
\label{sec:extrastress}

We begin with a Lemma:

\begin{lemma}
Let $f$ and $\varphi$: ${\mathbb S}^{n-1} \to {\mathbb R}$ be two smooth functions. Then, we have
\begin{equation}
\int_{{\mathbb S}^{n-1}} \nabla_\omega f \, \varphi \, d \omega = - \int_{{\mathbb S}^{n-1}}  f \, \nabla_\omega \varphi \, d \omega + (n-1) \int_{{\mathbb S}^{n-1}} f \, \varphi \, \omega \, d \omega. 
\label{eq:int_grad_om_phi}
\end{equation}
%\label{lem:int_grad_om_phi}
\end{lemma}

\noindent
\textbf{Proof:} Let $B \in {\mathbb R}^n$ be a fixed vector and denote by $X$ the left-hand side of \eqref{eq:int_grad_om_phi}. Then, using Stokes formula, we have
\begin{eqnarray*}
X \cdot B &=& \int_{{\mathbb S}^{n-1}} \nabla_\omega f \cdot B \, \varphi \, d \omega = \int_{{\mathbb S}^{n-1}} \nabla_\omega f \cdot P_{\omega^\bot} B \, \varphi \, d \omega = - \int_{{\mathbb S}^{n-1}}  f \, \nabla_\omega \cdot (P_{\omega^\bot} B \, \varphi) \, d \omega \\
&=& - \int_{{\mathbb S}^{n-1}}  f \, \nabla_\omega \cdot (P_{\omega^\bot} B) \, \varphi \, d \omega - \int_{{\mathbb S}^{n-1}}  f  \, P_{\omega^\bot} B \cdot \nabla_\omega\varphi \, d \omega. 
\end{eqnarray*}
We have 
$$
\nabla_\omega \cdot (P_{\omega^\bot} B) = \nabla_\omega \cdot  \nabla_\omega \, (\omega \cdot B) = \Delta_\omega \, (\omega \cdot B) = -(n-1) \,\omega \cdot B, 
$$
where the last identity follows from the fact that the function $\omega \mapsto \omega \cdot B$ is a spherical harmonic of degree~1.  
Thus, 
$$ X \cdot B = (n-1) \int_{{\mathbb S}^{n-1}}  f \, \varphi \, \omega \cdot B \,  d \omega - \int_{{\mathbb S}^{n-1}}  f  \, \, \nabla_\omega\varphi \cdot B\, d \omega, $$
which leads to \eqref{eq:int_grad_om_phi}.  \endproof

\medskip
\noindent
\textbf{Proof of \eqref{eq:express_sigmaR_1}:} Inserting \eqref{eq:muf} into the first equation of \eqref{eq:sigmae_Fe}, we have $\sigma_f^R = \Lambda \bar \sigma_s + \bar \sigma_a$ with
\begin{eqnarray} 
\bar \sigma &=& \int_{{\mathbb S}^{n-1}} f (\omega \otimes \nabla_\omega \mu_f^R ) \, d\omega \nonumber \\
&=& k_B T \int_{{\mathbb S}^{n-1}} \omega \otimes \nabla_\omega f \, d \omega + \int_{{\mathbb S}^{n-1}} f \, (\omega \otimes \nabla_\omega U_f^R) \, d\omega. 
\label{eq:sigma_comput1}
\end{eqnarray}
Using \eqref{eq:int_grad_om_phi} with $\varphi = \omega_i$, we get 
\begin{eqnarray*} 
\int_{{\mathbb S}^{n-1}} \omega_i \, \nabla_\omega f \, d \omega &=& - \int_{{\mathbb S}^{n-1}} \nabla_\omega \omega_i  \, f \, d \omega + (n-1) \int_{{\mathbb S}^{n-1}} f \, \omega \, \omega_i \, d\omega \\
&=& - \int_{{\mathbb S}^{n-1}} P_{\omega^\bot} e_i  \, f \, d \omega + (n-1) \int_{{\mathbb S}^{n-1}} f \, \omega \, (\omega \cdot e_i) \, d\omega \\
&=& - \int_{{\mathbb S}^{n-1}} \big(e_i - (e_i \cdot \omega) \omega \big)  \, f \, d \omega + (n-1) \int_{{\mathbb S}^{n-1}} f \, (e_i \cdot \omega) \omega \, d\omega \\
&=& n \int_{{\mathbb S}^{n-1}} f \big( \omega \, (\omega \cdot e_i) - \frac{1}{n} e_i \big) \, d\omega, 
\end{eqnarray*}
where $e_i$ denotes the $i$-th vector of the canonical basis of ${\mathbb R}^n$. In view of \eqref{eq:orienttensor}, it follows that $ \int_{{\mathbb S}^{n-1}} \omega \otimes \nabla_\omega f \, d \omega = n \rho_f Q_f$. Inserting this in \eqref{eq:sigma_comput1} leads to 
$$
\bar \sigma = n k_B T \rho_f Q_f + \int_{{\mathbb S}^{n-1}} f \, (\omega \otimes \nabla_\omega U_f^R) \, d\omega, 
$$
which, in turn, leads to \eqref{eq:express_sigmaR_1}. \endproof

\medskip
\noindent
\textbf{Proof of \eqref{eq:express_sigmaR_2}:} We multiply Doi's equation \eqref{eq:kinetic_2} by $\omega \otimes \omega - \frac{1}{n} \mathrm{Id}$ and integrate it with respect to $\omega$. This leads to 
\begin{eqnarray}
0&=&\int_{{\mathbb S}^{n-1}} (\partial_t f + \nabla_x \cdot (u f)) \, \big( \omega \otimes \omega - \frac{1}{n} \mathrm{Id} \big) \, d \omega \nonumber \\
&+&  \int_{{\mathbb S}^{n-1}} \nabla_\omega \cdot \big( f (\Lambda P_{\omega^\bot} E-W) \omega \big) \, \big( \omega \otimes \omega - \frac{1}{n} \mathrm{Id} \big) \, d \omega \nonumber \\
&-& \frac{D}{k_B T} \int_{{\mathbb S}^{n-1}} \nabla_\omega \cdot (f \, \nabla_\omega \mu_f^R) \, \big( \omega \otimes \omega - \frac{1}{n} \mathrm{Id} \big) \, d \omega \nonumber \\
&=:& \mathrm{I} + \mathrm{II} - \frac{D}{k_B T} \mathrm{III}. \label{eq:Doi_omom}
\end{eqnarray}
Using \eqref{eq:divu=0}, for any smooth function $g(x,t)$, we have $ \partial_t g + \nabla_x \cdot (u g) = D_t g$, where $D_t$ is given by \eqref{eq:Dt}. It follows that $\mathrm{I} = D_t (\rho_f Q_f)$ and, using \eqref{transport-mass}, that 
\begin{equation} 
\mathrm{I} = \rho_f \, D_t Q_f.
\label{eq:I}
\end{equation}
Using Stokes theorem, we get:
\begin{eqnarray}
\mathrm{III}_{ij} &=& \int_{{\mathbb S}^{n-1}} \nabla_\omega \cdot (f \, \nabla_\omega \mu_f^R) \, \omega_i \, \omega_j \, d \omega = 
- \int_{{\mathbb S}^{n-1}}  f \, \nabla_\omega \mu_f^R \cdot \nabla_\omega (\omega_i \, \omega_j) \, d \omega \nonumber \\
&=& - \int_{{\mathbb S}^{n-1}}  f \, \nabla_\omega \mu_f^R \cdot (\omega_j P_{\omega^\bot} e_i + \omega_i P_{\omega^\bot} e_j)  \, d \omega \nonumber \\
&=&  - \int_{{\mathbb S}^{n-1}}  f \, (\omega_j \nabla_\omega \mu_f^R \cdot e_i + \omega_i \nabla_\omega \mu_f^R \cdot e_j)  \, d \omega  = - \frac{2}{\Lambda} \big( (\sigma_f^R)_s \big)_{ij}, \label{eq:III}
\end{eqnarray}
where again, $e_i$ denotes the $i$-th vector of the canonical basis of ${\mathbb R}^n$. Now, similarly to $\mathrm{III}$, we have,  
\begin{eqnarray*}
\mathrm{II}_{ij} &=& \int_{{\mathbb S}^{n-1}} \nabla_\omega \cdot \big( f (\Lambda P_{\omega^\bot} E-W) \omega \big) \, \omega_i \, \omega_j  \, d \omega \\
&=& - \int_{{\mathbb S}^{n-1}} f \, \big( (\Lambda P_{\omega^\bot} E-W) \omega \big) \cdot (\omega_i e_j + \omega_j e_i)  \, d \omega \\
&=& - \int_{{\mathbb S}^{n-1}} f \, \big( (\Lambda E - W) (\omega \otimes \omega) + (\omega \otimes \omega) (\Lambda E + W) - 2 \Lambda \, \omega^{\otimes 4} : E \big)_{ij} \, d \omega, \\
\end{eqnarray*}
which leads to 
\begin{eqnarray}
\mathrm{II} = \rho_f \, \big( - \Lambda (E Q_f + Q_f E) + W Q_f - Q_f W - \frac{2 \Lambda}{n} E + 2 \Lambda \, {\mathbb T}_f:E \big). 
\label{eq:II}
\end{eqnarray}
Finally, using \eqref{eq:express_sigmaR_1}, the antisymmetric part of $\sigma_f^R$ is given by: 
\begin{equation}
(\sigma_f^R)_a = \frac{1}{2} \int_{{\mathbb S}^{n-1}} (\omega \otimes \nabla_\omega U_f^R - \nabla_\omega U_f^R \otimes \omega) \, f \, d \omega. 
\label{eq:sigfR_antisym}
\end{equation}
Now, inserting \eqref{eq:I}, \eqref{eq:III}, \eqref{eq:II} and \eqref{eq:sigfR_antisym} into \eqref{eq:Doi_omom} leads to \eqref{eq:express_sigmaR_2}. \endproof

%%%%%%%%%%%%%%%%%%%%%%%%%%%%%%%%%%%%%%%%%%%%%%%%%%%%%%%%%%%%%%%%
%%%%%%%%%%%%%%%%%%%%%%%%%%%%%%%%%%%%%%%%%%%%%%%%%%%%%%%%%%%%%%%%
%%%%%%%%%%%%%%%%%%%%%%%%%%%%%%%%%%%%%%%%%%%%%%%%%%%%%%%%%%%%%%%%
%%%%%%%%%%%%%%%%%%%%%%%%%%%%%%%%%%%%%%%%%%%%%%%%%%%%%%%%%%%%%%%%
\setcounter{equation}{0}
\section{Appendix to Section \ref{sec:main_result} on main result}
\label{sec:app_sec_main_res}

\subsection{Proof of Prop. \ref{prop:OP} on properties of $S_2$}
\label{subsec:proof_prop_3.6}

The proof uses Lemma 4.1 of \cite{Degond_Merino_M3AS20} which we recall here without proof. 

\begin{lemma}
Let $n \geq 2$. Define $\omega_\bot = P_{\Omega^\bot} \omega$. For any function $k$: $[-1,1] \to {\mathbb R}$, $r \mapsto k(r)$, we have:  
\begin{eqnarray}
&& \hspace{-1cm}
\int_{{\mathbb S}^{n-1}} k(\omega \cdot \Omega) \, \omega_\bot^{\otimes (2k+1)} \ d\omega =0, \quad \forall k \in {\mathbb N}, 
\label{eq:moments1} \\
&& \hspace{-1cm}
\int_{{\mathbb S}^{n-1}} k(\omega \cdot \Omega) \, \omega_\bot \otimes \omega_\bot \, d\omega = \frac{1}{n-1} \int_{{\mathbb S}^{n-1}} k(\omega \cdot \Omega) \, (1-(\omega \cdot \Omega)^2) \, d\omega \, \,  P_{\Omega^\bot}.
\label{eq:moments2} 
\end{eqnarray}
\label{lem:moments}
\end{lemma}

\noindent
\textbf{Proof of Proposition \ref{prop:OP}}
(i) The decomposition
\begin{equation}
\omega = (\omega \cdot \Omega) \, \Omega + \omega_\bot, 
\label{eq:omdecomp}
\end{equation} 
leads to 
\begin{equation}
\omega \otimes \omega = (\omega \cdot \Omega)^2 \, \Omega \otimes \Omega + (\omega \cdot \Omega) \,  (\omega_\bot \otimes \Omega + \Omega \otimes \omega_\bot) + \omega_\bot \otimes \omega_\bot. 
\label{eq:omotomdecomp}
\end{equation}
We insert \eqref{eq:omotomdecomp} into \eqref{eq:orienttensor} with $f = \rho G_{\eta A_\Omega}$. Thanks to \eqref{eq:GetaAom}, $\rho G_{\eta A_\Omega}$ is a function of $\omega \cdot \Omega$ only. So, the contribution of the middle term of \eqref{eq:omotomdecomp} vanishes thanks to \eqref{eq:moments1} and the contribution of the last term can be computed using \eqref{eq:moments2}. Using that $P_{\Omega^\bot} = \mathrm{Id} - \Omega \otimes \Omega$, we get 
$$ 
Q_{G_{\eta A_\Omega}} = \big\langle (\omega \cdot \Omega)^2 \big \rangle_{G_{\eta A_\Omega}} \, \Omega \otimes \Omega + \frac{1}{n-1} \big\langle 1 - (\omega \cdot \Omega)^2 \big \rangle_{G_{\eta A_\Omega}} \Big( \mathrm{Id} - \Omega \otimes \Omega \Big) - \frac{1}{n} \mathrm{Id}. 
$$
Rearranging these terms, we find \eqref{eq:Q_equib}.

\smallskip
\noindent
(ii) The leading eigenvalue of $Q_{G_{\eta A_\Omega}}$ is $\frac{n-1}{n} S_2(\eta)$ and is associated with the eigenvector $\Omega$. Thus, by virtue of \eqref{eq:OP}, the order parameter $\chi_{\rho G_{\eta A_\Omega}}$ is equal to $S_2(\eta)$.

\smallskip
\noindent
(iii) We first compute $S_2(0)$. When $\eta = 0$, we have $G_{\eta A_\Omega} = 1$. Thus, $
S_2(0) = \langle (n (\omega \cdot \Omega)^2 - 1)/(n-1) \rangle_1 =: r/s$, where, using the spherical coordinates as in the proof of Proposition \ref{prop:gibbs_uniaxial}, the numerator $r$ is given by 
$$
r = \int_0^\pi (n \cos^2 \theta - 1) \, \sin^{n-2} \theta \, d \theta = (n-1) W_{n-2} - n W_n. 
$$
Here, $W_n$ is twice the Wallis integral $W_n = \int_0^\pi \sin^n \theta \, d \theta$. From the well-known recursion formula for the Wallis integral (which can be easily proved by integration by parts): $W_n = \frac{n-1}{n} W_{n-2}$, we get that $r=0$ and thus, that $S_2(0) = 0$. 

We now show that $S_2'(\eta) \geq 0$, for all $\eta \geq 0$, where the prime denotes the derivative with respect to $\eta$. We have $S_2(\eta) = 1 - \frac{n}{n-1} \langle 1 - (\omega \cdot \Omega)^2 \rangle_{G_{\eta A_\Omega}}=:  1 - \frac{n}{n-1} F(\eta)$. We show that $F' \leq 0$. Using again the spherical coordinates, we have $F = I_n/I_{n-2}$ with $I_n(\eta) = \int_0^{\pi/2} \exp(\eta \cos^2 \theta) \, \sin^n \theta \, d \theta$ (by symmetry, we can reduce the interval of integration to $[0,\pi/2]$). Thus, $F' = (I'_n I_{n-2} - I'_{n-2} I_n)/I_{n-2}^2=:A/I_{n-2}^2$. We check the sign of the numerator $A$. We have  
\begin{eqnarray*}
&&\hspace{-1cm}
A(\eta)  = \int_{[0,\frac{\pi}{2}]^2} e^{\eta (\cos^2 \theta + \cos^2 \theta')} \, \sin^{n-2} \theta \, \sin^{n-2} \theta' \, \sin^2 \theta \, (\cos^2 \theta - \cos^2 \theta') \, d \theta \, d \theta'\\
&&\hspace{-1cm}
= \frac{1}{2} \int_{[0,\frac{\pi}{2}]^2} e^{\eta (\cos^2 \theta + \cos^2 \theta')} \, \sin^{n-2} \theta \, \sin^{n-2} \theta' \, (\sin^2 \theta - \sin^2 \theta') \, (\cos^2 \theta - \cos^2 \theta') \, d \theta\, d \theta', 
\end{eqnarray*}
where we pass from the first to the second line by exchanging $\theta$ and $\theta'$. Since $\sin$ is increasing and $\cos$ is decreasing on $[0,\frac{\pi}{2}]$, we have $A \leq 0$. 

Finally, when $\eta \to \infty$, the measure $G_{\eta A_\Omega} \, d \omega$ concentrates onto the sum of Dirac deltas $\frac{1}{2} (\delta_\Omega + \delta_{-\Omega})$. Since $P_2(\pm 1) = 1$, it follows that $S_2 \to 1$ when $\eta \to \infty$. This ends the proof. \endproof

%%%%%%%%%%%%%%%%%%%%%%%%%%%%%%%%%%%%%%%%%%%%%%%%%%%%%%%%%%%%%%%%
%%%%%%%%%%%%%%%%%%%%%%%%%%%%%%%%%%%%%%%%%%%%%%%%%%%%%%%%%%%%%%%%

\subsection{Proof of Eq. \eqref{eq:sigmae_equi} for the Leslie stresses}
\label{sec:extrastress_EL}

We have $f^\varepsilon \to f$ as $\varepsilon \to 0$ with $f$ given by \eqref{eq:localequ}. We will abbreviate $G_{\eta A_\Omega}$ into $G$ for simplicity. We define 
$$
\sigma = \lim_{\varepsilon \to 0} \big( \zeta \, \rho_{f^\varepsilon } {\mathbb T}_{f^\varepsilon } :E^\varepsilon + \sigma_{f^\varepsilon}^1 \big) = \zeta \, \rho \, {\mathbb T}_G:E + \sigma_{\rho G}^1. 
$$
From \eqref{eq:sigma1}, we get 
\begin{eqnarray} 
\sigma &=& \rho \, \Big\{ \frac{\Lambda^2}{2} (E Q_G + Q_G E) +  \frac{\Lambda}{2} (Q_G W - W Q_G) + \frac{\Lambda^2}{n} E  \nonumber \\
&& \hspace{0.5cm} + (\zeta - \Lambda^2) {\mathbb T}_G:E - \frac{\Lambda}{2} D_t Q_G + \alpha \beta \big[ \Delta_x (\rho Q_G) Q_G - Q_G  \Delta_x (\rho Q_G) \big] \Big\}. 
\label{eq:sigma_cpt1}
\end{eqnarray}
Now, for a generic distribution function $f$, we introduce the fourth-order tensorial order parameter given by 
\begin{equation}
{\mathbb Q}_f = {\mathbb T}_f - \frac{6}{n+4} (\langle \omega \otimes \omega \rangle_f \otimes \mathrm{Id})_s + \frac{3}{(n+2)(n+4)} (\mathrm{Id} \otimes \mathrm{Id})_s. 
\label{eq_4tensororder_0}
\end{equation}
Here, $(\langle \omega \otimes \omega \rangle_f \otimes \mathrm{Id})_s$ and  $(\mathrm{Id} \otimes \mathrm{Id})_s$ denote the symmetrizations of the fourth-order tensors $\langle \omega \otimes \omega \rangle_f \otimes \mathrm{Id}$ and $\mathrm{Id} \otimes \mathrm{Id}$ respectively. Specifically, 
\begin{eqnarray*}
6 \big( (\langle \omega \otimes \omega \rangle_f \otimes \mathrm{Id} \otimes \mathrm{Id})_s \big)_{ijk\ell} &=& \langle \omega_i \omega_j \rangle_f \, \delta_{k \ell} + \langle \omega_i \omega_k \rangle_f \, \delta_{j \ell} + \langle \omega_i \omega_\ell \rangle_f \, \delta_{jk} \\
&& \hspace{2cm} + \langle \omega_j \omega_k \rangle_f \, \delta_{i \ell} + \langle \omega_j \omega_\ell \rangle_f \, \delta_{ik} + \langle \omega_k \omega_\ell \rangle_f \, \delta_{ij}, \\
3 \big( (\mathrm{Id} \otimes \mathrm{Id})_s \big)_{ijk\ell} &=& \delta_{ij} \delta_{k \ell} + \delta_{ik} \delta_{j \ell} + \delta_{i\ell} \delta_{jk}, 
\end{eqnarray*} 
where $\delta$ denotes the Kronecker symbol. Eq. \eqref{eq_4tensororder_0} corresponds to the decomposition of ${\mathbb T}_f$ into irreducible tensors, i.e. invariant tensors under the action of the orthogonal group. The coefficients of the decomposition can be obtained by the requirement that the contraction of ${\mathbb Q}_f$ with respect to any two indices is zero. Owing to the fact that $\langle \omega \otimes \omega \rangle_f =Q_f + \frac{1}{n} \mathrm{Id}$, we get
\begin{equation}
{\mathbb Q}_f = {\mathbb T}_f - \frac{6}{n+4} (Q_f \otimes \mathrm{Id})_s - \frac{3}{n(n+2)} (\mathrm{Id} \otimes \mathrm{Id})_s, 
\label{eq_4tensororder}
\end{equation}
where the definition of $(Q_f \otimes \mathrm{Id})_s$ is similar to that of $(\langle \omega \otimes \omega \rangle_f \otimes \mathrm{Id} \otimes \mathrm{Id})_s$. Then, using \eqref{eq_4tensororder}, we have 
$$ 
{\mathbb T}_f:E = {\mathbb Q}_f:E + \frac{2}{n+4} (E Q_f + Q_f E) + \frac{2}{n(n+2)} E + \frac{1}{n+4} (Q_f:E) \mathrm{Id}. 
$$
Inserting this identity (with $f=G$) into \eqref{eq:sigma_cpt1}, we get 
\begin{eqnarray}
\sigma &=&  \rho \, \Big\{ \big(\frac{\Lambda^2}{2} +   \frac{2 (\zeta - \Lambda^2)}{n+4} \big) (E Q_G + Q_G E) +  \frac{\Lambda}{2} (Q_G W - W Q_G) + \frac{1}{n} \big(\Lambda^2 + \frac{2 (\zeta - \Lambda^2)}{n+2} \big) E  \nonumber \\
&& \hspace{0.5cm} + (\zeta - \Lambda^2) \, {\mathbb Q}_G:E  + \frac{\zeta - \Lambda^2}{n+4} (Q_G:E) \, \mathrm{Id} - \frac{\Lambda}{2} D_t Q_G \nonumber \\
&& \hspace{0.5cm} + \alpha \beta \big[ \Delta_x (\rho Q_G) Q_G - Q_G  \Delta_x (\rho Q_G) \big] \Big\}. 
\label{eq:sigma_cpt2}
\end{eqnarray}

Now, we state two lemmas whose proofs are deferred to the end of the present proof 

\begin{lemma}
We have
\begin{equation}
{\mathbb Q}_{G_{\eta A_\Omega}} = S_4(\eta) \, {\mathbb A}_{\Omega}, 
\label{eq:Q4_equi}
\end{equation}
where $S_4(\eta)$ is given by \eqref{eq:S2}
and where 
\begin{equation}
{\mathbb A}_{\Omega} = \Omega^{\otimes 4} - \frac{6}{n+4} (\Omega \otimes \Omega \otimes \mathrm{Id})_s + \frac{3}{(n+2)(n+4)} (\mathrm{Id} \otimes \mathrm{Id})_s. 
\label{eq:A4_def}
\end{equation}
\label{lem:Q4_equi}
\end{lemma}

\begin{lemma}
We have 
\begin{eqnarray}
\rho \alpha \beta \big[ \Delta_x (\rho Q_G) Q_G - Q_G  \Delta_x (\rho Q_G) \big] &=& \rho \frac{\Lambda S_2(\eta)}{2} \Big[ \frac{1}{c} (N \otimes \Omega - \Omega \otimes N ) \nonumber \\
& & \hspace{1.cm} - \big( E (\Omega \otimes \Omega) - (\Omega \otimes \Omega) E \big) \Big], 
\label{eq:antisym_stress}
\end{eqnarray}
with $N$ given by \eqref{eq:defN}. 
\label{lem:antisym_stress}
\end{lemma}

From \eqref{eq:Q4_equi} and \eqref{eq:A4_def}, it follows that 
\begin{eqnarray}
{\mathbb Q}_G:E &=& S_4 \Big\{ \big( E:(\Omega \otimes \Omega) \big) \Omega \otimes \Omega - \frac{2}{n+4} \big[ (\Omega \otimes \Omega) E + E (\Omega \otimes \Omega) \big] \nonumber \\
&& \hspace{2cm} - \frac{1}{n+4} \big( E:(\Omega \otimes \Omega) \big) \mathrm{Id} + \frac{2}{(n+2)(n+4)} E \big\},
\label{eq:Q4:E}
\end{eqnarray}
where the dependence of $S_4$ on $\eta$ is omitted for simplicity. Likewise, with \eqref{eq:Q_equib}, we get 
\begin{eqnarray}
E Q_G + Q_G E &=& S_2 \big[ (\Omega \otimes \Omega) E + E (\Omega \otimes \Omega) - \frac{2}{n} E \big], \label{eq:EQ+QE} \\
Q_G W - W Q_G &=& S_2 \big[ (\Omega \otimes \Omega) W - W (\Omega \otimes \Omega) \big], \label{eq:WQ+QW} \\
D_t Q_G &=& S_2 \big[ D_t \Omega \otimes \Omega + \Omega \otimes D_t \Omega) \big]. \label{eq:DtQG}
\end{eqnarray}
In \eqref{eq:DtQG}, we have used that $D_t S_2(\eta(\rho)) = \frac{dS_2}{d\eta}(\eta(\rho)) \, \frac{d\eta}{d\rho}(\rho) \, D_t \rho = 0$ thanks to \eqref{transport-mass}. Inserting Eqs. \eqref{eq:antisym_stress} to \eqref{eq:DtQG} into \eqref{eq:sigma_cpt2}, we get $\sigma = \sigma_L + \nabla_x \varphi$ where $\varphi$ is a scalar function which can be absorbed in the pressure, and $\sigma_L$ is given by \eqref{eq:sigmae_equi} with the constants, $\alpha_k$, $k=1, \ldots, 6$ given by \eqref{eq:al123}-\eqref{eq:al6}. This ends the proof. \endproof

\medskip
\noindent
\textbf{Proof of Lemma \ref{lem:Q4_equi}.} Using \eqref{eq_4tensororder}, \eqref{eq:Q_equib}, \eqref{eq:S2} and \eqref{eq:P2}, we get that 
\begin{eqnarray}
&&\hspace{-1cm}
{\mathbb Q}_{G_{\eta A_\Omega}} = {\mathbb T}_{G_{\eta A_\Omega}} - \frac{6 S_2}{n+4} \, (\Omega \otimes \Omega \otimes \mathrm{Id})_s + \Big( \frac{6 S_2}{n(n+4)}  - \frac{3}{n (n+2)} \Big) \, (\mathrm{Id} \otimes \mathrm{Id})_s \nonumber \\
&& \hspace{-1cm} = {\mathbb T}_{G_{\eta A_\Omega}} - \frac{6 \big( n \langle X^2 \rangle - 1 \big)}{(n-1)(n+4)}   \, (\Omega \otimes \Omega \otimes \mathrm{Id})_s 
\nonumber \\
&& \hspace{3cm} 
+ \Big( \frac{6 \big( n \langle X^2 \rangle  - 1 \big)}{(n-1)n(n+4)}  - \frac{3}{n (n+2)} \Big) \,  (\mathrm{Id} \otimes \mathrm{Id})_s , 
\label{eq:Q4_comput_1}
\end{eqnarray}
where $X = \omega \cdot \Omega$ and where we drop the index $G_{\eta A_\Omega}$ on the brackets $\langle \cdot \rangle$. Now, using the decomposition \eqref{eq:omdecomp}, we get 
$$
{\mathbb T}_{G_{\eta A_\Omega}} = \langle X^4 \rangle \, \Omega^{\otimes 4} + \Big( \langle X^2 \omega_\perp \otimes \omega_\perp \rangle \otimes (\Omega \otimes \Omega) \Big)_s + \big\langle \omega_\perp^{\otimes 4} \big\rangle. 
$$
We use \eqref{eq:moments2} to compute $\langle X^2 \omega_\perp \otimes \omega_\perp \rangle$. To evaluate $\big\langle \omega_\perp^{\otimes 4} \big\rangle$ we recall the last part of Lemma 4.1 of \cite{Degond_Merino_M3AS20} without proof: with the notations of Lemma \ref{lem:moments}, we have 
$$
\int_{{\mathbb S}^{n-1}} k(\omega \cdot \Omega) \, \omega_\perp^{\otimes 4} \, d\omega 
= \int_{{\mathbb S}^{n-1}} \frac{3 \, k(\omega \cdot \Omega) \, (1-(\omega \cdot \Omega)^2)^2}{(n-1)(n+1)}  \, d\omega \, \,  \big( P_{\Omega^\bot} \otimes P_{\Omega^\bot} \big)_s.
$$
This leads to 
$$
{\mathbb T}_{G_{\eta A_\Omega}} = \langle X^4 \rangle \, \Omega^{\otimes 4} + \frac{6 \, \langle X^2 (1 - X^2) \rangle}{n-1}  \,  \big( \Omega \otimes \Omega \otimes P_{\Omega^\bot} \big)_s 
+ \frac{3 \, \langle (1 - X^2)^2 \rangle}{(n-1)(n+1)}  \, \big( P_{\Omega^\bot} \otimes P_{\Omega^\bot} \big)_s. 
$$
Using that $P_{\Omega^\bot} = \mathrm{Id} - \Omega \otimes \Omega$, we obtain
\begin{eqnarray}
&& \hspace{-1cm}
{\mathbb T}_{G_{\eta A_\Omega}} = \Big( \langle X^4 \rangle - \frac{6 \, \langle X^2 (1 - X^2) \rangle}{n-1}  + \frac{3\, \langle (1 - X^2)^2 \rangle}{(n-1)(n+1)}  \Big) \, \Omega^{\otimes 4} \nonumber \\
&& \hspace{-1.8cm}
+ \Big( \frac{6\, \langle X^2 (1 - X^2) \rangle}{n-1}   - \frac{6 \, \langle (1 - X^2)^2 \rangle}{(n-1)(n+1)}  \Big) \,  \big( \Omega \otimes \Omega \otimes \mathrm{Id} \big)_s  
+\frac{3\, \langle (1 - X^2)^2 \rangle}{(n-1)(n+1)}  \, \big( \mathrm{Id} \otimes \mathrm{Id} \big)_s. 
\label{eq:Q4_comput_2}
\end{eqnarray}
Now, inserting \eqref{eq:Q4_comput_2} into \eqref{eq:Q4_comput_1}, we get \eqref{eq:Q4_equi}. \endproof

\medskip
\noindent
\textbf{Proof of Lemma \ref{lem:antisym_stress}.}  Thanks to \eqref{eq:alpharhoQ=etaA} and \eqref{eq:DeltaetaA}, we have
\begin{eqnarray}
&&\hspace{-1cm}
\alpha^2 \rho \Delta_x (\rho Q_G) Q_G = \eta \, \Delta_x ( \eta A_\Omega) A_\Omega \nonumber \\
&&\hspace{-0.5cm}
= \eta \Big[ \Delta_x \eta \, A_\Omega^2 + \frac{2(n-1)}{n} \,  (\nabla_x \eta \cdot \nabla_x) \Omega \otimes \Omega - \frac{2}{n} \,  \Omega \otimes (\nabla_x \eta \cdot \nabla_x) \Omega \nonumber \\
&&\hspace{-0.0cm}
 - \frac{2 \eta}{n} \nabla_x \Omega^T \nabla_x \Omega + \frac{\eta (n-1)}{n} \Delta_x \Omega \otimes \Omega - \frac{\eta}{n} \Omega \otimes \Delta_x \Omega + \eta (\Omega \cdot \Delta_x \Omega) \, \Omega \otimes \Omega \Big]. 
\label{eq:expressM0}
\end{eqnarray} 
Let $M$ be the tensor given by the left-hand side of \eqref{eq:antisym_stress}. Using \eqref{eq:Omega} and \eqref{eq:eta}, it follows from \eqref{eq:expressM0} that 
\begin{eqnarray}
M &=& \frac{\beta \eta}{\alpha} \, \Big[  2 \, \big(  (\nabla_x \eta \cdot \nabla_x) \Omega \otimes \Omega - \Omega \otimes (\nabla_x \eta \cdot \nabla_x) \Omega \big) +  \eta \,  \big( \Delta_x \Omega \otimes \Omega -  \Omega \otimes \Delta_x \Omega \big) \Big] \nonumber \\
&=&  \frac{\beta}{\alpha} \, \Big[ \Delta_x (\eta \Omega) \otimes (\eta \Omega) -  (\eta \Omega) \otimes \Delta_x (\eta \Omega)  \Big] \nonumber \\
&=&  \frac{\Lambda}{2\alpha} \, \Big[ \Big( \frac{N}{c} - P_{\Omega^\bot} E \Omega \Big) \otimes (\eta \Omega) -  (\eta \Omega) \otimes \Big( \frac{N}{c} - P_{\Omega^\bot} E \Omega \Big) \Big] \nonumber \\
&=&  \frac{\rho \Lambda S_2}{2} \, \Big[ \Big( \frac{N}{c} - P_{\Omega^\bot} E \Omega \Big) \otimes \Omega -  \Omega \otimes \Big( \frac{N}{c} - P_{\Omega^\bot} E \Omega \Big) \Big]. 
\label{eq:expressM}
\end{eqnarray} 
Then, we note that there exists a real number $z$ such that 
$$(P_{\Omega^\bot} E \Omega) \otimes \Omega = (E \Omega) \otimes \Omega + z \Omega \otimes \Omega = E (\Omega \otimes \Omega) + z \Omega \otimes \Omega, $$
and that the same real number $z$ is involved in the expression of $\Omega \otimes (P_{\Omega^\bot} E \Omega)$, so that we get 
$$ (P_{\Omega^\bot} E \Omega) \otimes \Omega - \Omega \otimes (P_{\Omega^\bot} E \Omega) = E (\Omega \otimes \Omega) - (\Omega \otimes \Omega) E. $$
Inserting this expression into \eqref{eq:expressM}, we get \eqref{eq:antisym_stress} which ends the proof of the Lemma. \endproof

%%%%%%%%%%%%%%%%%%%%%%%%%%%%%%%%%%%%%%%%%%%%%%%%%%%%%%%%%%%%%%%%
%%%%%%%%%%%%%%%%%%%%%%%%%%%%%%%%%%%%%%%%%%%%%%%%%%%%%%%%%%%%%%%%
\subsection{Proof of Eq. \eqref{eq:sigma_elas_equi} for the Ericksen stresses}
%\label{sec:extrastress_Erick}

We now compute $\lim_{\varepsilon \to 0} F^1_{f^\varepsilon} = F^1_{\rho G_{\eta A_\Omega}}$. Thanks to 
\eqref{eq:F1}, \eqref{eq:muf1} and \eqref{eq:def_U0_0}, we have 
\begin{eqnarray}
F^1_{\rho G_{\eta A_\Omega}} &=& - \rho \, \big\langle \nabla_x \mu^1_{\rho G_{\eta A_\Omega}} \big\rangle_{G_{\eta A_\Omega}}  \nonumber \\
&=& \beta \rho \big\{ \big\langle \nabla_x \Delta_x (\eta (\omega \cdot \Omega)^2 \big\rangle_{G_{\eta A_\Omega}} - \frac{1}{n} \nabla_x \Delta_x \big[ \eta + (n-1) \alpha \rho \big] \big\}. 
\label{eq:F1rhoG}
\end{eqnarray}
We compute, using the repeated index summation convention: 
\begin{eqnarray}
\partial_{x_i} \big[ \Delta_x \big( \eta (\omega \cdot \Omega)^2 \big) \big] &=& 
2 (\omega \cdot \Omega) \, \partial_{x_i} \Omega_k \, \omega_{\perp k} \, \partial^2_{x_j x_j} \eta 
+ (\omega \cdot \Omega)^2 \, \partial^3_{x_i x_j x_j} \eta \nonumber \\
&+& 4 \, \partial_{x_i} \Omega_k \, \omega_{\perp k} \, \partial_{x_j} \Omega_\ell \, \omega_{\perp \ell} \, \partial_{x_j} \eta 
+ 4 \, (\omega \cdot \Omega) \, \partial^2_{x_i x_j} \Omega_k \, \omega_k \, \partial_{x_j} \eta \nonumber \\
&+& 4 \, (\omega \cdot \Omega) \, \partial_{x_j} \Omega_k \, \omega_{\perp k} \, \partial^2_{x_i x_j} \eta 
+ 2 \, \partial_{x_j} \Omega_k \, \omega_{\perp k} \, \partial_{x_j} \Omega_\ell \, \omega_{\perp \ell} \, \partial_{x_i} \eta \nonumber \\
&+& 4 \, \partial^2_{x_i x_j} \Omega_k \, \omega_k \, \partial_{x_j} \Omega_\ell \, \omega_{\perp \ell} \, \eta 
+ 2 (\omega \cdot \Omega) \,  \partial^2_{x_j x_j} \Omega_k \, \omega_k \, \partial_{x_i} \eta \nonumber \\
&+& 2 \, \partial^2_{x_j x_j} \Omega_k \, \omega_k \, \partial_{x_i} \Omega_\ell \, \omega_{\perp \ell} \, \eta 
+ 2 (\omega \cdot \Omega) \,  \partial^3_{x_i x_j x_j} \Omega_k \, \omega_k \, \eta . 
\label{eq:computnablap}
\end{eqnarray}
Thanks to \eqref{eq:S2} and \eqref{eq:P2}, we have the following identities
$$ \langle (\omega \cdot \Omega)^2 \rangle_{G_{\eta A_\Omega}} = \frac{(n-1) S_2 + 1}{n}, \qquad \langle 1- (\omega \cdot \Omega)^2 \rangle_{G_{\eta A_\Omega}} = \frac{n-1}{n} (1-S_2). $$
Furthermore, the decomposition \eqref{eq:omdecomp} and the fact that $|\Omega|^2 = 1$, lead to the following identities
\begin{eqnarray*}
\partial_{x_i} \Omega_k \, \Omega_k &=& 0, \qquad \partial^2_{x_i x_j} \Omega_k \, \Omega_k  = - \partial_{x_i} \Omega_k \, \partial_{x_j} \Omega_k , \\
(P_{\omega^\bot})_{k \ell} \partial_{x_j} \Omega_\ell &=& \partial_{x_j} \Omega_k, \qquad \partial^2_{x_i x_j} \Omega_k \, \partial_{x_j} \Omega_k  = \frac{1}{2} \partial_{x_i} (\partial_{x_j} \Omega_k \, \partial_{x_j} \Omega_k ) \\
\partial^3_{x_i x_j x_j} \Omega_k \, \Omega_k  &=& - \partial_{x_j} ( \partial_{x_i} \Omega_k \, \partial_{x_j} \Omega_k ) - \frac{1}{2} \partial_{x_i} (\partial_{x_j} \Omega_k \, \partial_{x_j} \Omega_k ), \\
\partial^2_{x_j x_j} \Omega_k \, \partial_{x_i} \Omega_k  &=& \partial_{x_j} ( \partial_{x_i} \Omega_k \, \partial_{x_j} \Omega_k ) - \frac{1}{2} \partial_{x_i} (\partial_{x_j} \Omega_k \, \partial_{x_j} \Omega_k ). 
\end{eqnarray*}
Thus, taking the bracket $\langle  \cdot \rangle_{G_{\eta A_\Omega}}$ of \eqref{eq:computnablap}, noting that all odd powers of $\omega \cdot \Omega$ or of $\omega_\perp$ vanish by antisymmetry and using \eqref{eq:moments2} and the previous identities, we finally get
\begin{eqnarray*}
\big\langle  \nabla_x \Delta_x \big( \eta (\omega \cdot \Omega)^2 \big)  \big\rangle_{G_{\eta A_\Omega}} &=& \frac{(n-1) S_2 + 1}{n} \nabla_x \Delta_x \eta \\
&-& 2 S_2 \big( 2 \nabla_x \Omega \, \nabla_x \Omega^T + |\nabla_x \Omega|^2 \mathrm{Id} \big) \nabla_x \eta \\
&-& S_2 \nabla_x \cdot \big( 2 \nabla_x \Omega \, \nabla_x \Omega^T + |\nabla_x \Omega|^2 \mathrm{Id} \big) \eta . 
\end{eqnarray*}
Inserting this equation into \eqref{eq:F1rhoG}, using \eqref{eq:eta} and noting  that for a $n \times n$ tensor $S$ and a scalar $\varphi$, we have $\nabla_x \cdot (S \varphi) = (\nabla_x \cdot S) \varphi + S^T \nabla_x \varphi$, we get
\begin{eqnarray}
F^1_{\rho G_{\eta A_\Omega}} &=& - \frac{\beta}{\alpha} \nabla_x \cdot \Big[ \eta^2 \big( 2 \nabla_x \Omega \, \nabla_x \Omega^T + |\nabla_x \Omega|^2 \mathrm{Id} \big) \Big] \nonumber \\
&+& \beta \big\{ \frac{n-1}{n \alpha} \eta \nabla_x \Delta_x \eta - \frac{(n-1) \alpha}{n} \rho \nabla_x \Delta_x \rho \big\}. 
\label{eq:F1rhoG_comput}
\end{eqnarray}
The first of the following identities follows again from the fact that $|\Omega|^2 = 1$ and the second one is just straightforward algebra (which will also be applied with $\rho$ replacing $\eta$): 
\begin{eqnarray*}
\eta^2 \nabla_x \Omega \, \nabla_x \Omega^T &=& \nabla_x (\eta \Omega) \, \nabla_x (\eta \Omega)^T - \nabla_x \eta \otimes \nabla_x \eta, \\
\eta \nabla_x \Delta_x \eta  &=& - \nabla_x \cdot (\nabla_x \eta \otimes \nabla_x \eta) + \nabla_x \big( \eta \Delta_x \eta + \frac{1}{2} |\nabla_x \eta|^2 \big).  
\end{eqnarray*}
Inserting these identities into \eqref{eq:F1rhoG_comput}, we get $ F^1_{\rho G_{\eta A_\Omega}} = \nabla_x \cdot \sigma_E + \nabla_x \varphi$, where $\sigma_E$ is given by \eqref{eq:sigma_elas_equi} and $\varphi$ is a scalar function (different from the one appearing at the end of Section \ref{sec:extrastress_EL}) which can be absorbed in the pressure $p$. This ends the proof. \endproof

%%%%%%%%%%%%%%%%%%%%%%%%%%%%%%%%%%%%%%%%%%%%%%%%%%%%%%%%%%%%%%%%
%%%%%%%%%%%%%%%%%%%%%%%%%%%%%%%%%%%%%%%%%%%%%%%%%%%%%%%%%%%%%%%%
\subsection{Proof of the energy identity \eqref{eq:ener_EL}}
\label{sec:ener_EL}

Taking the dot product of \eqref{eq:NS_final} with $u$, integrating with respect to $x$ on ${\mathbb R}^n$ and 
using Stokes formula assuming that the spatial boundary terms vanish at infinity, we get 
\begin{equation}
 \frac{d}{dt} \int_{{\mathbb R}^n} \frac{|u|^2}{2} \, dx +  \frac{1}{\mathrm{Re}} \int_{{\mathbb R}^n} |\nabla_x u|^2 \, dx +  \frac{1}{\mathrm{Re} \mathrm{Er}} \int_{{\mathbb R}^n} (\sigma_L + \sigma_E) : \nabla_x u \, dx = 0. 
\label{eq:ddt_int_u2}
\end{equation}

We first compute the contribution of the Leslie stresses. Using the symmetry of $E$ and $\Omega \otimes \Omega$, we first have 
\begin{equation}
\big[ \alpha_1 \big( E:(\Omega \otimes \Omega) \big) \Omega \otimes \Omega + \alpha_4 E \big] : \nabla_x u = \alpha_1 \big( E:(\Omega \otimes \Omega) \big)^2 + \alpha_4 |E|^2. 
\label{eq:al1al4}
\end{equation}
Then, we remark that 
\begin{eqnarray*}
\big( (\Omega \otimes \Omega) E \big):\nabla_x u &=& |E \Omega|^2 - (E \Omega) \cdot (W \Omega), \\
\big( E (\Omega \otimes \Omega) \big):\nabla_x u &=& |E \Omega|^2 + (E \Omega) \cdot (W \Omega),
\end{eqnarray*}
which, with the second equation \eqref{eq:gamma_EL}, gives 
\begin{equation}
\big[ \alpha_5 (\Omega \otimes \Omega) E + \alpha_6 E (\Omega \otimes \Omega) \big] : \nabla_x u = (\alpha_5 + \alpha_6) |E \Omega|^2 + \gamma_2 (E \Omega) \cdot (W \Omega). 
\label{eq:al5al6}
\end{equation}
Also, with \eqref{eq:Omega_alter}, we have 
$$ 
N = - \frac{\gamma_2}{\gamma_1} P_{\Omega^\bot} E \Omega + \frac{1}{\gamma_1} P_{\Omega^\bot} H  :=N_1 + N_2. 
$$
Remarking that $(\Omega \otimes \Omega):W=0$ by the antisymmetry of $W$, we get
\begin{eqnarray*}
 (\Omega \otimes N_1): \nabla_x u &=& - \frac{\gamma_2}{\gamma_1} \big[ \Omega \otimes \big( E \Omega - (\Omega \cdot E \Omega) \Omega \big) \big]: (E+W) \\
&=& - \frac{\gamma_2}{\gamma_1} \big[ |E \Omega|^2 - \big( E:(\Omega \otimes \Omega) \big)^2 - (E \Omega) \cdot (W \Omega) \big], 
\end{eqnarray*}
and similarly
$$  (N_1 \otimes \Omega): \nabla_x u = - \frac{\gamma_2}{\gamma_1} \big[ |E \Omega|^2 - \big( E:(\Omega \otimes \Omega) \big)^2 + (E \Omega) \cdot (W \Omega) \big], $$
which, using \eqref{eq:gamma_EL} gives 
\begin{equation}
\big[ \alpha_2 \Omega \otimes N_1 + \alpha_3 N_1 \otimes \Omega \big] : \nabla_x u = - \frac{\gamma_2^2}{\gamma_1} \big[ |E \Omega|^2 - \big( E:(\Omega \otimes \Omega) \big)^2 \big] - \gamma_2 (E \Omega) \cdot (W \Omega). 
\label{eq:al2al3N1}
\end{equation}
Then, using \eqref{eq:gamma_EL}, we compute 
\begin{eqnarray}
&&\hspace{-1cm}
\big[ \alpha_2 \Omega \otimes N_2 + \alpha_3 N_2 \otimes \Omega \big] : \nabla_x u = \frac{1}{\gamma_1} \big[ \alpha_2 \Omega \otimes P_{\Omega^\bot} H + \alpha_3 P_{\Omega^\bot} H \otimes \Omega \big] : (E+W) \nonumber \\
&&\hspace{0cm}
= \frac{1}{\gamma_1} \big[ (\alpha_2 + \alpha_3) (P_{\Omega^\bot} H \otimes \Omega):E + (\alpha_3  - \alpha_2) (P_{\Omega^\bot} H \otimes \Omega):W \big] \nonumber \\
&&\hspace{0cm}
= P_{\Omega^\bot} H \cdot \big[ \frac{\gamma_2}{\gamma_1} E \Omega + W \Omega \big] = P_{\Omega^\bot} H \cdot P_{\Omega^\bot} \big[ \frac{\gamma_2}{\gamma_1} E \Omega + W \Omega \big] \nonumber \\
&&\hspace{0cm}
= \frac{1}{\gamma_1} |P_{\Omega^\bot} H |^2 - H \cdot \big( \partial_t \Omega + u \cdot \nabla_x \Omega), 
\label{eq:al2al3N2}
\end{eqnarray}
where, for the last equality, we have used \eqref{eq:Omega_alter} and \eqref{eq:defN} and the fact that $\partial_t \Omega + u \cdot \nabla_x \Omega$ is normal to $\Omega$. Then, collecting \eqref{eq:al1al4} to \eqref{eq:al2al3N2} and using \eqref{eq:sigmae_equi} leads to
\begin{eqnarray}
&&\hspace{-1cm}
\int_{{\mathbb R}^n} \sigma_L : \nabla_x u \, dx = \int_{{\mathbb R}^n} \rho \Big\{ \Big( \alpha_1 + \frac{\gamma_2^2}{\gamma_1} \Big) \big( E:(\Omega \otimes \Omega) \big)^2 + \alpha_4 |E|^2 \nonumber \\
&&\hspace{0cm}
+ \Big( \alpha_5 + \alpha_6 - \frac{\gamma_2^2}{\gamma_1} \Big) |E \Omega|^2 + \frac{1}{\gamma_1} | P_{\Omega^\bot} H |^2 - H \cdot \big( \partial_t \Omega + u \cdot \nabla_x \Omega) \Big\} \, dx . 
\label{eq:int_sigL_nau}
\end{eqnarray}

Expression \eqref{eq:sigma_elas_equi} for the Ericksen stresses involves three terms which we will denote by $\sigma_E^\Omega$, $\sigma_E^\eta$, $\sigma_E^\rho$ in the order in which they appear in this expression. We compute the contribution of each term successively. We have, using Stokes's formula, \eqref{eq:rel_H_EF}, \eqref{eq:divu=0_final} and assuming that the boundary terms vanish at infinity:
\begin{eqnarray} 
&&\hspace{-0.5cm}
\int_{{\mathbb R}^n} \sigma_E^\Omega : \nabla_x u \, dx =  - \frac{2 \beta}{\alpha} \int_{{\mathbb R}^n}  \big( \nabla_x (\eta \Omega) \nabla_x (\eta \Omega)^T \big):\nabla_x u  \, dx \nonumber \\
&&\hspace{-0.5cm}
= \frac{2 \beta}{\alpha} \int_{{\mathbb R}^n} \Big\{ \Delta_x (\eta \Omega) \cdot \big( (u \cdot \nabla_x) (\eta \Omega) \big) + \nabla_x(\eta \Omega) \big( (u \cdot \nabla_x) ( \nabla_x (\eta \Omega))^T \big) \Big\} \, dx \nonumber \\
&&\hspace{-0.5cm}
= \frac{2 \beta}{\alpha} \int_{{\mathbb R}^n} \Big\{ \eta \Delta_x (\eta \Omega) \cdot \big( (u \cdot \nabla_x) \Omega \big) + \Delta_x (\eta \Omega) \cdot \Omega \, (u \cdot \nabla_x) \eta +  \nabla_x \cdot \Big( u \frac{|\nabla_x (\eta \Omega)|^2}{2} \Big) \Big\} \, dx \nonumber \\
&&\hspace{-0.5cm}
= \int_{{\mathbb R}^n} \rho H \cdot \big( (u \cdot \nabla_x) \Omega \big) \, dx  + \frac{2 \beta}{\alpha} \int_{{\mathbb R}^n} \Delta_x (\eta \Omega) \cdot \Omega \, (u \cdot \nabla_x) \eta \, dx.
\label{eq:int_sigOm_nau}
\end{eqnarray}
A similar computation gives  
\begin{eqnarray}
\int_{{\mathbb R}^n} \sigma_E^\eta : \nabla_x u \, dx &=& -  \frac{(n+1) \beta}{n \alpha} \int_{{\mathbb R}^n} \Delta_x \eta \, (u \cdot \nabla_x)  \eta \, dx, 
\label{eq:int_sigeta_nau} \\
\int_{{\mathbb R}^n} \sigma_E^\rho : \nabla_x u \, dx &=&  - \frac{(n-1) \alpha \beta}{n} \int_{{\mathbb R}^n} \Delta_x \rho \, (u \cdot \nabla_x)  \rho \, dx. 
\label{eq:int_sigrho_nau}
\end{eqnarray}

Now, we consider the Oseen-Franck energy and successively compute the time derivative of each of the terms in \eqref{eq:Franck_ener}. We first have, thanks to Stokes's formula: 
$$
\frac{d {\mathcal E}_F^\Omega}{dt} = \frac{2 \beta}{\alpha} \int_{{\mathbb R}^n} \mathrm{Tr} \big\{ \nabla_x (\eta \Omega) \big( \partial_t \nabla_x (\eta \Omega) \big)^T \big\} \, dx  = - \frac{2 \beta}{\alpha} \int_{{\mathbb R}^n} \Delta_x (\eta \Omega) \cdot \partial_t (\eta \Omega) \, dx, 
$$
With\eqref{eq:rel_H_EF}, this leads to:

\begin{equation}
\frac{d {\mathcal E}_F^\Omega}{dt} + \int_{{\mathbb R}^n} \rho H \cdot \partial_t \Omega\, dx + \frac{2 \beta}{\alpha} \int_{{\mathbb R}^n} \Delta_x (\eta \Omega) \cdot \Omega \, \partial_t \eta \, dx = 0. 
\label{eq:dtEFOm}
\end{equation}
Straightforwardly, we get 
\begin{eqnarray}
\frac{d {\mathcal E}_F^\eta}{dt} - \frac{(n+1) \beta}{n \alpha} \int_{{\mathbb R}^n} \Delta_x \eta \cdot  \, \partial_t \eta \, dx &=& 0, \label{eq:dtEFeta} \\
\frac{d {\mathcal E}_F^\rho}{dt} - \frac{(n-1) \alpha \beta}{n} \int_{{\mathbb R}^n} \Delta_x \rho \cdot  \, \partial_t \rho \, dx &=& 0. 
\label{eq:dtEFrho}
\end{eqnarray}

Now, adding \eqref{eq:ddt_int_u2}, \eqref{eq:dtEFOm}, \eqref{eq:dtEFeta}, \eqref{eq:dtEFrho} together, using \eqref{eq:int_sigL_nau}, \eqref{eq:int_sigOm_nau}, \eqref{eq:int_sigeta_nau}, \eqref{eq:int_sigrho_nau} to eliminate $\sigma_L$ and $\sigma_E$ and finally using that $D_t \rho = 0$ and $D_t \eta = \frac{d \eta}{d \rho} D_t \rho = 0$, we get Eq.~\eqref{eq:ener_EL}. \endproof

%%%%%%%%%%%%%%%%%%%%%%%%%%%%%%%%%%%%%%%%%%%%%%%%%%%%%%%%%%%%%%%%
%%%%%%%%%%%%%%%%%%%%%%%%%%%%%%%%%%%%%%%%%%%%%%%%%%%%%%%%%%%%%%%%
%%%%%%%%%%%%%%%%%%%%%%%%%%%%%%%%%%%%%%%%%%%%%%%%%%%%%%%%%%%%%%%%
%%%%%%%%%%%%%%%%%%%%%%%%%%%%%%%%%%%%%%%%%%%%%%%%%%%%%%%%%%%%%%%%
\setcounter{equation}{0}
\section{Appendix to Section \ref{sec:GCI} on GCI} 
\label{sec:app_GCI}

\subsection{Proof of Proposition \ref{prop:altern_express_h}} 
\label{sec:proofpropalternh}

We first note that Eq. \eqref{eq:eq_eta} which defines $h_\eta$ can be alternately written as (dropping the index $\eta$ for simplicity):
\begin{equation}
(1-r^2) \, h'' + \big( 2 \eta (1 - r^2) - (n+1) \big) \, r h' - \big( 2 \eta r^2 + n - 1 \big) \, h = r. 
\label{eq:equationh}
\end{equation}
With \eqref{eq:defg}, we have 
$$ h(r) = - \frac{1}{2 \eta} \frac{1}{\sqrt{1 - r^2}} \, g (\cos^{-1} r). $$
Then, 
\begin{eqnarray*}
h'(r) &=& - \frac{1}{2 \eta} \Big[ - \frac{g' (\cos^{-1} r)}{1-r^2} + \frac{r \, g (\cos^{-1} r)}{(1-r^2)^{3/2}} \Big], \\
h''(r) &=& - \frac{1}{2 \eta} \Big[ \frac{g'' (\cos^{-1} r)}{(1-r^2)^{3/2}} - \frac{3r \, g' (\cos^{-1} r)}{(1-r^2)^2} + \frac{(1+2r^2) \, g (\cos^{-1} r)}{(1-r^2)^{5/2}} \Big].
\end{eqnarray*}
Inserting these expressions in \eqref{eq:equationh} and changing $r$ into $\cos \theta$, we get
\begin{equation}
g'' + \frac{\cos \theta \, \big( n-2 - 2 \eta \sin^2 \theta \big)}{\sin \theta} \, g' - \frac{n-2}{\sin^2 \theta} \, g = - 2 \eta \, \cos \theta \, \sin \theta, 
\label{eq:equationg}
\end{equation}
But, we have 
$$ \frac{1}{\sin^{n-2} \theta} \big( \sin^{n-2} \theta \, g' \big)' = g'' + (n-2) \frac{\cos \theta}{\sin \theta} g'. $$
With this and \eqref{eq:dU0dthet}, we realize that \eqref{eq:equationg} is nothing but \eqref{eq:g}. \endproof

%%%%%%%%%%%%%%%%%%%%%%%%%%%%%%%%%%%%%%%%%%%%%%%%%%%%%%%%%%%%%%%%
%%%%%%%%%%%%%%%%%%%%%%%%%%%%%%%%%%%%%%%%%%%%%%%%%%%%%%%%%%%%%%%%
\subsection{Proof of Lemma \ref{lem:a1a2a3}} 
\label{sec:prooflem:a1a2a3}

We use the same notations as in the proof of Lemma \ref{lem:Q4_equi}. From \eqref{eq:Q4_comput_2}, we get 
\begin{empheq}[left=\empheqlbrace]{align}
a_1 & = \langle X^4 \rangle - \frac{6 \, \langle X^2 (1 - X^2) \rangle}{n-1}  + \frac{3\, \langle (1 - X^2)^2 \rangle}{(n-1)(n+1)} , \label{eq:defa1} \\
a_2 &= \frac{\langle X^2 (1 - X^2) \rangle}{n-1}   - \frac{\langle (1 - X^2)^2 \rangle}{(n-1)(n+1)}  , \label{eq:defa2} \\
a_3 &= \frac{\langle (1 - X^2)^2 \rangle}{(n-1)(n+1)}  ,  \nonumber
\end{empheq}
Thus, with the change to spherical coordinates used in the proof of Prop. \ref{prop:gibbs_uniaxial}, an integration by parts, and Eqs. \eqref{eq:S2}, \eqref{eq:P2} and \eqref{eq:eta}, we get 
\begin{eqnarray} 
a_2 + a_3 &=& \frac{\langle X^2 (1 - X^2) \rangle}{n-1} = \frac{C_n}{(n-1) Z_\eta} \int_0^\pi e^{\eta \cos^2 \theta} \cos^2 \theta \, \sin^n \theta \, d \theta \nonumber \\
&=& \frac{C_n}{2 (n-1) \eta Z_\eta} \int_0^\pi e^{\eta \cos^2 \theta} (n\cos^2 \theta -1) \, \sin^{n-2} \theta \, d \theta \nonumber \\
&=& \frac{1}{2 \eta} \frac{\langle n X^2 -1 \rangle}{n-1} = \frac{S_2(\eta)}{2\eta} = \frac{1}{2 \alpha \rho}, \label{eq:a2+a3}
\end{eqnarray}
which shows the equality in \eqref{eq:alrhoa23}. 

Now, we have, thanks to \eqref{eq:a2+a3}%
$$
(n-1)(n+1) a_2 = (n+2) \langle X^2 (1-X^2) \rangle - \langle 1-X^2 \rangle = \frac{n+2}{2 \eta} \langle n X^2 - 1 \rangle - \langle 1-X^2 \rangle. 
$$
Thanks to \eqref{eq:S2}, \eqref{eq:P2}, we have $\langle X^2 \rangle = \frac{1}{n} (1 + (n-1) S_2(\eta))$. So,
\begin{eqnarray*} 
2 n (n+1) \eta a_2 &=& \frac{n}{n-1} \big[ \big( n(n+2) + 2 \eta \big) \langle X^2 \rangle - (n+2+2\eta) \big] \\
&=&  ( n(n+2) + 2 \eta)  S_2 (\eta) - 2 \eta. 
\end{eqnarray*}
Thus, with the change to spherical coordinates used in the proof of Prop. \ref{prop:gibbs_uniaxial}, we have
\begin{eqnarray*} 
&&\hspace{-1cm}
2 n (n-1) (n+1) \eta Z_\eta C_n^{-1} \, a_2 = \\
&&\hspace{0cm}
= \int_0^\pi e^{\eta \cos^2 \theta} \, \Big[ ( n(n+2) + 2 \eta) \, (n \cos^2 \theta - 1) - 2 (n-1) \eta \big] \sin^{n-2} \theta \, d \theta  \\
&&\hspace{0cm}
= \int_0^\pi e^{\eta \cos^2 \theta} \, \Big[ n(n+2) (n \cos^2 \theta - 1) + 2 n \eta (\cos^2 \theta - 1) \big] \sin^{n-2} \theta \, d \theta  \\
&&\hspace{0cm}
= \int_0^\pi e^{\eta \cos^2 \theta} \, \Big[ 2 n(n+2) \eta \cos^2 \theta \, \sin^2 \theta - 2 n \eta \sin^2 \theta \big] \sin^{n-2} \theta \, d \theta  \\
&&\hspace{0cm}
= 2 \eta n \int_0^\pi e^{\eta \cos^2 \theta} \, \big( (n+2) \cos^2 \theta - 1 \big) \sin^n \theta \, d \theta 
\end{eqnarray*}
The passage between the third and fourth lines uses the same integration by parts as in \eqref{eq:a2+a3}. The other equalities are just simple algebraic rearrangements. Comparing with \eqref{eq:S2}, \eqref{eq:P2}, we notice that the integral of the last line is equal to the quantity $S_2^{(n+2)}$ which is the quantity $S_2$ in dimension $n+2$ up to a prefactor $(n+1) Z_\eta C_n^{-1}$. Thus, we have $a_2 = S_2^{(n+2)}/(n-1)$. Now, we can apply Prop. \ref{prop:OP} (iii) and conclude that $0 < a_2 < \frac{1}{n-1}$. In particular, $a_2 \not = 0$, which finishes to show \eqref{eq:alrhoa23}. 

Finally, it is a simple algebra, using \eqref{eq:defa1} and \eqref{eq:defa2} to show that 
$$ 
a_1 + (n+4) a_2 = \frac{\langle n X^2 - 1 \rangle}{n-1} = S_2(\eta), $$
showing \eqref{eq:a1A2}. This ends the proof. \endproof

%%%%%%%%%%%%%%%%%%%%%%%%%%%%%%%%%%%%%%%%%%%%%%%%%%%%%%%%%%%%%%%%
%%%%%%%%%%%%%%%%%%%%%%%%%%%%%%%%%%%%%%%%%%%%%%%%%%%%%%%%%%%%%%%%
\subsection{Proof of Prop. \ref{prop:CnonequiisnotkerDC*}.} 
\label{sec:CnonequiisnotkerDC*}

Let $f = \rho G_{\eta A_\Omega}$ with $\eta \not = \eta(\rho)$. This means that \eqref{eq:eta} is not satisfied. In other words, 
\begin{equation}
\eta' = \alpha \rho S_2(\eta) \not = \eta.
\label{eq:eta'not=eta}
\end{equation} 
From \eqref{eq:Q_equib}, it follows that $\alpha \rho Q_{\rho G_{\eta A_\Omega}} = \eta' A_\Omega$. So, with \eqref{eq:adjlinear_gene}, we get
\begin{equation}
D_{\rho G_{\eta A_\Omega}}C^* g (\omega) = L_{\eta' A_\Omega}^* g (\omega) - \alpha \rho \,  (\rho Q)_{G_{\eta A_\Omega} L_{\eta A_\Omega}^* g} : \omega \otimes \omega  .
\label{eq:adjlinear_gene_CE}
\end{equation}
Suppose that $g$ is a GCI associated with $(\eta, A_\Omega)$. Then, by \eqref{eq:GCI1}, there exists $V \in \{\Omega\}^\bot$ such that 
\begin{equation}
G_{\eta A_\Omega} L_{\eta A_\Omega}^* g = (\omega \cdot \Omega) \, (\omega \cdot V) \, G_{\eta A_\Omega}.  
\label{eq:GCI1_CE}
\end{equation}
By a similar computation (using the same notations) to what was done in the proof of Lemma \ref{lem:equivX}, we get
\begin{eqnarray}
\alpha \rho \,  (\rho Q)_{G_{\eta A_\Omega} L_{\eta A_\Omega}^* g} : \omega \otimes \omega &=& 2 \alpha \rho (a_2 + a_3) (\omega \cdot \Omega) \, (\omega \cdot V) \nonumber \\
&=& \frac{\eta'}{\eta} (\omega \cdot \Omega) \, (\omega \cdot V).  
\label{eq:rhoQ_CE}
\end{eqnarray}
For the second equality, we have used that $a_2 + a_3 = \frac{S_2(\eta)}{\eta}$ (see the proof of Lemma \ref{lem:a1a2a3} in Appendix \ref{sec:prooflem:a1a2a3}) and \eqref{eq:eta'not=eta}. 
On the other hand, simple algebraic manipulations and the use of \eqref{eq:GCI1_CE} show that 
\begin{eqnarray}
L_{\eta' A_\Omega}^* g (\omega) &=& L_{\eta A_\Omega}^* g (\omega) + 2 (\eta' - \eta) (\omega \cdot \Omega) P_{\omega^\bot} \Omega \cdot \nabla_\omega g \nonumber \\
&=& (\omega \cdot \Omega) \, (\omega \cdot V) + 2 (\eta' - \eta) (\omega \cdot \Omega) P_{\omega^\bot} \Omega \cdot \nabla_\omega g. \label{eq:Leta'AOM}
\end{eqnarray}
Inserting \eqref{eq:rhoQ_CE} and \eqref{eq:Leta'AOM} into \eqref{eq:adjlinear_gene_CE} gives
$$
D_{\rho G_{\eta A_\Omega}}C^* g (\omega) = (\eta' - \eta) (\omega \cdot \Omega) \Big[ - \frac{1}{\eta} (\omega \cdot V) + 2 P_{\omega^\bot} \Omega \cdot \nabla_\omega g \Big]  .
$$
Suppose now that $g$ is also an element of ker$\, (D_{\rho G_{\eta A_\Omega}}C^*)$. This implies that
\begin{equation} 
2 P_{\omega^\bot} \Omega \cdot \nabla_\omega g = \frac{1}{2 \eta} (\omega \cdot V). 
\label{eq:gintersect}
\end{equation}
From now on, we restrict to dimension $n=3$ and use the spherical coordinates $(\theta, \varphi)$ associated to the cartesian basis $(V, W, \Omega)$  with pole at $\Omega$ (defining $W = \Omega \times V$, using the symbol $\times$ for the cross product). In these coordinates, \eqref{eq:gintersect} is written in terms of $\tilde g (\theta, \varphi) = g (\omega)$ according to
$$ \partial_\theta \tilde g =  \frac{1}{2 \eta} \, \cos \varphi. $$
Thus,
$$
\tilde g(\theta, \varphi) = \frac{1}{2 \eta} \, \theta \, \cos \varphi + h(\varphi), 
$$
where $h$ is an arbitrary function. The smoothness of $g$ at $\omega = \Omega$ requires $h=0$. However, we see that $g$ cannot be smooth at $\omega = - \Omega$ (i.e. for $\theta = \pi$) because the function $\theta \, \cos \varphi$ does not tend to a constant when $\theta \to \pi$. However, by the elliptic regularity theorem, $g \in C^\infty({\mathbb S}^{n-1})$. This is a contradiction. This means that the only possible solution is when $V=0$, i.e. 
$$
{\mathcal C}_{\eta A_\Omega} \cap \mathrm{ker}\, (D_{\rho G_{\eta A_\Omega}}C^*) = \{0\}.
$$ 
Since ${\mathcal C}_{\eta A_\Omega} \not = \{0\}$, this shows \eqref{eq:CnonequiisnotkerDC*} (with $\Sigma = A_{\Omega}$) and ends the proof. \endproof

%%%%%%%%%%%%%%%%%%%%%%%%%%%%%%%%%%%%%%%%%%%%%%%%%%%%%%%%%%%%%%%%
%%%%%%%%%%%%%%%%%%%%%%%%%%%%%%%%%%%%%%%%%%%%%%%%%%%%%%%%%%%%%%%%
%%%%%%%%%%%%%%%%%%%%%%%%%%%%%%%%%%%%%%%%%%%%%%%%%%%%%%%%%%%%%%%%
%%%%%%%%%%%%%%%%%%%%%%%%%%%%%%%%%%%%%%%%%%%%%%%%%%%%%%%%%%%%%%%%
\setcounter{equation}{0}
\section{Appendix to Section \ref{sec:omega_abstract} on the derivation of the equation for $\Omega$} 
\label{sec:app_deriv_Om}

\subsection{Proof of Eq. \eqref{eq:gamalter}} 
\label{sec:proofgamalter}

We first consider $\tilde \gamma_3$. With \eqref{eq:gamma3}, \eqref{eq:gamma2} and \eqref{eq:gamma1}, and using the spherical coordinates and the notations $C_n$ and $Z_\eta$ described in the proof of Prop. \ref{prop:gibbs_uniaxial} as well as the change $r = \cos \theta$, we have (dropping the indices $\eta A_\Omega$ to $G$ and $\eta$ to $h$ for simplicity):
\begin{eqnarray}
\tilde \gamma_3 &=& \frac{\rho}{n-1} \int_{{\mathbb S}^{n-1}} G \, h \, (\omega \cdot \Omega) \big(1 - (\omega \cdot \Omega)^2 \big) \Big( 2 \eta \big(1 - 2 (\omega \cdot \Omega)^2 \big) - 2n \Big) \, d \omega \nonumber \\
&=& \frac{\rho}{n-1} \frac{C_n}{Z_\eta} \int_0^1 (1-r^2)^{\frac{n-1}{2}} \, e^{\eta r^2} \, r \, \big( 2 \eta (1-2r^2) -2n \big) \, h \, dr, 
\label{eq:gam3_comput1}
\end{eqnarray}
Besides, multiplying Eq. \eqref{eq:eq_eta} by $r$,  integrating with respect to $r \in [0,1]$, and noting that, thanks to two successive integration by parts we have 
$$ \int_0^1 r \, \big( (1-r^2)^{\frac{n+1}{2}} \, e^{\eta r^2} \, h' \big)' dr = \int_0^1 (1-r^2)^{\frac{n-1}{2}} \, e^{\eta r^2} \, r \, \big( 2 \eta (1-r^2) -(n+1) \big) \, h \, dr, $$
we get
\begin{equation}
\int_0^1 (1-r^2)^{\frac{n-1}{2}} \, e^{\eta r^2} \, r \, \big( 2 \eta (1-2r^2) -2n \big) \, h \, dr =  \int_0^1 (1-r^2)^{\frac{n-1}{2}} \, e^{\eta r^2} \, r^2 \, dr. 
\label{eq:gam3_comput2}
\end{equation}
Inserting \eqref{eq:gam3_comput2} into \eqref{eq:gam3_comput1} and integrating by parts once more, we get
\begin{eqnarray*}
\tilde \gamma_3 &=& \frac{\rho}{n-1} \frac{C_n}{Z_\eta} \int_0^1 (1-r^2)^{\frac{n-1}{2}} \, e^{\eta r^2} \, r^2 \, dr \\
&=& \frac{\rho}{n-1} \frac{C_n}{Z_\eta} \frac{1}{2 \eta} \int_0^1 (1-r^2)^{\frac{n-3}{2}} \, e^{\eta r^2} \, (n r^2 - 1) \, dr \\
&=&  \frac{\rho}{2 \eta} \int_{{\mathbb S}^{n-1}} G \, \frac{n (\omega \cdot \Omega)^2 - 1}{n-1} \, d \omega = \frac{\rho \, \big\langle P_2(\omega \cdot \Omega) \big\rangle_G}{2 \eta} = \frac{\rho \, S_2(\eta)}{2 \eta}, 
\end{eqnarray*}
where, in the last line, we have reverted back to the variable $\omega$ and used \eqref{eq:S2} and \eqref{eq:P2}. This shows the first equation in Formula \eqref{eq:gamalter}.

We now consider $\tilde \gamma_1$. Changing to spherical coordinates in \eqref{eq:gamma1}, we get
$$
\tilde \gamma_1 = \frac{2 \eta \rho}{n-1} \frac{C_n}{Z_\eta} \int_0^\pi e^{\eta \cos^2 \theta} \, h(\cos \theta) \, \cos \theta \, \sin^n \theta \, d \theta. 
$$
Using \eqref{eq:defg} and \eqref{eq:dU0dthet}, this can be changed into
$$
\tilde \gamma_1 = - \frac{\rho}{2 \eta (n-1)} \frac{C_n}{Z_\eta} \int_0^\pi e^{\eta \cos^2 \theta} \, g(\theta) \, \frac{d \tilde U_0}{d \theta} \, \sin^{n-2} \theta \, d \theta \\
$$
But from \eqref{eq:GetaAom}, we have $ \frac{Z_\eta}{C_n} = \int_0^\pi e^{\eta \cos^2 \theta} \, \sin^{n-2} \theta \, d \theta$, which leads to the second equation in Formula \eqref{eq:gamalter}. \endproof

\end{document}